\documentclass[showpacs,preprintnumbers,amsmath,amssymb,twocolumn,nofootinbib,nobibnotes,longbibliography]{revtex4-1}
\usepackage{color}
\usepackage[colorlinks,urlcolor=blue]{hyperref}
\usepackage{epstopdf}
\usepackage{graphicx}
\usepackage[normalem]{ulem}

\usepackage[english]{babel}
\usepackage{tabularx}

\usepackage{dcolumn}
\usepackage{multirow}
\usepackage{bm}
\usepackage{upgreek}
\usepackage{physics}

\begin{document}

\title{Valley and spin accumulation in ballistic and hydrodynamic channels
}

\author{M.~M.~Glazov}
\affiliation{Ioffe Institute,  	194021 St.~Petersburg, Russia}


\begin{abstract}
A theory of the valley and spin Hall effects and resulting accumulation of the valley and spin polarization is developed for ultraclean  channels made of two-dimensional semiconductors where the electron mean free path due to the residual disorder or phonons exceeds the channel width. Both ballistic and hydrodynamic regimes of the electron transport are studied. The polarization accumulation is determined by interplay of the anomalous velocity, side-jump and skew scattering effects. In the hydrodynamic regime, where the electron-electron scattering is dominant, the valley and spin current generation and dissipation by the electron-electron collisions are taken into account. The accumulated  polarization magnitude and its spatial distribution depend strongly on the transport regime. The polarization is much larger in the hydrodynamic regime as compared to the ballistic one. Significant valley and spin polarization arises in the immediate vicinity of the channel edges due to the side-jump and skew scattering mechanisms.
\end{abstract}

\maketitle

\section{Introduction}\label{sec:intro}

Transport of spin and valley degrees of freedom is of particular interest owing to fascinating fundamental physics~\cite{dyakonov_book,dyakonov71a,dyakonov71,hirsch99,murakami03,sinova04,wunderlich05,kavokin05prl,Leyder:2007ve,Mak27062014,Ubrig:2017aa,Lundt:2019aa}. Spin-orbit interaction drives anomalous transport where the  electric current transversal to the external electric field is generated in spin-polarized media~\cite{Hall:1881aa,RevModPhys.82.1539} and related phenomena, particularly, the spin Hall effect (SHE) and valley Hall effect (VHE). Because of the VHE, the particles in different valleys of the Brillouin zone propagate in opposite directions and accumulate at the opposite edges of the sample. The VHE can be used for non-magnetic valley manipulation.

The rise of a novel material system, transition metal dichalcogenide monolayers~\cite{Mak:2010bh,Splendiani:2010a,2053-1583-2-2-022001,Kolobov2016book,RevModPhys.90.021001} with multivalley band structure, spin-valley locking and chiral selection rules for optical transitions~\cite{Xiao:2012cr,Mak:2012qf,Xu:2014cr} boosted experimental and theoretical studies of the VHE~\cite{Xiao:2012cr,Mak27062014,PhysRevB.90.075430,Jin893,Onga:2017aa,Unuchek:2019aa,Lundt:2019aa,PhysRevLett.120.207401,PhysRevLett.122.256801,Ubrig:2017aa,2020arXiv200405091G,Glazov2020b}. The microscopic mechanisms of the effect have been debated in the literature until recently. However, in Refs.~\cite{2020arXiv200405091G,Glazov2020b} we have shown that regardless the origin of the drag force -- caused by the real or synthetic electric field, phonon or photon drag, or the Seebeck effect -- which results in the direct flow of the particles, the VHE is related solely to the scattering. Namely, the asymmetric or skew scattering by impurities and phonons and the side-jumps in the course of the scattering result in the valley current transversal to the direct current and eventually lead to the valley accumulation at the sample edges. The anomalous velocity arising in the presence of the external potential field~\cite{RevModPhys.82.1959,PhysRevB.78.205201,PhysRevLett.101.106401,Xiao:2012cr,PhysRevLett.115.166804,Onga:2017aa,PhysRevLett.122.256801,PhysRevB.100.121405,Gianfrate:2020aa}, e.g., where the current is induced by the electric field, is compensated by the part of the side-jump contribution, that is by the anomalous velocity resulting from defect or phonon-induced force field. Similar compensation takes place for the spin Hall effect in semiconductors as well~\cite{dyakonov_book,PhysRevB.75.045315,Sinitsyn_2007,Ado_2015}.

Recent progress in nanotechnology has made it possible to create two-dimensional electron systems with ultrahigh mobility where the electron mean free path $l$ exceeds by far the width of the channel $w$~\cite{PhysRevB.51.13389,PhysRevLett.111.166601,Bandurin1055,Moll1061,Gusev:2018tg,Krishna-Kumar:2017wn,Ku:2020ue,Sulpizio:2019uc,Berdyugin162,Gusev:2020vd}. In this case, the electron momentum dissipation takes place mainly at the channel edges. Moreover, the electron-electron interaction can give rise to a novel -- hydrodynamic  -- regime of the electron transport where electrons behave collectively as a viscous fluid~\cite{gurzhi63,Gurzhi_1968,PhysRevLett.103.025301,PhysRevLett.106.256804,PhysRevB.92.165433,PhysRevB.96.195401,PhysRevB.91.035414,PhysRevLett.117.166601,Levitov:2016aa,PhysRevB.97.205129,PhysRevB.95.115425}, see Refs.~\cite{Narozhny:2017vc,Lucas2018} for review and Refs.~\cite{PhysRevB.97.085109,PhysRevB.98.165412,PhysRevB.100.125419} for discussion of other non-diffusive regimes of the electron transport in ultraclean channels. Description of anomalous transport in such ultrahigh mobility systems is a pressing problem. In recent works~\cite{PhysRevB.100.115401,PhysRevB.103.125106,funaki2021vorticityinduced,tatara1} some aspects of the electron anomalous transport have been addressed, however, the consistent theory of the VHE and SHE in the ballistic and hydrodynamic regimes is absent to the best of our knowledge. Our paper aims to fill this gap.

Here we present the theory of the valley and spin Hall effects and accumulation of polarization in electron channels with long mean free path, $l \gg w$, addressing both ballistic (where electron-electron collisions are unimportant) and hydrodynamic (where the electron-electron scattering length $l_{ee} \ll w$) transport regimes. We calculate all contributions to the VHE and SHE: skew-scattering, side-jump, and anomalous velocity, and discuss their interplay. We uncover the role of the scattering by the sample edges  and introduce the impurity stripe model  to address the edge scattering microscopically. We study a role of electron-electron collisions in the valley current generation and dissipation. We demonstrate that different mechanisms of the VHE and SHE can manifest themselves in the different parts of the channel. 

The paper is organized as follows. After brief introduction in Sec.~\ref{sec:intro}, we present in Sec.~\ref{sec:normal} the basic description of the normal electron transport in ultraclean channels outlining the ballistic and hydrodynamic regimes. We also present a model of the impurity stripes (Sec.~\ref{sec:imp}) enabling simple  microscopic approach to the edge scattering. Section~\ref{sec:vhe} contains the theory of the VHE and SHE and  polarization accumulation inside the channel. Both ballistic (Sec.~\ref{sec:ball}) and hydrodynamic (Sec.~\ref{sec:hydro}) regimes are addressed. The electron-electron scattering effects on the spin and valley current relaxation and generation are addressed in Sec.~\ref{sec:ee}. In Sec.~\ref{sec:imp:str} the spin and valley accumulation in the impurity stripes is studied. The obtained results are summarized in Sec.~\ref{sec:concl} and the paper is concluded with Sec.~\ref{sec:concl:new}.

\section{Electron transport in ultraclean channels}\label{sec:normal}

This section contains preliminaries about the normal electron transport in narrow channels. In particular, we present the model of the diffusive scattering at the channel edges and calculate the electron distribution function in the presence of the longitudinal electric field and the longitudinal conductivity both in the ballistic and in  hydrodynamic regimes.

\begin{figure}[t]
\includegraphics[width=\linewidth]{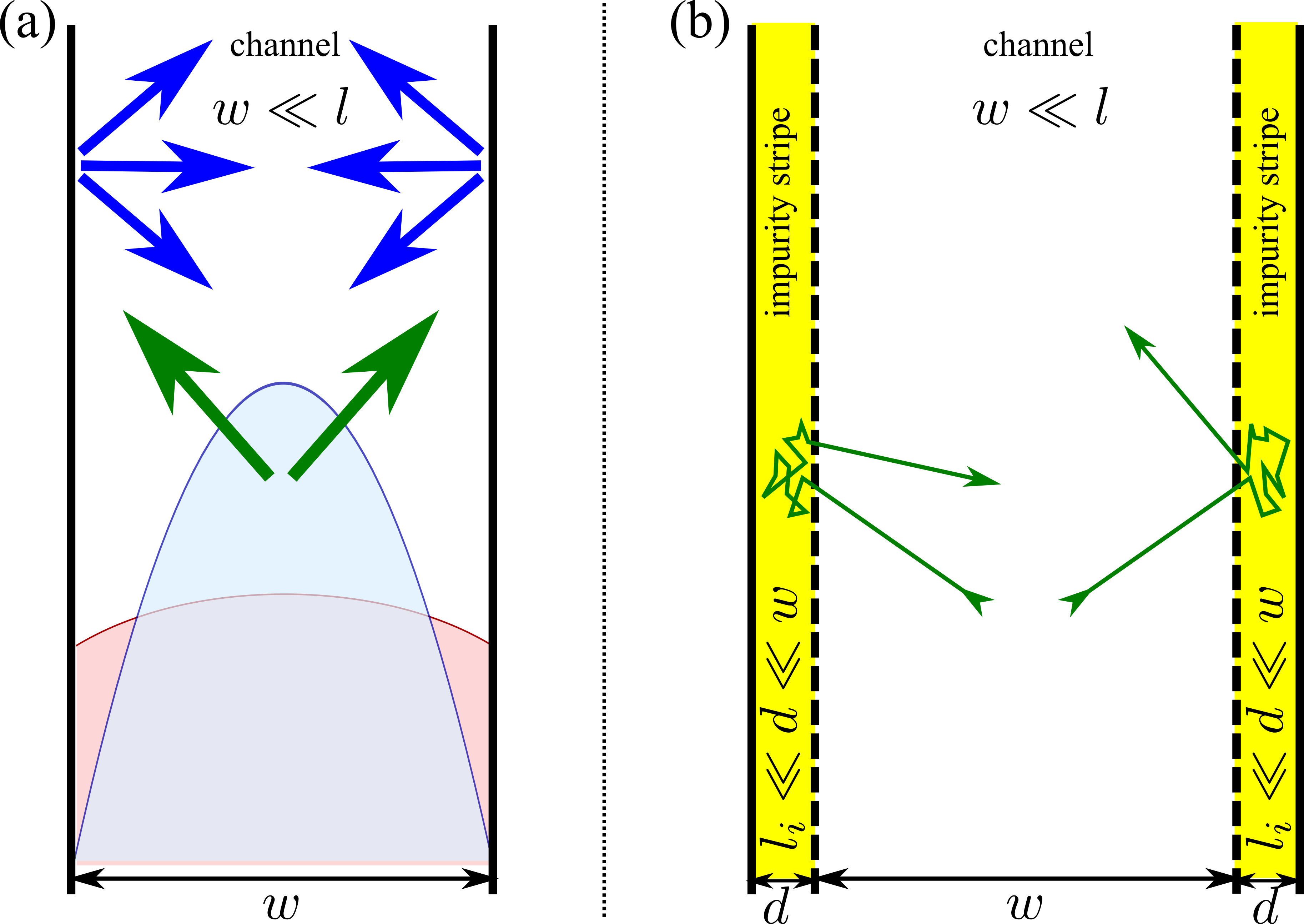}
\caption{(a) Schematic illustration of the narrow channel and diffusive scattering at the edges. Green arrows show the directions of incident electron velocities and blue arrows show the directions of the scattered electrons velocities. Light red and blue profiles show the steady-state distribution of the electron velocity in the ballistic and hydrodynamic regimes, respectively. (b) Model of the impurity stripes shown by yellow areas at the channel edges. Green lines show typical electron trajectories: ballistic inside the channel and diffusive in the impurity stripes.}\label{fig:channel}
\end{figure}

\subsection{Diffusive scattering}

We consider a channel with the two-dimensional electron gas, Fig.~\ref{fig:channel}(a). The motion of electrons is bounded along the $x$-axis and the motion along the $y$-axis is free. We assume that the channel width $w$ is by far smaller than the electron mean free path $l$ caused by the scattering by a static short-range disorder. In this regime, the main source of the electron momentum relaxation is the scattering by the edges of the channel. Usually, a phenomenological description of diffusive edge scattering is employed where the following boundary condition for the electron distribution function $f_{\bm p}(x)$ is used~\cite{Fuchs:1938tn,sondheimer48,Falkovskii70}:
\begin{equation}
\label{BC:diff}
f_{\bm p}(\pm w/2) = \begin{cases}
\mathrm{const},~~p_x>0,~~x=-w/2,\\
\mathrm{const},~~p_x<0,~~x=w/2.
\end{cases}
\end{equation}
Here $\bm p$ is the electron wavevector. This condition means that the electrons travelling from the edge, i.e., scattered by the edge, have random directions of their momentum as shown in Fig.~\ref{fig:channel}(a). 


In the presence of the static electric field $\bm E\parallel y$ the electron distribution function  acquires a linear-in-$\bm E$ correction which can be found from the kinetic equation~\cite{ll10_eng,Falkovskii70}
\begin{equation}
\label{kin:E}
v_x \frac{\partial \delta f_{\bm p}(x)}{\partial x}+ \frac{\delta f_{\bm p}(x)}{\tau} -e E_y v_y f_0'=Q_{ee}\{\delta f_{\bm p}\},
\end{equation}
where $\bm v = (v_x,v_y) = \hbar \bm p/m$ is the electron velocity, $m$ is its effective mass, $\tau$ is the momentum relaxation time (related to the residual impurities or phonons), $f_0$ is the equilibrium Fermi-Dirac distribution function, and prime denotes the derivative over the energy.  The quantity $Q_{ee}\{\delta f_{\bm p}\}$ in the right-hand side is the electron-electron collision integral.

We start from the \textit{\textbf{ballistic regime}} where the electron-electron scattering can be neglected, $Q_{ee}\{\delta f_{\bm p}\}=0$. The solution of Eq.~\eqref{kin:E} with the boundary condition \eqref{BC:diff} readily  yields
\begin{multline}
\label{d_f0}
\delta f_{\bm p}(x)
=eE_yl(-f_0')\sin{\varphi_{\bm p}} \\
\times \qty{1 - \exp[-{{x+\text{sgn}(\cos{\varphi_{\bm p}})w/2} \over l\cos{\varphi_{\bm p}}}]}.
\end{multline}
Here $\varphi_{\bm p}$ is the angle between the momentum and $x$-axis,  $l = v\tau$ is the mean free path in respect to the scattering in the ``bulk'' of the sample.

In the narrow ultraclean channels with $w\ll l$, we have from Eq.~\eqref{d_f0} up to the first order in $w/l$:
\begin{equation}
\label{d_f}
\delta f_{\bm p}(x)\approx eE_y(-f_0')\sin{\varphi_{\bm p}} {{x+\text{sgn}(\cos{\varphi_{\bm p}})w/2} \over \cos{\varphi_{\bm p}}}.
\end{equation}
In this regime the channel conductivity relating the two-dimensional current density ($\mathcal L$ is the normalization length along the $y$-axis)
\begin{equation}
\label{2Dcurrent}
j_y = \frac{1}{w\mathcal L}\int_{-w/2}^{w/2} \sum_{\bm p} v_y \delta f_{\bm p}(x) dx,
\end{equation}
with the electric field component $E_y$ as 
\begin{equation}
j_y = \sigma_{yy} E_y
\end{equation}
 takes the following form:
\begin{equation}
\label{sigma:yy}
\sigma_{yy}^{b} = \frac{2}{\pi} \sigma_0 \frac{w}{l} \ln{\left(\frac{l}{w}\right)}, \quad \sigma_0 = \frac{e^2 N \tau}{m},
\end{equation}
Here and in what follows we omit the factor describing the spin and valley degeneracy of the electron states, $N$ is the total two-dimensional electron density.

Equation~\eqref{sigma:yy} shows that the channel width $w$ plays a role of the effective mean free path $l_{\rm eff}$. Strictly speaking, $l_{\rm eff} \sim w \ln{(l/w)}$. The logarithmic factor $\ln{(l/w)}$ accounts for the contribution of the electrons at the grazing incidence at the edges: These electrons travel large distance $\sim w/\cos{\varphi_{\bm p}}$ between the consecutive collisions between the edges.

The analysis above is valid for a ballistic channel. Let us now turn to the opposite limit of very strong electron-electron scattering where the \textit{\textbf{hydrodynamic regime}} is realized. Here we assume that (i) the relaxation of the second angular harmonic of the distribution function is controlled by frequent electron-electron collisions and (ii) the corresponding relaxation length $l_2 \equiv l_{ee} = {v\tau_{ee}}${, where $\tau_{ee}$ is the electron-electron scattering time}, is the smallest parameter of the problem: $l_2\ll w \ll l$. In this situation we can keep only first and second angular harmonics in the electron distribution function $\delta f_{\bm p}(x)$~\cite{Gurzhi_1968}:\footnote{Hereafter we consider only $\bm E$-linear response. It allows us to disregard any higher order harmonics of the distribution function which result from the fact that, due to the electron-electron collisions, the distribution function relaxes to the `shifted' by the drift velocity Fermi-Dirac function.}
\begin{equation}
\label{decomp}
\delta f_{\bm p}(x) =  \sin{\varphi_{\bm p}} \,F_{1}^s(x)+\sin{2\varphi_{\bm p}} \, F_2^s(x),
\end{equation}
where $F_{n}^s(x)$ are the coordinate-dependent harmonics of the distribution function whose dependence on the electron energy is omitted for brevity. In the presence of electric field $\bm E\parallel y$ we have the following set of coupled equations 
\begin{equation}
\label{Ey:12}
\frac{l }{2}\frac{\partial F_{2}^s}{\partial x}+ F_1^s = -e E_y l f_0' , \quad \frac{l_2}{2} \frac{\partial F_{1}^s}{\partial x} + F_2^s=0.
\end{equation} 
We stress that $l$ here is due to the residual disorder and $l_2$ is due to the electron-electron scattering. Equations~\eqref{Ey:12} can be conveniently combined into one equation for $F_1^s(x)$:
\begin{equation}
\label{F1s:hydro}
-\eta \frac{\partial^2F_1^s}{\partial x^2} + \frac{F_1^s}{\tau} = - e v E_y f_0',
\end{equation}
where the viscosity 
\begin{equation}
\label{visc}
\eta = \frac{v l_2}{4}.
\end{equation}
In what follows we completely neglect the disorder in the channel setting $\tau \to \infty$. Hence, following Refs.~\cite{Gurzhi_1968,PhysRevLett.117.166601} we obtain 
\begin{equation}
\label{df_hydro}
\delta f_{\bm p}(x) = - \frac{e  E_y v_y f_0'}{2\eta} \left\{\left[\left(\frac{w}{2}\right)^2-x^2\right] +\frac{2l_2 v_x}{v} x \right\}.
\end{equation} 
Naturally, Eq.~\eqref{df_hydro} contains the first and second angular harmonics of the distribution function; as expected, the second harmonic is much smaller than the first one. The electron velocity distribution corresponds in this case to the Poiseuille flow.
Accordingly, the longitudinal conductivity can be written as~\cite{PhysRevLett.117.166601}
\begin{equation}
\label{sigma:yy:hydro}
\sigma_{yy}^{h} = \sigma_0 \frac{w^2}{12\eta \tau}.
\end{equation}
Here the role of grazing electrons is drastically reduced because of the electron-electron collisions: Although conserving the total momentum of the colliding pair, they efficiently change the incidence angle of the electron at the edge. In the hydrodynamic regime the role of the effective mean free path is played by $l_{\rm eff} \sim w^2/l_2 \gg w$ yielding $\sigma^h_{yy} \gg \sigma^b_{yy}$. This relation is natural, because due to the efficient electron-electron collisions for the most of electrons the momentum dissipation is indirect and, hence, inefficient: Via multiple collisions, the momentum is transferred from electrons in the bulk of the channel to those near the edges where the dissipation occurs. The smaller $l_2$ is, the more efficient is the suppression of the edge scattering resulting in larger conductivity $\sigma^h_{yy}$.

\subsection{Impurity stripe model}\label{sec:imp}

Microscopic derivation of the diffusive boundary condition~\eqref{BC:diff} is quite involved~\cite{PhysRev.141.687,Falkovskii70,Andreev:1972wt,PhysRevB.96.195401,PhysRevB.99.035430}. On the other hand, as we demonstrate below, the edge scattering is crucial for the anomalous transport and VHE. Thus, it is instructive to introduce a simple model of the edge which allows us to derive the results above without explicit use of Eq.~\eqref{BC:diff}. To that end we introduce the \emph{impurity stripe model} of the edge: We assume that there are narrow stripes of the width $d \ll w$ in the vicinity of each edge with high concentration of the scattering centers, i.e., impurities, so that the mean free path in these impurity stripes $l_i = v\tau_i \ll d$ ($\tau_i$ is the electron scattering time in the impurity stripe), Fig.~\ref{fig:channel}(b). We also assume that the electron-electron scattering is irrelevant in the impurity stripe, $\tau_i \ll \tau_{ee}$. Thus, the electron motion in the impurity stripe is diffusive rather than ballistic. At the same time, the boundary conditions at the outer edges of the impurity stripes are unimportant and, for simplicity, we can assume specular reflection from these edges. Corresponding distribution functions can be found solving the kinetic equation~\eqref{kin:E} in different regions of the channel and joining the solutions at $x=\pm w/2$.

In the limit $l_i \to 0$ the solution for the $\delta f_{\bm p}$ within the channel, Eq.~\eqref{d_f0} (in the ballistic regime) or Eq.~\eqref{df_hydro} (in the hydrodynamic regime), is not affected by the presence of the impurity stripes. By contrast, within the impurity stripes we obtain 
\begin{multline}
\label{df_stripe}
\delta f_{\bm p}(x) = - eE_y l_if_0' \sin{\varphi_{\bm p}} \\
+ 
\begin{cases} 
{\delta} f_{\bm p}^<(x), \quad  -d-\frac{w}{2}\leqslant x\leqslant -\frac{w}{2},\\
{\delta} f_{\bm p}^>(x), \quad \frac{w}{2}\leqslant x\leqslant \frac{w}{2}+d,
\end{cases}
\end{multline}
where for the ballistic channel ($w\ll l$)
\begin{subequations}
\label{df<res}
\begin{multline}
\label{df<b}
{\delta} f_{\bm p}^{<,b}(x) =  eE_y l_if_0' \frac{\sin{\varphi_{\bm p}}}{\cos{\varphi_{\bm p}}}
\\
 \times  
\begin{cases}
0, \quad v_x>0,\\
  \frac{w}{l_i}  \exp[\left(- \frac{x+w/2}{l_i\cos{\varphi_{\bm p}}}\right)] , \quad v_x<0,
\end{cases}
\end{multline}
while for the hydrodynamic channel 
\begin{multline}
\label{df<h}
{\delta} f_{\bm p}^{<,h}(x) =  eE_y l_if_0' \sin{\varphi_{\bm p}}\cos{\varphi_{\bm p}}
\\
 \times  
\begin{cases}
0, \quad v_x>0,\\
  \frac{4w}{l_i}  \exp[\left(- \frac{x+w/2}{l_i\cos{\varphi_{\bm p}}}\right)] , \quad v_x<0,
\end{cases}
\end{multline}
\end{subequations}
Within the stripe at the right edge of the channel the solution $\delta f^>_{\bm p}(x)$ can be obtained from Eq.~\eqref{df<res} by natural replacements $w/2\to -w/2$, $v_x \to -v_x$:
\begin{subequations}
\label{df>res}
\begin{multline}
\label{df>b}
{\delta} f_{\bm p}^{>,b}(x) =  -eE_y l_if_0' \frac{\sin{\varphi_{\bm p}}}{\cos{\varphi_{\bm p}}}
\\
 \times  
\begin{cases}
0, \quad v_x<0,\\
  \frac{w}{l_i}  \exp\left(- \frac{x-w/2}{l_i\cos{\varphi_{\bm p}}}\right) , \quad v_x>0,
\end{cases}
\end{multline}
and
\begin{multline}
\label{df>h}
{\delta} f_{\bm p}^{>,h}(x) =  -eE_y l_if_0' \sin{\varphi_{\bm p}}\cos{\varphi_{\bm p}}
\\
 \times  
\begin{cases}
0, \quad v_x<0,\\
  \frac{4w}{l_i}  \exp\left(- \frac{x-w/2}{l_i\cos{\varphi_{\bm p}}}\right) , \quad v_x>0,
\end{cases}
\end{multline}
\end{subequations}
 Up to $l_i/l \ll 1$ corrections, the distribution function is continuous at $x=\pm w/2$. One can check that the contributions of the impurity stripes to the channels conductivity are negligible at $d \ll w$. Note that in the case of ballistic channel corresponding expressions~\eqref{df<b} and \eqref{df>b} are inapplicable for grazing angles where $|\cos{\varphi_{\bm p}}| \lesssim w/l$. In such situation the ratio $\sin{\varphi_{\bm p}}/\cos{\varphi_{\bm p}}$ should be replaced by the full expression [cf. Eq.~\eqref{d_f0}]
 \begin{equation}
 \label{rep:b}
\frac{\sin{\varphi_{\bm p}}}{|\cos{\varphi_{\bm p}}|} \to \sin{\varphi_{\bm p}}\left[1 - \exp(-{w \over l|\cos{\varphi_{\bm p}}|})\right].
 \end{equation}

The first term in Eq.~\eqref{df_stripe} is responsible for the field-induced electric current in the stripe. The second contribution, ${\delta} f_{\bm p}^<(x)$, is related to the electrons entering the impurity stripe from the channel. One can check that this contribution is responsible for the momentum dissipation in the channel. Indeed, the momentum loss per unit length in both impurity stripes takes the form
\begin{equation}
\label{loss}
\dot{P}_y = \frac{2}{\mathcal L} \int_{-w/2-d}^{-w/2} \sum_{\bm p\bm p'} {\delta} f_{\bm p}^<(x) (p'_y - p_y) W^{(i)}_{\bm p',\bm p}  dx,
\end{equation}
where the factor $2$ accounts for both left and right stripes, Fig.~\ref{fig:channel}(b), and $W^{(i)}_{\bm p',\bm p}$ is the rate of the transitions from the state $\bm p\to \bm p'$ in the impurity stripe due to the scattering. Evaluating the integral and sum in Eq.~\eqref{loss} we obtain 
\begin{equation}
\label{loss:1}
\dot{P_y} = - e E_y N w,
\end{equation}
which equals to the electric force (per unit length) applied to the electrons in the channel taken with the opposite sign.\footnote{In the ballistic regime one has to use next to the leading order in $w/l$ correction to $\delta f_{\bm p}(x)$ in order to obtain the momentum balance with account for the $\ln{(w/l)}$ of the ``bulk'' electrons.}

\section{Valley Hall effect in the channel}\label{sec:vhe}

\subsection{Mechanisms of the VHE and SHE}

We now turn to the anomalous transport of electrons and study the valley and spin Hall effects. We employ the model discussed in Ref.~\cite{2020arXiv200405091G} and consider the transition metal dichalcogenide monolayer with two valleys $\bm K_\pm$ in the Brillouin zone. In the case of the spin Hall effect,  the spin-up, $\uparrow$, and spin-down, $\downarrow$, states can be mapped onto the valleys $\bm K_+$ and $\bm K_-$ because of the same properties of the corresponding Bloch functions under time-reversal. For specificity, in what follows,   we use the term valley and discuss valley currents and valley polarization accumulation bearing in mind that the same results hold for the spin currents and spin polarization accumulation.  The effective Hamiltonian in the corresponding valley in the simplest two-band approximation (two-dimensional Dirac model) takes the form~\cite{Xiao:2012cr,2053-1583-2-2-022001}
\begin{equation}
\label{H:el}
\mathcal H_\pm = \begin{bmatrix}
0 & \pm\gamma k_\mp\\
\pm\gamma k_\pm & - E_g
\end{bmatrix}, \quad \gamma = \frac{\hbar p_{cv}}{m_0} 
{\in \mathbb R.}
\end{equation}
Here $E_g$ is the bandgap, $p_{cv}$ is the interband momentum matrix element, $k_\pm = k_x \pm \mathrm ik_y$, $x$ and $y$ are the axes in the 2D plane. The same model can be used for the SHE since the Hamiltonians $\mathcal H_\pm$ in Eq.~\eqref{H:el}  describes two spin branches in a quantum well. Importantly, the band mixing is chiral and depends on the valley or spin branch: For the one state ($\bm K_+$ or spin-$\uparrow$) the coupling is $\propto k_x - \mathrm i k_y$, and for the other state ($\bm K_-$ or spin-$\downarrow$) it is $\propto k_x + \mathrm i k_y$. In quantum wells such form of the two-band Hamiltonian is a direct consequence of the spin-orbit coupling in the valence band, heavy-light hole splitting, and the $\bm k\cdot \bm p$-mixing of the bands~\cite{ivchenko05a}.  In the case of transition-metal dichalcogenides it results from the specific form of the Bloch functions at the $\bm K_\pm$ points stemming from the broken sublattice symmetry and can be considered as an effective valley-orbit coupling. The generalization of the model to account for both spin and valley degrees of freedom is straightforward. In what follows we consider degenerate electrons with the Fermi energy $\varepsilon_F$ reckoned from the conduction band mininum being small as compared to $E_g$ and spin-orbit splitting of the bands.\footnote{The finite temperature $T$ is needed to active the electron-electron collisions. Accordingly,  we also assume that $k_B T\ll \varepsilon_F \ll E_g$.} Hence, we include the interband mixing in the lowest non-vanishing order in $\gamma k/E_g$. We also assume that the electrons scatter by a residual static short-range disorder with the potential in the conduction band, $V_c(\bm r)$, and in the valence band, $V_v(\bm r)$, given by
\begin{equation}
\label{v:dis}
V_{c,v}(\bm r) = \sum_{i} U_{c,v} \delta(\bm r - \bm R_i),
\end{equation}
where $\bm R_i$ are the random positions of the defects, and $U_{c,v}$ are the parameters describing the scattering potential in the conduction $U_c$ and valence $U_v$ bands, respectively. For the screened Coulomb impurities $U_c = U_v$. The disorder-induced momentum relaxation rate can be calculated using the Fermi's golden rule~\cite{ll3_eng} $\tau^{-1} = (2\pi/\hbar) N_{imp}\sum_{\bm k'} |U_c|^2 \delta (\varepsilon_k - \varepsilon_k')$ with $\varepsilon_k = {\hbar^2 k^2}/{2m}$ being the electron kinetic energy, $m$ its effective mass, and reads~\cite{daviesBOOK}
\begin{equation}
\label{tau:dis}
\frac{1}{\tau} = \frac{2\pi}{\hbar} g|U_c|^2 N_{imp},
\end{equation}
where $g=m/(2\pi\hbar^2)$ is the density of states (per valley) and $N_{imp}$ is the density of the residual impurities.

The analysis shows that there are the following contributions to the VHE and SHE~\cite{PhysRevB.75.045315,Sinitsyn_2007,Ado_2015,2020arXiv200405091G} (we present the expressions for the $\bm K_+$ valley or spin-$\uparrow$, the expressions for the $\bm K_-$ valley or spin{-}$\downarrow$ differ by the sign):
\begin{enumerate}
\item Anomalous velocity contribution
\begin{equation}
\label{v:anom}
\bm v_a = -{2\xi e \over \hbar} \hat{\bm{z}} \times \bm E,\quad \xi = \frac{\gamma^2}{E_g^{{2}}},
\end{equation}
arising in the external electric field. 
\item Side-jump accumulation contribution related to the electron wavepacket shift at the scattering ~\cite{PhysRev.95.1154,belinicher82,Sturman2019,2020arXiv200405091G}.
\begin{multline}
\label{electron:shifts}
\bm R_{\bm k'\bm k} 
\equiv (X_{\bm k'\bm k}, Y_{\bm k'\bm k}) \\
= \xi \frac{U_v}{U_c} [ (\bm k'-\bm k) \times \hat{\bm z}] + \xi[ (\bm k'-\bm k) \times \hat{\bm z}], 
\end{multline}
resulting in the anomalous side-jump current $\propto \sum_{\bm k\bm k'} W_{\bm k\bm k'} \bm R_{\bm k'\bm k}$ with $W_{\bm k\bm k'}$ being the electron scattering rate. The first term in Eq.~\eqref{electron:shifts} is associated with the wavevector dependence of the matrix element phase and the second term in responsible for the Bloch electron coordinate.
\item Side-jump contribution related the anomalous distribution formed as a result of the work of the electric field $e\bm E \bm R_{\bm k'\bm k}$ in the course of scattering which results in the scattering asymmetry.
\item The skew (asymmetric) scattering by single impurities or impurity pairs (coherent scattering) characterized by the asymmetric scattering probability
\begin{equation}
\label{skew:imp}
W_{\bm k' \bm k}^{as} = \xi S_{imp}[\bm k \times \bm k']_z  \delta(\varepsilon_{k}-\varepsilon_{k'}) ,
\end{equation}
with 
\begin{equation}
\label{Simp}
S_{imp} = {{2\pi}U_v\over \tau} -\frac{U_v}{U_c} \frac{\hbar}{g\varepsilon_k \tau^2}.
\end{equation}
Here the first term is responsible for the skew-scattering by single impurity, while the second term describes the two-impurity coherent scattering~\cite{2020arXiv200405091G,Ado_2015}.
\end{enumerate}
Importantly, the parameter $\xi$ introduced in Eq.~\eqref{v:anom} controls the strength of the effective valley- or spin-orbit coupling; accordingly the spin and valley Hall effects are proportional to $\xi$. Hereafter we consider only $\xi$-linear regime.

While in diffusive systems the anomalous velocity contribution 1 is compensated by a part of the side-jump contributions 2, 3~\cite{dyakonov_book,PhysRevB.75.045315,2020arXiv200405091G,Glazov2020b}, for narrow channels, as we show below, such compensation generally is absent and all effects should be taken into account.

We stress that the mechanisms 1--4 result in the valley or spin current flowing in the bulk of the sample (along the $x$-axis in the geometry of Fig.~\ref{fig:channel}). However, the current cannot flow through the sample edges, and it should be compensated by the ``diffusive'' current which arises due to the accumulated valley polarization.  The theory for the ``bulk'' spin or valley Hall conductivity developed in Ref.~\cite{2020arXiv200405091G} and standard model of the polarization accumulation at the sample edges~\cite{dyakonov71,dyakonov71a,dyakonov_book} cannot be directly applied to our situation because the motion of the electrons is ballistic rather than diffusive. Hence, our goal is to find the steady-state distribution of the electron valley polarization across the narrow channel accounting for all the contributions to the valley Hall effect introduced above. 

To find the valley (spin) polarization we apply the generalized kinetic equation for the distribution function which accounts for also the anomalous contributions to the electron velocity. Namely, we introduce $\Delta f_{\bm p}^\pm(x)$, the non-equilibrium part of the distribution function related to the valley Hall effect in the $\bm K_\pm$ valley. In the case of SHE signs $\pm$ denote the spin branches. Note that this part of the distribution function is even at $k_y \to -k_y$ in our geometry, in contrast to the part $\delta f_{\bm p}$ introduced in Sec.~\ref{sec:normal} and arising as a linear (normal) response to the electric field $\bm E\parallel y$, see Fig.~\ref{fig:channel}. Naturally, $\Delta f_{\bm p}^+(x) = -\Delta f_{\bm p}^-(x)$. Since we are interested in the effects at sufficiently low temperatures where the electron gas is degenerate, the energy dependence of $\Delta f_{\bm p}^\pm(x)$ is unimportant. Thus, it is convenient to introduce the energy-integrated distribution function which depends on the direction of the momentum $\varphi$ and coordinate $x$ only:
\begin{equation}
\label{energ:int}
\Delta F_{\varphi}^\pm(x) = g \int_0^\infty \Delta f_{\bm p}^\pm(x) d\varepsilon_p .
\end{equation}
Accordingly, the kinetic equation takes the form~[cf. Refs. \cite{PhysRevB.59.14915,PhysRevB.101.155204,2021arXiv210205675K}]
\begin{multline}
\label{kin:A}
\frac{\partial}{\partial x} \left[v_x\Delta F_{\varphi}^\pm(x) + \left(v_a^\pm + v_{sj}^\pm\right) N_1\right] + \frac{\Delta F_{\varphi}^\pm - \overline{\Delta F_{\varphi}^\pm}}{\tau}  \\
= Q_{ee}\{\Delta F^\pm_{\varphi}\} + G^\pm(\varphi,x).
\end{multline}
Here $v_x = v\cos{\varphi}$ where $v$ is the Fermi velocity,  $N_1$ is the electron density per valley, $v_{sj}^\pm {\equiv v_{sj}^\pm(x)}$ is the contribution to the anomalous velocity caused by the side-jump, overline denotes the averaging over angle $\varphi$ of the electron momentum,  and 
\begin{equation}
{G^\pm(\varphi,x) = G_{sk}^\pm(\varphi,x) + G_{adist}^\pm(\varphi,x),}
\end{equation}
is the generation rate due to (i) the skew scattering, $G^\pm_{sk}$, and (ii) the side-jump induced anomalous distribution, $G^\pm_{adist}$. We assume that both valleys are equally occupied, $2N_1 = N$, and the system is electrically neutral, thus the valley Hall current generated in the bulk of the channel is compensated by the diffusive current arising due to the valley polarization gradient rather than by Hall electric field. In Eq.~\eqref{kin:A} the valley (or spin) relaxation processes are disregarded assuming that the corresponding valley (spin) relaxation time $\tau_v \gg \tau$  and the corresponding valley relaxation length $l_v = v\tau_v$ exceeds by far both the mean free path and channel width. Equation~\eqref{kin:A} clearly demonstrates the continuity of the electron flow: Integrating it over $\varphi$ we immediately see that the divergence of the total velocity (including the anomalous contributions) is zero, as expected for $l_v \to \infty$, i.e., where the valley polarization is conserved.

Equation~\eqref{kin:A} is valid both inside the channel and in the impurity stripes, in the latter case with the replacement $\tau\to \tau_i$. It is, however, instructive to consider separately the effects arising in the bulk of the channel and in the impurity stripes. In this section, we focus on the valley Hall effect in the bulk of the channel. The theory of the valley accumulation in the impurity stripes is presented below in Sec.~\ref{sec:imp:str}. 

Let us now derive explicit expressions for the particular terms in Eq.~\eqref{kin:A}. Here we present only the scattering-unrelated contributions and contributions due to the impurities within the channel; the VHE and SHE due to the electron-electron collisions are studied in Secs.~\ref{sec:ee} and \ref{sec:hydro}. We start from the anomalous contributions to the electron velocity.  We recall that the field-induced anomalous velocity $v_a^+$ is given by Eq.~\eqref{v:anom}, $v_a^-=-v_a^+$; naturally it is independent of the scattering mechanisms. The anomalous velocity contribution due to the side-jump accumulation effect (at the impurity scattering) can be derived following Ref.~\cite{2020arXiv200405091G} with the result
\begin{equation}
\label{v:sj}
v_{sj}^\pm(x) = \pm \frac{2\pi}{\hbar N_1} \sum_{\bm p\bm p'} X_{\bm p'\bm p} |M_{\bm p'\bm p}|^2 \delta( \varepsilon_{p'} -  \varepsilon_{p}) \delta f_{\bm p}(x).
\end{equation}
Here $M_{\bm p'\bm p}$ is the matrix element of the electron scattering by the static disorder described by Eq.~\eqref{v:dis}, $\delta f_{\bm p}(x)$ is the non-equilibrium distribution function arising in the linear order in $\bm E$ and given by Eq.~\eqref{d_f} for ballistic channels and Eq.~\eqref{df_hydro} for the hydrodynamic channels.

We turn next to the description of the generation terms. The generation rate of the anomalous distribution related to the side-jump effect can be recast as~\cite{2020arXiv200405091G}
\begin{multline}
\label{G:adist}
G_{adist}^\pm(\varphi,x) 
= \pm {2\pi\over \hbar } g\int_0^\infty d\varepsilon_p
 \sum_{\bm p'}|M_{\bm p'\bm p}|^2 
\\
\times (-eE_y Y_{\bm p'\bm p}) \delta'(\varepsilon_{p'}-\varepsilon_p) [f_0(\varepsilon_{p'})-f_0(\varepsilon_p)].
\end{multline}
Finally, the generation rate due to the skew scattering takes the form
\begin{multline}
\label{G:sk}
G_{sk}^\pm(\varphi,x) 
= \pm  \xi S_{imp} g \\
\times \int_0^\infty d\varepsilon_p\sum_{\bm p'} [\bm p' \times \bm p]_z  \delta(\varepsilon_{p}-\varepsilon_{p'})  \delta f_{\bm p'}(x).
\end{multline}
Equation~\eqref{kin:A} together with expressions Eqs.~\eqref{v:anom}, \eqref{v:sj}, \eqref{G:adist}, \eqref{G:sk}, and the boundary conditions of vanishing currents at the channels edges allow us to determine the distribution function $\Delta F_{\varphi}^\pm(x)$, and, in particular, the profile of the electron density in the valley
\begin{equation}
\label{N:val:def}
\Delta N^\pm(x) = \overline{ \Delta F^\pm_{\varphi}(x)}.
\end{equation}
Naturally, the total, valley averaged, electron distribution $\Delta F^+_{\varphi}(x) + \Delta F^-_{\varphi}(x)$ is unaffected. Thus, two equations~\eqref{kin:A} for $\Delta F^\pm_{\varphi}(x)$ can be reduced to a single equation for the valley (spin) imbalance distribution or valley pseudospin (spin) $$S_\varphi(x) = \frac{1}{2} \left[\Delta F^+_{\varphi}(x) - \Delta F^-_{\varphi}(x)\right]$$ as 
\begin{multline}
\label{kin:AA}
\frac{\partial}{\partial x} \left[v_x S_{\varphi}(x) + \left(v_a^+ + v_{sj}^+\right) N_1\right] + \frac{S_{\varphi} - \overline{S_{\varphi}}}{\tau}  \\
= Q_{ee}\{S_{\varphi}\} + G^+(\varphi,x).
\end{multline}
Since we are interested in $\xi$-linear contributions to $\Delta N^\pm(x) = \pm \overline{S_\varphi(x)}$, it is instructive to calculate the skew-scattering, anomalous velocity, and side-jump effects separately because these effects are additive.
Below we present the results for the ballistic and hydrodynamic channels.

\subsection{Ballistic regime}\label{sec:ball}

In this section we study the anomalous transport of electrons in ballistic channels neglecting the electron-electron scattering, i.e., at $Q_{ee}\equiv 0$.

Let us start with anomalous contributions. The anomalous velocity results in the flux of electrons in a given valley or with a given spin along the $x$-axis. As mentionned above, at the channel edges the flux should vanish. Therefore, the density gradient should appear in such a way, that the ``diffusive'' current caused by this density gradient compensates the anomalous one. One can readily check that the distribution function $ S_{\varphi}(x)$ in the form
\begin{equation}
\label{DF:va:ball}
 S^{a}_{\varphi}(x) =  2\frac{v_a^+  x}{vl} N_1 - 2\cos{\varphi} \frac{v_a^+}{v} N_1,
\end{equation}
satisfies Eq.~\eqref{kin:AA} with the boundary conditions of vanishing current. Indeed, the density gradient results, due to the drift term $v\cos{\varphi} \partial/\partial x$, in the anisotropic current-carrying distribution of electrons which compensates the current due to the anomalous velocity.
Correspondingly, the anomalous velocity results in the valley accumulation profile in the form 
\begin{equation}
\label{N:val}
\Delta N^{\pm,a}(x) =\pm\frac{4\xi e}{v\hbar} E_y N_1 \frac{x}{l}.
\end{equation}

The side-jump accumulation also results in the anomalous velocity inside the channel. Evaluating sum in Eq.~\eqref{v:sj} we arrive at the coordinate-independent side-jump accumulation current. Kinetic equation~\eqref{kin:A} can be solved exactly in the same way as above with the result
\begin{equation}
\label{N:val:sj:acc}
\Delta N^{\pm,sj}(x) =\mp\left(1+ \frac{U_v}{U_c}\right) \frac{w}{\pi l} \ln{(l/w)} \frac{4\xi e}{v\hbar} E_y N_1 \frac{x}{l}.
\end{equation}

The remaining contributions to the valley accumulation stem from the generation terms in the right-hand side of Eq.~\eqref{kin:A}. Calculation shows that in the leading order in $l/w$ the functions $G_{adist}^\pm(\varphi,x)$ and $G_{sk}^\pm(\varphi,x)$ are the coordinate-independent and contain only the first angular harmonics resulting in $G^\pm(x) = \pm G \cos{\varphi}$. Hence, the solution of Eq.~\eqref{kin:A} can be recast as
\begin{equation}
\label{G:cos:ball}
 S_{\varphi}(x) =  G \frac{x}{v}.
\end{equation}
Particularly, for the side-jump anomalous distribution effect we arrive at
\begin{equation}
\label{N:val:adist:ball}
\Delta N^{\pm,adist}(x) =\mp\left(1+ \frac{U_v}{U_c}\right)\frac{2\xi e}{v\hbar} E_y N_1 \frac{x}{l},
\end{equation}
while for the skew scattering 
\begin{multline}
\label{N:val:skew:ball}
\Delta N^{\pm,sk}(x) =\\
\pm  4{S_{imp}g\tau \frac{\varepsilon_F\tau}{\hbar}} \frac{w}{l}\frac{\ln{(l/w)}}{\pi} \frac{\xi e}{v\hbar} E_y N_1 \frac{x}{l}.
\end{multline}
Here we took into account that the electron Fermi energy $\varepsilon_F = N_1/g$.

Let us now briefly discuss the obtained results. We note that the density profile is a linear function of the coordinate. This is general feature of the narrow ballistic channels: Since the characteristic decay length of the distribution function in the kinetic theory is the mean free path, for the channels with $w\ll l,l_v$ it is sufficient to keep in the distribution function only the coordinate-independent and $x$-linear terms.

Furthermore, the magnitude of the $\Delta N^\pm$ strongly depends on the mechanism of the valley Hall effect. It is particularly clear if one compares the two side-jump contributions: the side-jump accumulation and the anomalous distribution. Indeed, the former  is smaller than the latter by a factor $(w/l)\ln{l/w} \ll 1$, cf. Eqs.~\eqref{N:val:sj:acc} and \eqref{N:val:adist:ball}. This small factor accounts for the fraction of the momentum lost within the channel. Correspondingly, the smallness of the side-jump accumulation effect is because there is almost no momentum dissipation in the channel: The field-induced anisotropic current-carrying distribution, $\delta f_{\bm p}(x) \propto \sin{\varphi} E_y$, Eq.~\eqref{d_f} is determined mainly by the momentum loss at the sample edges, while the side-jump accumulation current is further proportional to the scattering rate in the bulk of the channel, Eq.~\eqref{v:sj}. At the same time, the anomalous distribution contribution, Eq.~\eqref{G:adist}, contains the equilibrium distribution function and, therefore, is parameterically larger.  For the same reason, unlike the case of diffusive two-dimensional systems, a compensation of the anomalous velocity contribution with the side-jump contributions is absent in the ultraclean channels. 

Naturally, the skew-scattering contribution also contains a small factor $(w/l)\ln{l/w} \ll 1$ since it is also determined by the scattering in the channel. Note that the ratio of the skew-scattering and side-jump accumulation contributions, 
\begin{equation}
{\frac{\Delta N^{\pm, sk}}{\Delta N^{\pm,sj}} \propto S_{imp} g\tau \frac{\varepsilon_F\tau}{\hbar} \propto g U_c\frac{\varepsilon_F\tau}{\hbar},} 
\end{equation}
can be either large or small depending on the strength of the scattering~\cite{2020arXiv200405091G}. At the same time, the ratio $\Delta N^{\pm, sk}/\Delta N^{\pm, adist}$ (as well as the ratio $\Delta N^{\pm, sj}/\Delta N^{\pm, adist}$) contains additional small factor $w/l\ln{(l/w)}$, as discussed above. 

To conclude this part, we present the  leading (in the small parameter $(w/l)\ln{l/w} \ll 1$) contribution to the spin or valley polarization accumulation in narrow ballistic channel caused by the anomalous velocity and side-jump induced anomalous distribution, Eqs.~\eqref{N:val} and \eqref{N:val:adist:ball}:
\begin{equation}
\label{N:val:ball:main}
\Delta N^{\pm,b} = \pm \left(1-\frac{U_v}{U_c}\right)\frac{2\xi e}{v\hbar} E_y N_1 \frac{x}{l}.
\end{equation}
Corresponding  valley polarization profile is shown in Fig.~\ref{fig:inchannel} by the red line.

The parametric difference between the side-jump accumulation and skew-scattering, on the one hand, and the side-jump anomalous distribution and the anomalous velocity, on the other hand, is even more pronounced in the hydrodynamic regime, see Sec.~\ref{sec:hydro} below. Remarkably, the side-jump accumulation and the skew scattering provide a non-trivial valley distribution in the impurity stripes, as shown in Sec.~\ref{sec:imp:str}.

\subsection{Electron-electron scattering effects on valley current}\label{sec:ee}

Here we address the effects of the electron-electron collisions on the anomalous transport properties of the electron gas. Before turning to the solution of Eq.~\eqref{kin:A} with account for frequent electron-electron collisions, which is presented in next Sec.~\ref{sec:hydro}, we briefly discuss the role of electron-electron collisions in the valley current generation and relaxation, leaving the presentation of the complete theory for a separate paper. Same analysis equally applies to the spin current generation and relaxation due to the electron-electron scattering.

\begin{figure*}[t!]
\includegraphics[width=0.7\linewidth]{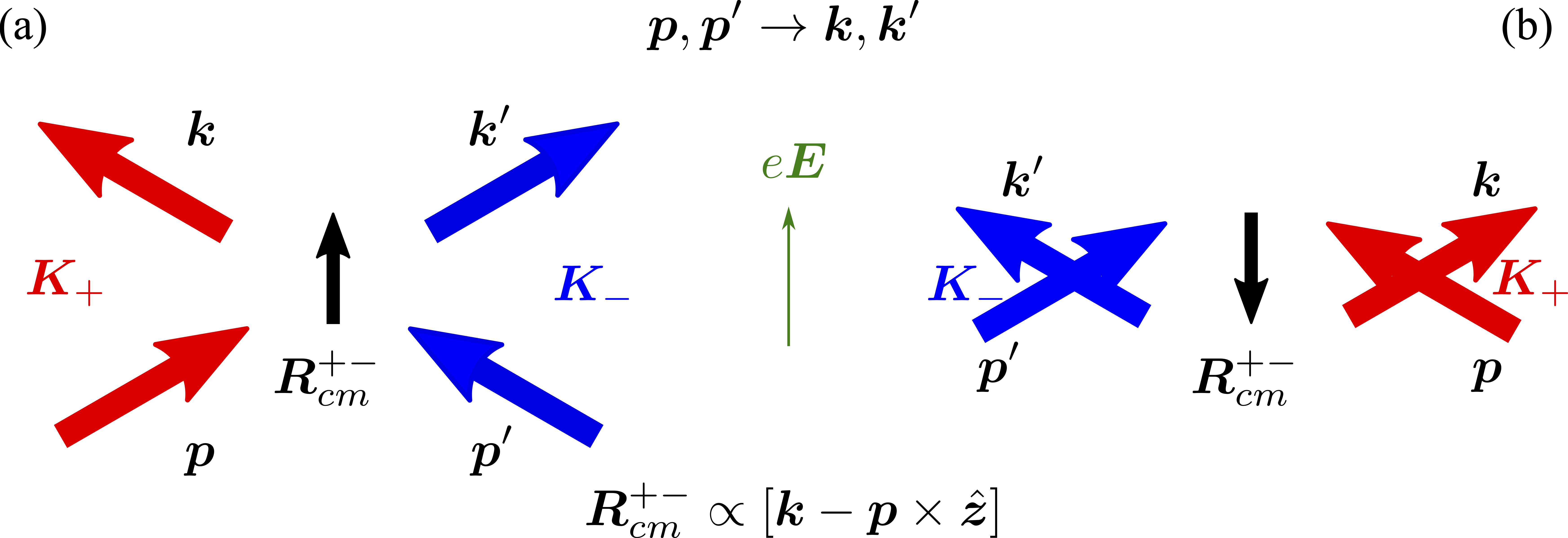}
\caption{Valley or spin current dissipation and generation at the electron-electron scattering. The process of the electron-electron collision from the initial state $\bm p, \bm p'$ to the final state $\bm k,\bm k'$ is shown. Thick solid arrows show the electron momenta before and after the collision, color denotes the valleys/spins (red corresponds to the $\bm K_+$ valley/spin-$\uparrow$ and blue corresponds to the $\bm K_-$ valley/spin-$\downarrow$). The processes in (a) and (b) differ by the direction of the transferred wavevector $\bm k - \bm p$ and, accordingly, by the direction of the center of mass shift, $\bm R^{+-}_{cm}$, Eq.~\eqref{R:tot:pair}. The collision with positive work of the field ${2}e\bm E\cdot \bm R^{+-}_{cm}>0$ [panel (a)] reduces the potential energy of the electron pair in the external field and is more efficient than the collision with  ${2}e\bm E\cdot \bm R^{+-}_{cm}<0$ [panel (b)].}\label{fig:ee:adist}
\end{figure*}

To start, let us analyze the role of the skew-scattering and side-jump effects at the electron-electron collisions. To that end, we consider the matrix element of the Coulomb scattering of an electron pair from the states $\bm k,\bm k'$ to the states $\bm p, \bm p'$. Let $\sigma_{i}$ be the valley index of the $i$th electron, $i=1,2$. The matrix element of the direct scattering process reads~\cite{boguslawski,badescu:161304,glazov2009,badalyan-2009}
\begin{multline}
\label{me:ee}
M_{\sigma_1\sigma_2}(\bm k,\bm k' \to \bm p,\bm p') = V(q) \delta_{\bm k+ \bm k', \bm p+ \bm p'} 
\\
\times 
\left( 1+ \mathrm i \xi \sigma_1[\bm p\times \bm k]_z + \mathrm i \xi \sigma_2[\bm p'\times \bm k']_z\right),
\end{multline}
where $V(q)$ is the matrix element of the appropriately screened Coulomb interaction. In Eq.~\eqref{me:ee} we took into account only $\xi$-linear contributions, see Refs.~\cite{glazov2009,glazov2010} for the full expression derived in the extended Kane model. The analysis of the skew scattering effect at the electron-electron collisions is presented in Appendix~\ref{app:ee-skew}. It is shown that  the skew scattering contribution is absent due to the energy and momentum conservation laws,\footnote{Strictly speaking this result is rigorously derived for the non-degenerate electrons, full analysis requires a separate study.} however, it can result in the valley Hall effect in bilayers, cf.~\cite{badalyan-2009,PhysRevB.84.033305}.

It is straightforward to derive, following Refs.~\cite{PhysRevLett.121.226601,PhysRevLett.113.182302}, the coordinate shifts of the electrons at the Coulomb scattering. In our case the individual shifts of the electronic wavepackages are given by [cf. Eq.~\eqref{electron:shifts}]
\begin{equation}
\label{ee:electron:shifts}
\bm R^{\sigma_1}_{\bm p\bm k} 
=2 \xi \sigma_1[ (\bm p-\bm k) \times \hat{\bm z}];~\bm R^{\sigma_2}_{\bm p'\bm k'} 
=2 \xi \sigma_2[ (\bm p'-\bm k') \times \hat{\bm z}].
\end{equation}
Particularly, for two electrons belonging to the same valley, $\sigma_1=\sigma_2$, the net shift of the pair, i.e., the shift of their center-of-mass, $$\bm R_{cm}^{\sigma\sigma} = \frac{1}{2}\left(\bm R^\sigma_{\bm p\bm k} + \bm R^\sigma_{\bm p'\bm k'}\right) \propto [\bm p+ \bm p' - \bm k -\bm k']\times \hat{\bm z}$$ vanishes  because of the momentum conservation. Such collisions contribute neither to the net electric current nor to the valley Hall effect. If the electrons are in the opposite valleys, the net shift of the pair can be non-zero, see below. Noteworthy, the side-jump accumulation-induced valley Hall current  is contributed by the difference of the shifts for particles at the opposite valleys, $\bm R_{diff} = \bm R^{\sigma}_{\bm p\bm k} - \bm R^{-\sigma}_{\bm p'\bm k'} \propto [\bm p+ \bm p' - \bm k -\bm k']\times \hat{\bm z}$, which  vanishes because of the momentum conservation.\footnote{Accounting for the exchange interaction results in the antisymmetrization of the matrix element and does not change the conclusions.}

Importantly, the electron-electron collisions in different valleys (or with opposite spins) result in the side-jump-induced anomalous distribution. It originates from the fact that the electron-electron collisions cause the relaxation of the valley and spin currents~\cite{glazov02,amico02,amico:045307,glazov04a,weber05}. The valley current relaxation is schematically shown in Fig.~\ref{fig:ee:adist}: In the course of collision between electrons in different valleys the $x$-component of the valley current changes despite the total momentum conservation. With account for the electron shifts in the external field the relaxation becomes different for the valley current flowing along the $+x$ and $-x$, cf. panels (a) and (b) of Fig.~\ref{fig:ee:adist}, and valley current can be generated.

To analyze the effect quantitatively, we consider the pair with $\sigma_1=+$ and $\sigma_2=-$ and calculate the shift of its center of mass at the scattering $\bm p, \bm p' \to \bm k, \bm k'$ as
\begin{equation}
\label{R:tot:pair}
\bm R^{+-}_{cm} = \frac{1}{2}\left(\bm R^+_{\bm k\bm p} + \bm R^-_{\bm k'\bm p'}\right)  = 2\xi [\bm k - \bm p \times \hat{\bm z}]. 
\end{equation}
Taking into account the work of the electric field at the scattering from $\bm p, \bm p' \to \bm k, \bm k'$ being equal to $2e\bm E \cdot  \bm R^{+-}_{cm}$ we obtain the generation rate of anomalous distribution in the $\bm K_+$ valley  as [cf. Ref.~\cite{PhysRevLett.121.226601}]
\begin{multline}
\label{adist:gen:ee}
G_{ee,adist}^+ = \sum_{\bm p\bm p'\bm k'} \frac{{2}e\bm E\cdot\bm  R^{+-}_{cm}}{k_BT}
W_0(\bm k,\bm k';\bm p,\bm p') \\
\times f(\varepsilon_k)f(\varepsilon_{k'})[1-f(\varepsilon_p)][1-f(\varepsilon_{p'})] ,
\end{multline}
where $W_0(\bm k,\bm k';\bm p,\bm p') = 2|V_{\bm k - \bm p}|^2$ is the symmetric scattering probability, cf. Appendix~\ref{app:ee-skew}.
In the $\bm K_-$ valley the generation rate of the anomalous distribution has the same form as Eq.~\eqref{adist:gen:ee} but the sign is opposite: $G_{ee,adist}^- = - G_{ee,adist}^+$: Physically the collisions with $e\bm E \cdot \bm R^{+-}_{cm}>0$ are accompanied by the reduction of the electron pair potential energy [Fig.~\ref{fig:ee:adist}(a)] and are more efficient as compared with the collisions with $e\bm E \cdot \bm R^{+-}_{cm}<0$ [Fig.~\ref{fig:ee:adist}(b)] where the potential energy of the pair is increased.\footnote{For the same reason the valley current can be converted to the electric current.}

In order to find $\Delta f_{\bm k}^{\pm, adist:ee}$ we need to determine its relaxation rate. The corresponding electron-electron collision integral reads~\cite{glazov02,glazovnato}:
\begin{multline}
\label{adist:rel:ee}
Q\{\Delta f^+,f_0\} = \sum_{\bm p\bm p'\bm k'} 
W_0(\bm k,\bm k';\bm p,\bm p')\\
\times [\Delta f^{+, ee,adist}_{\bm k} F(\bm k';\bm p,\bm p')  
\\
- \Delta f^{+, ee,adist}_{\bm p} F(\bm p';\bm k, \bm k')],
\end{multline}
with 
\[
F(\bm k';\bm p,\bm p') = f^0(\varepsilon_{k'})[1-f^0(\varepsilon_p)-f^0(\varepsilon_{p'})] + f^0(\varepsilon_p)f^0(\varepsilon_{p'}).
\]
Comparing Eqs.~\eqref{adist:gen:ee} and \eqref{adist:rel:ee} we obtain that the generation and dissipation can be balanced provided that 
\begin{equation}
\label{anomalous:ee}
\Delta f_{\bm k}^{\pm, ee,adist} =\mp {4}\xi e\bm E \cdot[\bm k\times\hat{\bm z}]f_0'
\end{equation}
yielding the bulk valley Hall current (per electron)
\begin{equation}
\label{v:adist:ee}
\bm v_{ee,adist}^{\pm} = \frac{{4}\xi e}{\hbar} \hat{\bm z}\times \bm E.
\end{equation}

For further calculations is it convenient to introduce the relaxation time approximation for the electron-electron scattering. In terms of previously introduced generation rates of anomalous distribution this process can be expressed as
\begin{equation}
\label{G:ee:adist:short}
G_{adist}^\pm = \mp \frac{8\xi e}{\hbar} E_y N_1 \frac{\cos{\varphi}}{v\tau_{ee}},
\end{equation}
while the electron-electron collision integral can be recast at the same approximation as
\begin{equation}
\label{Q:ee:short}
Q_{ee}\{S_\varphi\} = - \frac{S - S_{\varphi}}{\tau_{ee}}.
\end{equation}
We abstain from the detailed discussion of the applicability of the relaxation time approximation for the electron-electron collisions in the two-dimensional electron gas with arbitrary degeneracy since typical energy transferred in the collision is $\sim k_B T$ and corresponds to the thermal spread of the electron distribution~\cite{gantmakher87,glazov02,glazov04a}.

\subsection{Hydrodynamic regime}\label{sec:hydro}

Now we have all prerequisites to turn to the hydrodynamic regime of the VHE in ultraclean channels.
The electron-electron collisions result in the spin and valley current relaxation. Therefore, in the hydrodynamic regime in the kinetic equation~\eqref{kin:AA} for $ S_{\varphi}(x)$, the relaxation times of the first and second angular harmonics are comparable and controlled by the fast electron-electron collisions. It makes purely hydrodynamic description of the valley accumulation inapplicable, in general. Particularly, in the decomposition of $ S_{\varphi}(x)$ over the angular harmonics of $\varphi$
\begin{equation}
\label{decomp:S}
S_{\varphi}(x)  = S_0(x) + \cos{\varphi} \, S_{1} (x) + \cos{2\varphi} \,S_{2} (x),
\end{equation}
cf. Eq.~\eqref{decomp}, higher harmonics omitted above are not necessarily small. We make sure, however, that these omitted harmonics are unimportant in our situation due to the specific form of the field-induced $\delta f_{\bm p}(x)$, Eq.~\eqref{df_hydro}, where the second angular harmonic is already smaller than the first one.\footnote{Note that an  approximate hydrodynamic description is also possible in the case where the electron-electron collisions are irrelevant, but the scattering is caused by a smooth disorder with $\tau_n \propto n^{-2}$, i.e., where the relaxation of the high angular harmonics is much faster than that of the first one. However, the physics behind this approximation is different.} 

We stress that the simplified kinetic equation~\eqref{kin:AA} is valid at low enough temperatures where $k_B T \ll \varepsilon_F$ such that the thermal spread of the electron distribution function is practically unimportant. On the other hand, the electron-electron scattering free path $l_{ee} = v\tau_{ee}$ should be smaller than both the mean free path and the channel width: $$\tau_{ee} \ll \frac{w}{v} \ll \tau.$$ Taking into account that $\tau_{ee}^{-1} = \Lambda (k_B T)^2/\hbar\varepsilon_F $, where $\Lambda \sim 1$ is the dimensionless factor which just slightly depends on the temperature~\cite{chaplik71eng,Giuliani82,PhysRevB.53.9964,glazov04a,PhysRevB.102.241409}, we obtain the following conditions on the system parameters
\begin{equation}
\label{cond:T}
\sqrt{ \frac{\hbar v}{l} \varepsilon_F} \ll \sqrt{ \frac{\hbar v}{w} \varepsilon_F} \ll  k_B T \ll \varepsilon_F.
\end{equation}
The condition $k_B T \ll \varepsilon_F$ is, in fact, unessential for the basic physics described below but simplifies the kinetic equation and its solution.

Note that, on the one hand, we  neglected impurity scattering within the channel in calculation of the $\delta f_{\bm p}(x)$ as in Eq.~\eqref{df_hydro}, Sec.~\ref{sec:normal}. On the other hand, we account for the impurity scattering to evaluate the side-jump and skew-scattering effects. It is legitimate since the weak scattering gives rise to small corrections to the normal part of the distribution function, but its effect on anomalous transport can be considerable.

Hence, under approximations formulated above, we obtain from Eqs.~\eqref{kin:AA} and \eqref{decomp:S} the set of equations for the three functions $S_{0,1,2}(x)$:
\begin{subequations}
\label{Full:norm}
\begin{align}
&\frac{d}{d x} \left[\frac{v}{2}S_1  + \left(v_a^++v_{sj}^+\right) N_1 \right]  =0, \label{eq:S0} \\
&l_1 \frac{d}{d x}  \left(S_{0} + \frac{1}{2} S_{2}\right) + S_1 =C(x) , \label{eq:S1}\\
&\frac{l_2}{2} \frac{d  S_{1}}{ d x}  + S_2 = 0.\label{eq:S2}
\end{align}
\end{subequations}
Here, as above, the valley relaxation is completely neglected $l_v\to \infty$, $l_1$ is the relaxation length of the first angular harmonics of the valley distribution, which accounts for the electron-electron collisions,  and $l_2$ is the relaxation length of the second angular harmonics of the valley distribution function related to the viscosity by Eq.~\eqref{visc}.
In the simplest possible relaxation time approximation, Eq.~\eqref{Q:ee:short}, $l_1 = l_2 = v\tau_{ee}$, but generally $l_1$ and $l_2$ can be somewhat different. We recall that Eq.~\eqref{eq:S0} is just a continuity equation for the valley polarization which accounts for the anomalous velocity. The function
\begin{equation}
C(x) = \frac{2l_1}{v}\overline{\cos{\varphi} G_{\varphi}^+(x)},
\end{equation}
in Eq.~\eqref{eq:S1} describes the generation of the valley Hall current due to the side-jump anomalous distribution and the skew-scattering. We neglected the generation of the second angular harmonics in Eq.~\eqref{eq:S2} due to the smallness of the second harmonics in the normal distribution function $\delta f_{\bm p}$. Equations~\eqref{Full:norm} should be supplemented with the boundary conditions requiring the absence of the valley current at the channels edges:
\begin{equation}
\label{no:vcurrent:bc}
\left.\frac{v}{2}S_1  + \left(v_a^++v_{sj}^+\right) N_1 \right|_{x=\pm w/2}=0.
\end{equation}
As a result, the solution of Eq.~\eqref{eq:S0} takes a particularly simple form
\begin{equation}
\label{S1:sol}
S_1(x) = -2\frac{v_a^++v_{sj}^+}{v} N_1, 
\end{equation}
which means that the ``diffusive'' current in the channel compensates the anomalous current, see corresponding analysis for the ballistic case, Eq.~\eqref{DF:va:ball}. Combining Eqs.~\eqref{eq:S1} and \eqref{eq:S2} we obtain the first-order differential equation for the valley population imbalance $S_0(x)$ which can be readily solved as
\begin{equation}
\label{S0:sol}
S_0(x) = \int_0^x  \left[ C(x') - S_1(x')+ \frac{l_1l_2}{4} \frac{d^2 S_1(x')}{dx'^2}\right]\frac{dx'}{l_1},
\end{equation}
where we took into account that $S_0(x)$ is an odd function of the coordinate.

Now we present the results for the valley accumulation in the hydrodynamic regime calculated for all relevant mechanisms of the effect after Eq.~\eqref{S0:sol}. As in Sec.~\ref{sec:ball}, we start from the anomalous contributions to the valley accumulation. The anomalous velocity contribution can be readily derived as
\begin{equation}
\label{N:val:ee:hyd}
\Delta N^{\pm,a}(x) =\pm S_0(x) =\pm{4}\frac{\xi e}{v\hbar} E_y N_1 \frac{x}{l_1}.
\end{equation}
This expression has the same form as Eq.~\eqref{N:val} with the replacement of $l\to l_1 \ll l$, resulting in the enhancement of the effect. This is because of the electron-electron interaction which shortens the relaxation time of $S_1$: As a result, to compensate the same anomalous current a higher density gradient is needed. The side-jump accumulation effect also results in the contribution to the anomalous velocity which is derived from Eqs.~\eqref{df_hydro} and \eqref{v:sj} in the form
\begin{equation}
\label{v:sj:acc:ee}
v_{sj}^+(x) =  -{\xi\over \hbar}  \left(1+{U_v\over U_c}\right) eE_y \frac{v}{2\eta l}\left[\left(\frac{w}{2}\right)^2-x^2\right].
\end{equation}
Note that here $l$ is the mean free path due to the rare electron-impurity collisions. Making use of Eqs.~\eqref{S1:sol} and \eqref{S0:sol} we arrive at 
\begin{multline}
\label{N:val:ee}
\Delta N^{\pm,sj}(x) =\pm S_0(x) 
= 
\mp{2\xi e\over v\hbar}  \left(1+{U_v\over U_c}\right) E_y N_1 \frac{x}{l_1}  
\\
\times
\frac{v }{2\eta l}\left\{ \left[\left(\frac{w}{2}\right)^2-\frac{x^2}{3}\right] - \frac{l_1l_2}{2}\right\}.
\end{multline}
Provided that the electron-electron collisions control both $l_1$ and $l_2$, we have $l_1l_2 \ll w^2$ and the second term in curly brackets in negligible in the considered hydrodynamic regime. Using the definition of viscosity, Eq.~\eqref{visc}, Eq.~\eqref{N:val:ee} can be rewritten as
\begin{multline}
\label{N:val:ee:1}
\Delta N^{\pm,sj}(x)\\
 =
\mp  \left(1+{U_v\over U_c}\right){\xi e\over v\hbar} E_y N_1 \frac{x}{l_1}  
\frac{4}{ll_2}\left[\left(\frac{w}{2}\right)^2-\frac{x^2}{3}\right] .
\end{multline}
The profile of accumulated density thus deviates from the simple linear one, because of the parabolic coordinate dependence of $\delta f_{\bm p}(x)$ in the hydrodynamic regime.

Now we turn to the remaining contributions due to the anomalous distribution formed due to the side-jump effect and due to the skew scattering. First we make use of Eqs.~\eqref{anomalous:ee} and \eqref{v:adist:ee} to calculate the profile of the electron density due to the anomalous distribution generated at the electron-electron scattering. Straightforward calculation shows that
\begin{equation}
\label{N:val:adist:eeee}
\Delta N^{\pm,ee,adist}(x) =\mp{8}\frac{\xi e}{v\hbar} E_y N_1 \frac{x}{l_1}.
\end{equation}
It is twice larger than the contribution due to the anomalous velocity, Eq.~\eqref{N:val:ee:hyd}. Hence, a half of $\Delta N^{\pm,ee,adist}$ compensates $\Delta N^{\pm,a}$. Such compensation, surprising as it may seem, follows directly from comparison of Eqs.~\eqref{v:adist:ee} and \eqref{v:anom}. Thus, one half of electron-electron scattering induced anomalous distribution remains\footnote{If the electron-electron interaction potentials were different in the conduction band and in the valence band, the corresponding ratio $U_v^{ee}/U_c^{ee}$ appears as a prefactor in the total valley polarization, Eq.~\eqref{N:val:av+adist:eeee}, while the contributions with Bloch coordinates cancel the anomalous velocity.}
\begin{equation}
\label{N:val:av+adist:eeee}
\Delta N^{\pm,ee}(x) =\mp{4}\frac{\xi e}{v\hbar} E_y N_1 \frac{x}{l_1}.
\end{equation}

 The anomalous distribution effect due to the impurities can be calculated exactly as in Sec.~\ref{sec:ball} with the result
\begin{equation}
\label{N:val:adist:ee}
\Delta N^{\pm,adist}(x) =\mp\left(1+ \frac{U_v}{U_c}\right)\frac{{2}\xi e}{v\hbar} E_y N_1 \frac{x}{l}.
\end{equation}
Note that unlike the anomalous velocity contribution, Eq.~\eqref{N:val:ee:hyd}, here the profile is controlled by the impurity-induced mean free path exactly as in Eq.~\eqref{N:val:adist:ball}. Finally, for the skew scattering contribution we obtain
\begin{multline}
\label{N:val:skew:ee}
\Delta N^{\pm,sk}(x) =\\
\pm  S_{imp}g\tau \frac{\varepsilon_F \tau_1}{\hbar}\frac{\xi e}{v\hbar}  E_y N_1 \frac{x}{l_1} \frac{4}{l l_2}  \left[\left(\frac{w}{2}\right)^2-\frac{x^2}{3}\right].
\end{multline}
The profile of the valley polarization is the same as for the side-jump accumulation effect. Ratio of the skew-scattering to the side-jump accumulation can be estimated as (at $U_c \sim U_v$)
\[
\left|\frac{gS_{imp}\tau}{1+{U_v}/{U_c}} \frac{\varepsilon_F \tau_1}{\hbar}\right| \sim g |U_c| \frac{\varepsilon_F \tau_1}{\hbar},
\]
and is a product of small, $g|U_c|\ll 1$, and large, $\varepsilon_F \tau_1/\hbar$, factors. Thus, these contributions can be comparable. 

It is worth to note that contribution due to the impurity-induced side-jump accumulation, Eq.~\eqref{N:val:ee:1} and, generally, the contribution caused by the skew scattering, Eq.~\eqref{N:val:skew:ee} exceed by far the contribution due to the impurity-induced anomalous distribution, Eq.~\eqref{N:val:adist:ee} because $l_1l_2 \ll w$. On the other hand, the remaining contribution from the electron-electron anomalous distribution (with account for compensation of the anomalous velocity), Eq.~\eqref{N:val:av+adist:eeee}, can be larger or smaller as compared to Eqs.~\eqref{N:val:ee:1} and Eqs.~\eqref{N:val:skew:ee} depending on the parameter
\begin{equation}
\label{rsmall}
r = \frac{w^2}{ll_1},
\end{equation}
which is a product of a small $w/l\ll 1$ and large $w/l_1 \gg 1$ factors. If $r\ll 1$, the contribution due to the electron-electron scattering induced anomalous distribution, Eq.~\eqref{N:val:av+adist:eeee}, is dominant. On the contrary, if $r \gg 1$, the impurity-induced contributions, Eqs.~\eqref{N:val:ee:1} and \eqref{N:val:skew:ee} dominate. Combining Eqs.~\eqref{N:val:ee:1}, \eqref{N:val:av+adist:eeee}, and \eqref{N:val:skew:ee}  we obtain the leading contribution to the valley polarization in the hydrodynamic regime in the form
\begin{multline}
\label{N:val:hydro:main}
\Delta N^{\pm,h}(x)
 =
\mp {\xi e\over v\hbar} E_y N_1 \frac{x}{l_1}\\
\times
\left\{{4}+  \left(\mathcal C_{sj} + \mathcal C_{sk}\right)  
\frac{4}{ll_2}\left[\left(\frac{w}{2}\right)^2-\frac{x^2}{3}\right]\right\},
\end{multline}
with $\mathcal C_{sj} = 1+U_v/U_c$ and $\mathcal C_{sk} = -S_{imp} g\tau \varepsilon_F \tau_1/\hbar$. The valley (spin) polarization profile is shown in Fig.~\ref{fig:inchannel} by the blue line.

\subsection{Discussion}

\begin{figure}[tb]
\includegraphics[width=0.9\linewidth]{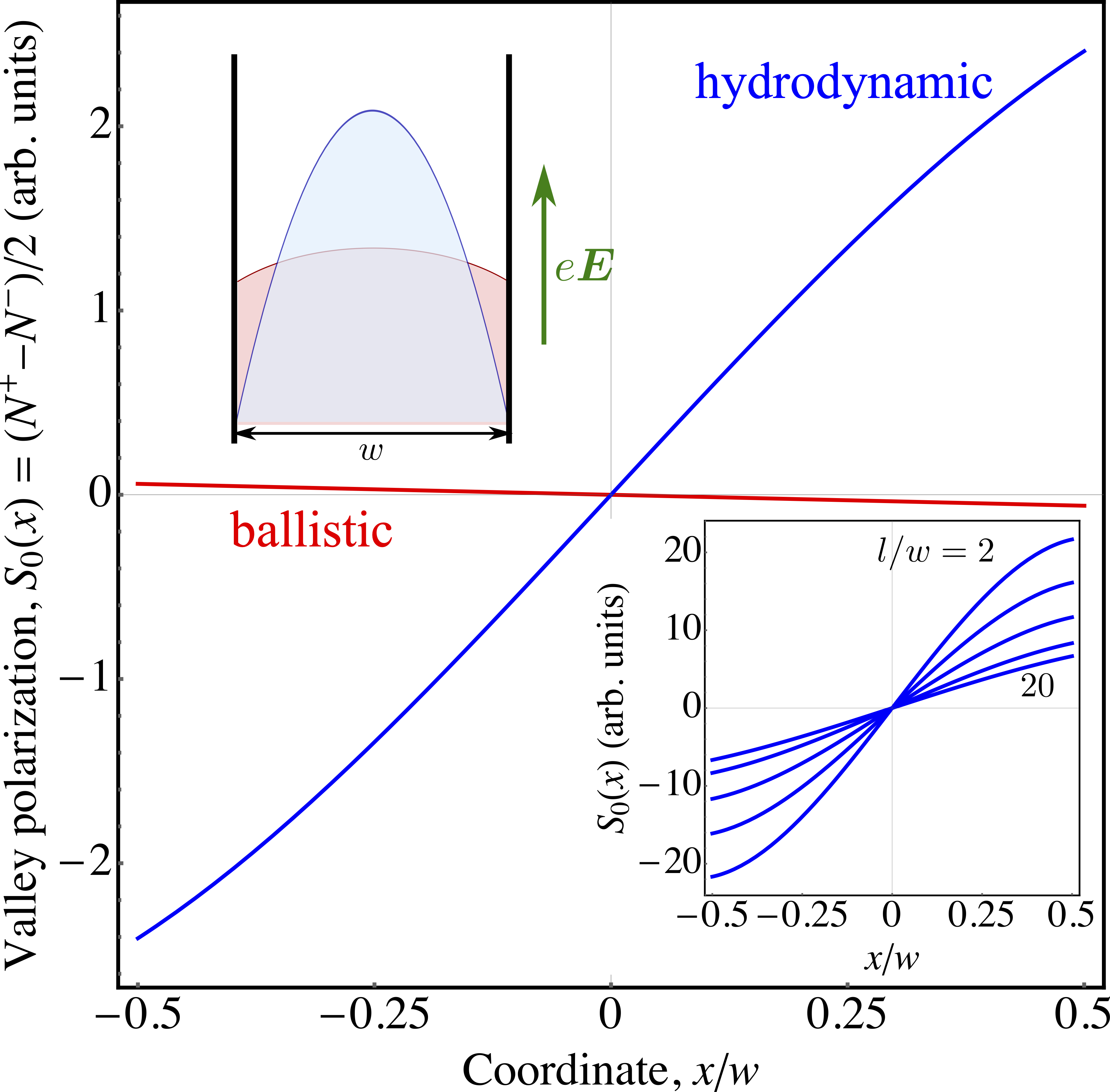}
\caption{Spin or valley polarization calculated after Eq.~\eqref{N:val:ball:main} (blue line, ballistic channel) and Eq.~\eqref{N:val:hydro:main} (red line, hydrodynamic channel). The parameters are $\mathcal C_{sj}=1.8$, $C_{sk}=2.2$, and $l/w=5$, $l_1=l_2=0.3w$. Top left inset shows the schematics of the channel, velocity profiles and direction of the applied field, cf. Fig.~\ref{fig:channel}. Bottom right inset shows the valley polarization in the hydrodynamic channel calculated after Eq.~\eqref{N:val:hydro:main} at $\mathcal C_{sj}=1.8$, $C_{sk}=2.2$, $l_1=l_2=0.1w$ and different ratios $l/w=2,3,5,10,20$. }\label{fig:inchannel}
\end{figure}

Here we summarize and briefly discuss the obtained results for the valley accumulation within the channel. The valley polarization $S_0(x) = [N^+(x) - N^-(x)]/2$ is plotted in Fig.~\ref{fig:inchannel} by the red line for the ballistic regime and blue line for the hydrodynamic regime. In the \textit{\textbf{ballistic regime}} where the electron-electron scattering is negligible the main contribution to the valley polarization is given by the  combination of the anomalous velocity and side-jump induced anomalous distribution effects due to the residual disorder, Eq.~\eqref{N:val:ball:main}. Corresponding valley polarization is linear in the coordinate, see the red curve in Fig.~\ref{fig:inchannel}. By contrast, in the \textit{\textbf{hydrodynamic regime}}, the anomalous velocity is compensated by the anomalous distribution induced by the electron-electron collisions. The resulting valley polarization is driven by the remaining part of the electron-electron scattering induced anomalous distribution, as well as by the  side-jump accumulation and skew scattering due to the residual disorder, Eq.~\eqref{N:val:hydro:main}. Here the dependence is more complicated, it is linear at $x\ll w$ and saturates near the edge boundaries, blue curve in Fig.~\ref{fig:inchannel}.

Let us compare the valley polarization magnitude for the ballistic and hydrodynamic channels in more detail. Generally, the distributions of electrons in $\bm K_\pm$ valleys can be recast in the form
\begin{equation}
\label{val:accum:gen}
N^+(x) = -N^-(x) = S_0(x) = \mathcal C \frac{\xi e}{v\hbar} E_y N_1 \mathcal D(x),
\end{equation}
where the coefficient $\mathcal C \sim 1$ depends on the mechanism, the factor $\xi e E_yN_1/(v\hbar)$ is the same for all mechanisms, and the function $\mathcal D(x)$ describes the coordinate distribution:
\begin{equation}
\label{val:accum:fun}
\mathcal D(x) = \frac{x}{l} \times
\begin{cases}
1, \quad \mbox{ball},\\
\frac{1}{l_1l_2} \left[\left(\frac{w}{2}\right)^2-\frac{x^2}{3}\right], \quad \frac{w^2}{l l_1} \gg 1, \quad \mbox{hyd},\\
\frac{l}{l_1}, \quad \quad \frac{w^2}{l l_1} \ll 1, \quad \mbox{hyd},
\end{cases}
\end{equation}
for the leading contributions in the ballistic (bal) and hydrodynamic (hyd) regimes. Importantly, the valley accumulation in the hydrodynamic regime is parametrically larger than in ballistic regime because $l_1$, $l_2\sim v\tau_{ee} \ll w$. The difference of the signs of the slopes of the accumulated polarization for the ballistic and hydrodynamic channels results from the fact that in the the anomalous velocity dominates in ballistic channel for the parameters of our calculation.

Interestingly, in the hydrodynamic regime the magnitude and shape of the valley polarization depend on the ratio $l/w$ (at fixed $l_1/w,l_2/w$), see Eqs.~\eqref{N:val:hydro:main}, \eqref{val:accum:fun} and the bottom right inset in Fig.~\ref{fig:inchannel}. This is because with increase in $l/w$ the role of the impurity scattering within the channel is suppressed. Thus, corresponding side-jump and skew-scattering contributions become less important. These contributions provide non-linear shape of the valley polarization arising from the specific shape of the electron distribution in the Poiseuille flow [Eq.~\eqref{df_hydro}], see the second term in curly brackets of Eq.~\eqref{N:val:hydro:main}. Hence, in accordance with the bottom right inset in Fig.~\ref{fig:inchannel} increase in $l/w$ makes the valley polarization smaller and its coordinate dependence closer to a linear one. Note that for fixed system parameters (mean free paths, $\xi$, $N_1$ and $\mathcal C_{sj}, \mathcal C_{sk}$) the valley polarization increases $\propto w$ in the ballistic case or as $w^3$ or $w$ in the hydrodynamic case depending on $w^2/ll_1$, Eq.~\eqref{val:accum:fun}.


\section{Valley and spin accumulation in impurity stripes}\label{sec:imp:str}

\subsection{Model}

Let us now turn to the valley and spin Hall and accumulation effects in the impurity stripes. Here the spin or valley imbalance has two sources: (i) the VHE or SHE induced by the electric field acting on the electrons within the stripes and (ii) the VHE or SHE resulting from the electrons which were driven by the field in the channel and enter the stripes to loose their momentum. The latter is associated with the $\delta f_{\bm p}^\lessgtr(x)$ contributions in the distribution function of electrons in the stripe, see Eqs.~\eqref{df_stripe}. The contribution (i) can be readily evaluated because the electron motion within the stripe is diffusive. Thus, the anomalous Hall current can be calculated after Ref.~\cite{2020arXiv200405091G} with the result
\begin{equation}
\label{VHE:diff:j}
j^\pm_x = \mp \sigma_{\rm VH} E_y, 
\end{equation}
where $\sigma_{\rm VH}$ for all relevant mechanisms is given in the Tab. I of Ref.~\cite{2020arXiv200405091G}. Since no current flows through the stripe boundaries\footnote{Strictly speaking, the current vanishes at $x=w/2+d$ for the right stripe and at $x=-w/2-d$ for the left stripe. These boundary conditions are sufficient to derive Eq.~\eqref{dN:str:i}.} the density gradient forms inducing the diffusive current and compensating the valley or spin Hall current.  To calculate the distribution function and  polarization we use the kinetic Eq.~\eqref{kin:AA} with $Q_{ee}\equiv 0$ and $\tau \to \tau_i$:
\begin{multline}
\label{kin:AA:i}
\frac{\partial}{\partial x} \left[v_x S_{\varphi}(x) + \left(v_a^+ + v_{sj}^+\right) N_1\right] + \frac{ S_{\varphi} - \overline{ S_{\varphi}}}{\tau_i}  \\
= G^+(\varphi,x).
\end{multline}
For the polarization distribution caused by the VHE or SHE within the stripe, Eq.~\eqref{VHE:diff:j}, we have [cf. Eq.~\eqref{DF:va:ball}]
\begin{equation}
\label{dN:str:i}
\Delta N^\pm_{i,1}(x) = \frac{2j_{\rm VH}}{ev l_i}\left(x-\frac{w}{2}\right) = \mp \frac{2\sigma_{\rm VH}}{evl_i}\left(x-\frac{w}{2}\right),
\end{equation}
for $w/2\leqslant x\leqslant w/2+d$  [right impurity stripe, see Fig.~\ref{fig:channel}(b)]; for the left impurity stripe where $-d-w/2\leqslant x\leqslant -w/2$ the distribution is the same with the replacement of $x-w/2$ by $x+w/2$. We stress that Eq.~\eqref{dN:str:i} is derived for the long spin or valley relaxation length in the impurity stripes, $l_{v,i} \gg d$.

We now address the contribution (ii) related to the electrons entering the impurity stripe from the channel. As previously, we use the term valley to denote both the valley and spin degrees of freedom. The contribution to the valley Hall effect and valley accumulation has two origins: the side-jump accumulation and the skew-scattering.\footnote{Note that both the anomalous velocity and anomalous distribution effects are related to the electric field effect on the electrons within the stripe and, thus, included in $\sigma_{\rm VH}$ in Eq.~\eqref{dN:str:i}. There is no need to account electric field action on $\bm E$-linear $\delta f^\lessgtr_{\bm p}(x)$.} Hereafter we consider the right stripe, $w/2\leqslant x\leqslant w/2+d$, for specificity, and the valley polarization in the left stripe can be obtained by the mirror reflection.

We use Eq.~\eqref{v:sj} to evaluate the velocity due to the side-jump accumulation effect with the result
\begin{equation}
\label{v:sj:i}
v_{sj}^\pm(x) = \pm V_a \times
\begin{cases}
 \mathcal F_{1}\left(\frac{x-w/2}{l_i}\right),~~\mbox{ball},\\
 \mathcal F_{-1}\left(\frac{x-w/2}{l_i}\right),~\mbox{hyd},
\end{cases}
\end{equation}
where `ball' and `hyd' refers to the ballistic and hydrodynamic regime of the electron propagation within the channel and the corresponding form of $\delta f^>_{\bm p}(x)$, Eqs.~\eqref{df>b} and \eqref{df>h}, respectively,
\begin{equation}
\label{v:sj:i:1}
V_a = -\left(1+ \frac{U_v}{U_c}\right)\frac{\xi e E_y}{\hbar}\frac{w}{l_i}\times \begin{cases}1,~~\mbox{ball},\\
4 ,~~\mbox{hyd},
\end{cases}
\end{equation}
and we introduced the set of functions (their properties are presented in Appendix~\ref{app:Cn})
\begin{equation}
\label{aux:Cn}
\mathcal F_n(z) = \int_{-\pi/2}^{\pi/2} \frac{\sin^2{\varphi}}{\cos^n{\varphi}} 
\exp(-\frac{z}{\cos{\varphi}}) \frac{d\varphi}{\pi}, \quad z>0.
\end{equation}
One can readily check that $ S_\varphi(x)$ in the form
\begin{multline}
\label{dS:sj:i}
 S^{sj,>}_\varphi(x) =  -\frac{2V_a}{v} N_1 \sin^2\varphi \exp(-\frac{x-w/2}{l_i\cos{\varphi}})\\
\times 
\begin{cases}
\cos^{-2}{\varphi}, \quad \mbox{ball},\\
1, \quad \mbox{hyd},
\end{cases}
\end{multline}
satisfies Eq.~\eqref{kin:AA:i} with $G^+\equiv 0$, $v_a^+\equiv 0$ and $v_{sj}^+$ given by Eq.~\eqref{v:sj:i}, see Appendix~\ref{app:kineq:imp} for details. Particularly, for the accumulated valley distribution we have
\begin{multline}
\label{impurity:stripe:sj}
\Delta N^{\pm,sj}_{i,>}(x) = \pm \overline{ S^{sj,>}_\varphi(x)} \\
= \mp \frac{V_a}{v} N_1\begin{cases}
\mathcal F_{2}\left(\frac{x-w/2}{l_i} \right), \quad \mbox{ball},\\
\mathcal F_0 \left(\frac{x-w/2}{l_i} \right), \quad \mbox{hyd}.\\
\end{cases}
\end{multline}

The generation term due to the skew scattering can be recast in the following form by virtue of Eq.~\eqref{G:sk}
\begin{equation}
\label{G:i:sk}
G^+(\varphi,x) =\cos{\varphi} \, G_{sk} \begin{cases}
\mathcal F_1\left(\frac{x-w/2}{l_i} \right), \quad \mbox{ball},\\
\mathcal F_{-1}  \left(\frac{x-w/2}{l_i} \right), \quad \mbox{hyd}.
\end{cases}
\end{equation}
where the parameter $G_{sk}$ is given by 
\begin{equation}
\label{Gsk:i}
G_{sk} =\frac{2\xi e}{\hbar} E_y N_1 S_{imp}' g\tau_i \frac{\varepsilon_F\tau_i}{\hbar} \frac{w}{l_i^2} \times
\begin{cases}1,~~\mbox{ball},\\
4 ,~~\mbox{hyd}.
\end{cases}
\end{equation}
Note that in Eq.~\eqref{Gsk:i} we introduced $S_{imp}'$ responsible for the skew scattering by the disorder in the impurity stripe. It has a form similar to Eq.~\eqref{Simp} but with the parameters of the impurity stripe:
\begin{equation}
\label{Simp'}
S_{imp}' = {{2\pi}U_v\over \tau_i} -\frac{U_v}{U_c} \frac{\hbar}{g\varepsilon_F \tau_i^2}.
\end{equation}

In this situation where $G^+\propto \cos{\varphi}$, the solution of Eq.~\eqref{kin:AA:i} reads [see Appendix~\ref{app:kineq:imp}]
\begin{equation}
\label{dS:sk:i}
S^{sj,>}_\varphi(x) = -\frac{1}{v\cos{\varphi}} \int_{x}^\infty G^+(\varphi,x') dx'.
\end{equation}
Accordingly,
\begin{multline}
\label{impurity:stripe:sk}
\Delta N^{\pm,sk}_{i,>}(x) = \pm \overline{ S^{sj,>}_\varphi(x)} \\
= \mp \frac{G_{sk} l_i}{v} \begin{cases}
\mathcal F_{0}\left(\frac{x-w/2}{l_i} \right), \quad \mbox{ball},\\
\mathcal F_{-2} \left(\frac{x-w/2}{l_i} \right), \quad \mbox{hyd}.
\end{cases}
\end{multline}
Equations~\eqref{dN:str:i}, \eqref{impurity:stripe:sj}, and \eqref{impurity:stripe:sk} give the valley accumulation in the right impurity stripe. As already mentioned, the results for the left stripe can be obtained by the mirror reflection.

\subsection{Discussion}

Let us analyse the magnitudes of the effects in more detail. To that end we present the valley Hall conductivity $\sigma_{\rm VH}$ in Eqs.~\eqref{VHE:diff:j} and \eqref{dN:str:i} as~\cite{2020arXiv200405091G}
\begin{equation}
\label{sigma:vh}
\sigma_{\rm VH} = \mathcal C_{\rm VH} \frac{\xi e^2}{\hbar} \xi N_1,
\end{equation}
where $\mathcal C_{\rm VH}$ is the dimensionless coefficient  on the order of unity determined by the details of the scattering. As a result, for $\Delta N^\pm_{i,1}$ we obtain
\begin{equation}
\label{dN:str:i:1}
\Delta N_{i,1}^\pm(x) = \mp 2\mathcal C_{\rm VH} \frac{\xi e}{v\hbar} E_y N_1 \frac{x-w/2}{l_i}.
\end{equation}
Naturally, this contribution related to the VHE on the electrons resident in the impurity stripe depends linearly on the coordinate, and its slope is determined by $1/l_i$. The contributions due to the electrons entering the stripe, Eqs. \eqref{impurity:stripe:sj}, and \eqref{impurity:stripe:sk}, can be combined in the similar form
\begin{equation}
\label{impurity:stripe:1}
\Delta N_{i,2}^\pm(x) = \pm \mathcal C' \frac{\xi e}{v\hbar} E_y N_1 \frac{w}{l_i} \mathcal F\left(\frac{x-w/2}{l_i}\right).
\end{equation}
Here $\mathcal C'$ is the coefficient (typically, on the order of unity) describing the side-jump accumulation and skew-scattering effects [cf. Eq.~\eqref{N:val:hydro:main}] and $\mathcal F(x)$ is the appropriate coordinate-dependent function. Since the impurity stripes are short and characterized by small mean free path $l_i \ll d \ll w$ the factor $w/l_i\gg 1$ in Eq.~\eqref{impurity:stripe:1}. Therefore the dominant contribution to the valley accumulation in the stripes comes from the electrons entering the stripe from the channel and loose their momentum in the stripe, $|\Delta N_{i,2}^\pm(x)| \gg |\Delta N_{i,1}^\pm(x)|$.

\begin{figure}[tb]
\includegraphics[width=0.95\linewidth]{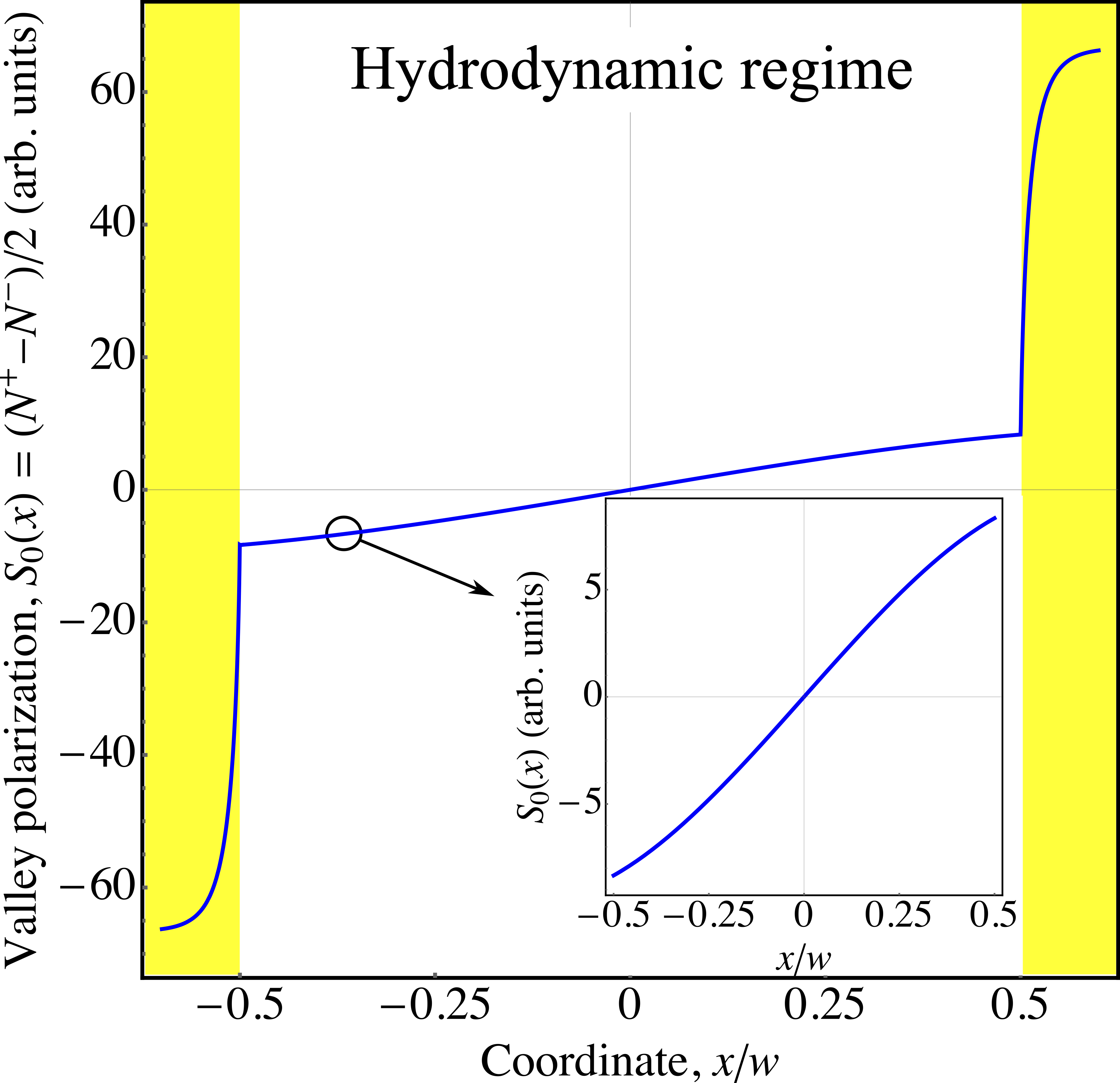}
\caption{Spin or valley polarization in the ultraclean channel in the hydrodynamic regime calculated after Eq.~\eqref{N:val:hydro:main} (inside the channel) and Eq.~\eqref{impurity:stripe:sj} (in the impurity stripes shown by yellow shaded areas). Shaded areas show the impurity stripes. For the illustrative purposes we took into account only the side-jump contributions. Inset shows the zoom-in of the valley polarization inside the channel, cf. Fig.~\ref{fig:inchannel}.  The parameters are $l/w=10$, $l_1=l_2=w/10$, $d=w/10$, $l_i=d/3$. }\label{fig:total}
\end{figure}

We compare now the results for the valley accumulation within the channel and the impurity stripes assuming that the dimensionless coefficients $\mathcal C$, $\mathcal C'$, \ldots are about the same.  In the ballistic regime the comparison is straightforward: Making use of Eqs.~\eqref{val:accum:gen} and \eqref{val:accum:fun} we obtain that the valley accumulation in the channel is suppressed by the factor $l_i/l$ as compared with the valley accumulation in the impurity stripe. In the hydrodynamic regime the situation is more complex: The profile of the valley polarization within the channel is controlled by the parameter $r$, Eq.~\eqref{rsmall}. The corresponding ratio of the valley density in the channel and in the impurity edge is determined by another ratio
\[
R=
\begin{cases}
\frac{l_i}{l} \left(\frac{w}{l_{ee}}\right)^2, \quad r\gg 1,\\
\frac{l_i}{l_{ee}}, \quad r\ll 1.
\end{cases}
\]
This ratio $R$ can, in general, be large or small. 
For very long impurity-induced mean free path $r\ll 1$ and $R\ll 1$ as well, since in any case $l_i \ll l_{ee}$ and the accumulation in the impurity stripes dominates. For moderate mean free paths where $r \gg 1$ the parameter $R\gtrsim 1$ and the valley polarization in the channel can be  dominant.

\begin{figure}[tb]
\includegraphics[width=0.9\linewidth]{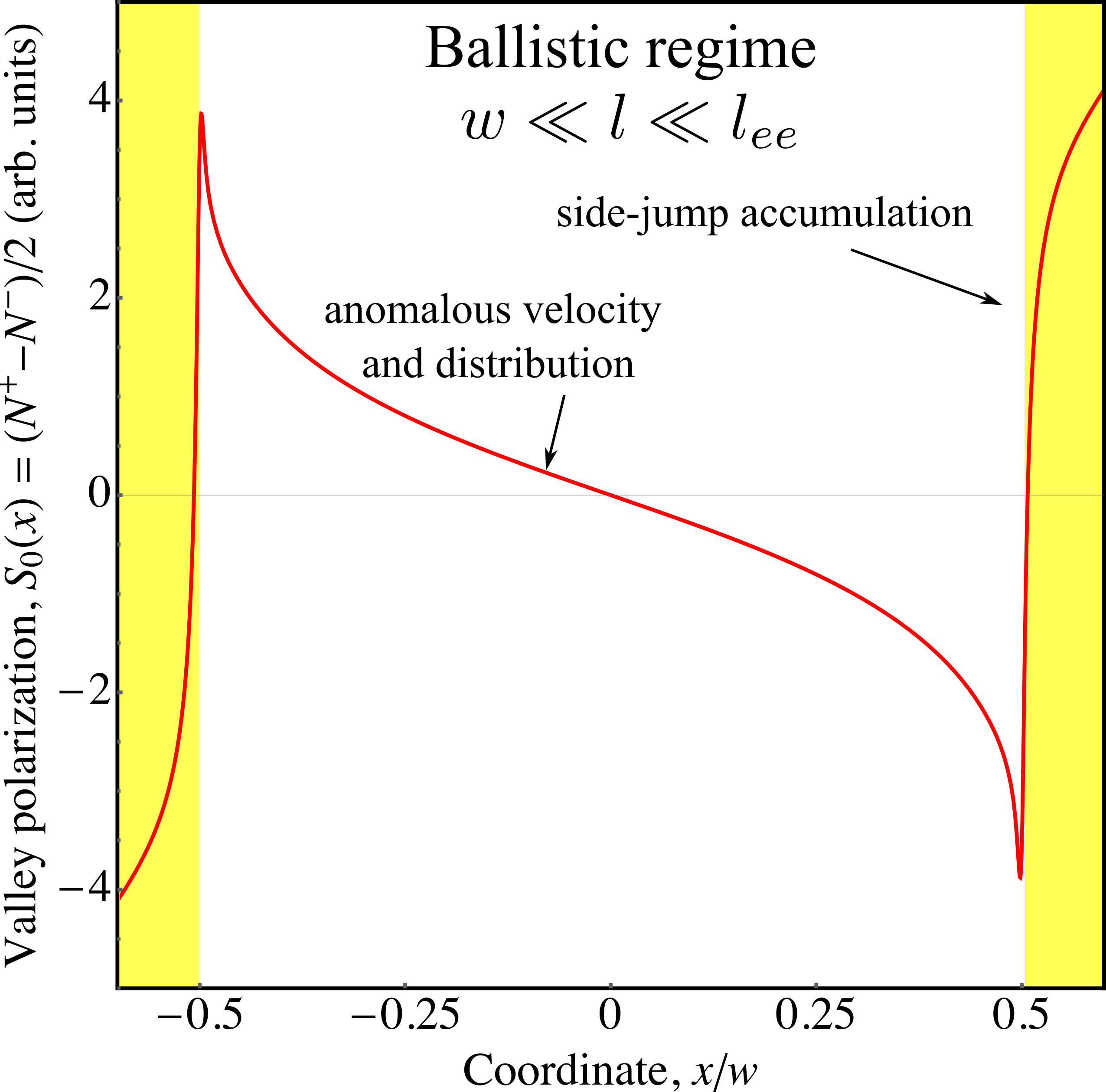}
\caption{Spin or valley polarization in the ultraclean channel in the ballistic regime calculated after Eq.~\eqref{sol}, see Appendix~\ref{app:lambda}. For illustrative purposes we disregard skew scattering and take $U_v=U_c$ which results in the cancellation of the anomalous velocity in the channel. Parameters are $w/l = 1/5$, $l_i/l = 1/10$, the mean free path changes between  $l$ and $l_i$ at the channel edges on the length $\chi=l/20$. Yellow shaded areas demonstrate impurity stripes.}\label{fig:total:ball}
\end{figure}

The dependence of the valley polarization on the coordinate $x$ should be continuous across the whole system ``channel+impurity stripes''. Thus, to construct the complete solution the constants should be added to $\Delta N^\pm_>(x)$ and $\Delta N^\pm_<(x)$ to match the solutions within the channels and in the impurity stripes. Such a procedure works in the case of hydrodynamic regime where the solutions within the impurity stripe are given by the smooth functions $\mathcal F_0$ and $\mathcal F_{-2}$, see Eqs.~\eqref{impurity:stripe:sj} and \eqref{impurity:stripe:sk} and Appendix~\ref{app:Cn}. Corresponding valley polarization shown in Fig.~\ref{fig:total} is calculated using Eq.~\eqref{N:val:hydro:main} (inside the channel) and Eq.~\eqref{impurity:stripe:sj} (in the impurity stripes) and matching the valley polarization $S_0$ at the channel-stripe boundaries in the hydrodynamic regime. In this case, the parameter $R \ll 1$, and the valley accumulation in the impurity stripes is largest.

In the ballistic regime such simple matching of solutions is impossible: If follows from Eq.~\eqref{impurity:stripe:sj} that the side-jump contribution diverges at the boundary between the impurity stripe and the channel. Indeed, function $\mathcal F_2(|x\pm w/2|/l) \propto |x\pm w/2|^{-1}$, Eq.~\eqref{aux:Cn:small}, making it impossible to match the solutions. Partially, this divergence results from using the simplified form of the distribution functions inside the impurity stripes, Eqs.~\eqref{df<b} and \eqref{df>b}. These simplified solutions are inapplicable for $|\cos{\varphi}|\ll w/l$. Replacing the approximate solutions by their exact form, see Eq.~\eqref{rep:b}, we weaken the singularity: For $|x\pm w/2|/l \ll w/l$ function $\mathcal F_2$ is replaced by $\mathcal F_1$ which has only logarithmic divergence. Eq~\eqref{aux:C1:small}. This logarithmic divergence comes from the simplification of our model where the momentum relaxation time abruptly changes between the channel and impurity stripe. This divergence, together with the jump of the side-jump accumulation contribution to the velocity, results in the divergence of the valley occupancy at the edge.\footnote{In hydrodynamic regime the electron-electron scattering provide small length-scale $l_{ee}$ and efficiently smooths-out possible divergencies} To overcome the divergence, we need to take into account that the transition between the channel and impurity stripe is not abrupt, but the scattering time $\tau$ (and the mean free path $l$) continuously changes between $\tau$ and $\tau_i$ ($l$ and $l_i$) as a function of coordinate. While the detailed analysis of the effects in ballistic channels will be reported elsewhere, we  briefly summarize the calculations in Appendix~\ref{app:lambda}. The calculated of the valley polarization in the ballistic channel are presented in Fig.~\ref{fig:total:ball}. One clearly see the pronounced features in the valley polarization arising in the vicinity of the channel edges, confirming our qualitative analysis presented above.

\section{Summary}\label{sec:concl}

We developed the theory of the valley and spin Hall effects and polarization accumulation in the ultraclean channels made from two-dimensional semiconductors where the mean free path of the electron $l$ exceeds by far the channel width $w$. In this case the electron conductivity is controlled by the edge scattering. We considered two regimes of the electron transport: \emph{ballistic}, where the electrons rarely scatter off residual disorder, and \emph{hydrodynamic}, where the electron-electron collisions are efficient such that the electron-electron scattering free path $l_{ee} \ll w$. Correspondingly, in the former regime the electrons are practically ballistic, while in the latter regime the electrons can be considered as a viscous fluid. 

Under conditions of the valley or spin Hall effect the electric field applied along the channel results in the transversal valley or spin current. The latter gives rise to the  gradient of the valley or spin polarization -- inhomogeneous polarization accumulation -- which compensates the transversal valley current. 

We identified the key mechanisms of the polarization accumulation related to the (i) anomalous velocity in the external electric field, (ii) side-jump, and (iii) skew-scattering and analyzed their contributions to the valley and spin accumulation effect. These mechanisms stem from the effective spin-orbit coupling resulting from the mixing of the conduction and valence bands. Generally, all these mechanisms are equally important and manifest themselves at different parts of the channel. 

At the \emph{ballistic} regime, the valley or spin accumulation inside the channel is mainly controlled by the anomalous velocity and side-jump induced anomalous distribution, while the contributions of the side-jump accumulation and skew scattering are smaller by the factor $\sim w/l$. 

At the \emph{hydrodynamic} regime, additional contributions to the effect result from the electron-electron collisions which both provide efficient dissipation of the valley current, and contribute to the side-jump anomalous distribution effect. As a result, the anomalous velocity contribution inside the channel is compensated. We show that the valley or spin accumulation in hydrodynamic channels is controlled by an interplay of the side-jump accumulation and skew scattering by the residual impurities in the channel, as well as by the electron-electron scattering induced anomalous distribution. At the same time, the anomalous distribution effect due to the impurities is suppressed by the factor $w^2/l_{ee}^2$.

Importantly, the significant accumulation of the valley or spin polarization takes place in the immediate vicinity of the channel edges. The physical reason is related to the fact, that the electron momentum relaxation is provided by the edges. To calculate the valley accumulation near the edge we suggested the impurity stripe model, which assumes that in the narrow stripes near the channels edges the impurity density is so high that the electron transport becomes diffusive. The calculated distribution of the polarization near the edge depends on the transport regime in the channel.

Note that additional contribution to the valley or spin accumulation in the channel may result from the rotational viscosity of the electrons which transforms a curl of the velocity to the valley or spin polarization~\cite{de2013non,PhysRevB.96.020401,Takahashi:2020um}. This effect requires spin or valley relaxation and deserves further studies.

Regarding the observation of spin or valley Hall effect in ultraclean channels, we expect that the effect is easier to observe in conventional GaAs-based Hall bar samples where the mean free path due to the residual disorder exceeds tens of microns. However, the progress in 2D material technology makes us confident that the structures based on transition metal dichalcogenide monolayers with $l \gg w$ will also be available opening up prospects to study ballistic and hydrodynamic transport in such systems. 

\section{Conclusion}\label{sec:concl:new}

To conclude, the theory of the spin and valley accumulation in ultraclean two-dimensional semiconductor channels is proposed. The theory is considers both ballistic and hydrodynamic regimes of the electron transport and the model takes into account all relevant contributions to the valley and spin Hall effect. We demonstrate that the main contribution to the valley and spin accumulation appear in the vicinity of the channel edges where the momentum relaxation occurs.

\acknowledgements

The author is grateful to L.E. Golub  and P.S. Alekseev for valuable discussions. The financial support of the Russian Science Foundation (Project No. 17-12-01265) is acknowledged.

\appendix

\section{Skew scattering at the electron-electron collisions}\label{app:ee-skew}

The absence of the skew scattering induced VHE for the collisions of the electrons belonging to the same valley, $\sigma_1=\sigma_2$, is straightforward to show. Indeed, despite the presence of asymmetric terms in Eq.~\eqref{me:ee}, the valley Hall current cannot be generated because of the momentum conservation law: The transversal to the  $y$-axis component of the electron momentum cannot appear due to the electron-electron scattering.

\begin{widetext}
Thus, the case of interest for us is then the electrons are in the opposite valleys. To analyze the effect in this case, we introduce the antisymmetrized states as
\begin{subequations}
\label{anti}
\begin{align}
&|\bm k+,\bm k'-\rangle = \frac{1}{2\sqrt{\mathcal S}}\left[e^{\mathrm i \bm k\bm r_1 + \mathrm i \bm k'\bm r_2} \mathcal U_{\bm k, +}(\bm r_1)\mathcal U_{\bm k', -}(\bm r_2) - e^{{\mathrm i \bm k\bm r_2 + \mathrm i \bm k'\bm r_1}}  \mathcal U_{\bm k', -}(\bm r_1)\mathcal U_{\bm k, +}(\bm r_2)\right],\\
&|\bm k-,\bm k'+\rangle = \frac{{-1}}{2\sqrt{\mathcal S}}\left[e^{\mathrm i \bm k\bm r_1 + \mathrm i \bm k'\bm r_2} \mathcal U_{\bm k, -}(\bm r_1)\mathcal U_{\bm k', +}(\bm r_2) - e^{{\mathrm i \bm k\bm r_2 + \mathrm i \bm k'\bm r_1}}  \mathcal U_{\bm k', +}(\bm r_1)\mathcal U_{\bm k, -}(\bm r_2)\right].\label{second}
\end{align}
\end{subequations} 
Here $\mathcal S$ is the normalization area, $\pm$ in the brackets denote the valley, $\mathcal U_{\bm k,\sigma}(\bm r)$ is the corresponding Bloch function, and the overall minus sign in Eq.~\eqref{second} is introduced for convenience.
Making use of Eq.~\eqref{me:ee} we derive the matrix elements of the electron-electron scattering
\begin{subequations}
\label{me:ee:anti}
\begin{align}
M\begin{pmatrix}
\bm p+, \bm k+\\
\bm p'-, \bm k'-
\end{pmatrix} = 
\delta_{\bm k+ \bm k',\bm p+ \bm p'} V(\bm p - \bm k) \{1+ \mathrm i \xi [\bm p\times \bm k]_z - \mathrm i \xi[\bm p'\times \bm k']_z\},\\
M\begin{pmatrix}
\bm p-, \bm k+\\
\bm p'+, \bm k'-
\end{pmatrix} = 
\delta_{\bm k+ \bm k',\bm p+ \bm p'} V(\bm p' - \bm k) \{1+ \mathrm i \xi [\bm p'\times \bm k]_z - \mathrm i \xi[\bm p\times \bm k']_z\}.
\end{align}
\end{subequations}
\end{widetext}
The asymmetry of the scattering rates appears, similarly to the case of impurities or phonons, beyond the Born approximation. The derivation of the full collision integral is quite involved. Here, for illustrative purposes we  consider the non-degenerate electron gas in order to neglect the occupancies of the final and intermediate states and greatly simplify the analysis.

To calculate the in-scattering rate to the state $|\bm k+,\bm k-\rangle$ we need to take into account that the transitions are possible from both $|\bm p+,\bm p'-\rangle$ and $|\bm p-,\bm p'+\rangle$ states. As an intermediate states one has two possible pairs $|\bm k_1+,\bm k_1'-\rangle$ or $|\bm k_1-,\bm k_1'+\rangle$. Following Ref.~\cite{Sturman_1984} we obtain
\begin{equation}
\label{Q:in}
Q_{|\bm k+,\bm k-\rangle}^{in} = \sum_{\bm p,\bm p',\sigma=\pm} W\begin{pmatrix}
\bm k+, &\bm p\sigma\\
\bm k'-, &\bm p-\sigma
\end{pmatrix} f_{\bm p,\sigma} f_{\bm p',-\sigma},
\end{equation}
where $f_{\bm p,\sigma}$ is the electron distribution function in the valley $\sigma$, and
\begin{widetext}
\begin{multline}
\label{W:ee}
W
\begin{pmatrix}
\bm k+, &\bm p\sigma\\
\bm k'-, &\bm p'-\sigma
\end{pmatrix}
= \frac{2\pi}{\hbar} \delta_{\bm k+ \bm k',\bm p+ \bm p'} \delta(E_{\bm k} + E_{\bm k'} - E_{\bm p} - E_{\bm p'}) \times
\\
\left[\left|M\begin{pmatrix}
\bm k+, &\bm p\sigma\\
\bm k'-,&\bm p'-\sigma
\end{pmatrix}\right|^2+
 {2\pi} \sum_{\bm k_1, \bm k_1',\sigma_1} \delta_{\bm k+\bm k',\bm k_1+\bm k_1'}\delta(E_{\bm k} +E_{\bm k'} - E_{\bm k_1} - E_{\bm k_1'})
 \right.
 \\
 \left.
 \times \Im\left\{M\begin{pmatrix}
\bm k+,&\bm k_1\sigma_1\\
\bm k'-,& \bm k_1'-\sigma_1
\end{pmatrix}
M\begin{pmatrix}
\bm k_1\sigma_1, &\bm p\sigma\\
\bm k_1'-\sigma_1,& \bm p'-\sigma
\end{pmatrix}
M\begin{pmatrix}
\bm  p\sigma, &\bm k+\\
\bm p-\sigma, &\bm k'-
\end{pmatrix} \right\}
 \right].
\end{multline}
\end{widetext}
Analogously, the out-scattering term takes the form
\begin{equation}
\label{Q:out}
Q_{|\bm k+,\bm k-\rangle}^{out} = \sum_{\bm p,\bm p',\sigma=\pm} W\begin{pmatrix}
 \bm p\sigma,&\bm k+\\
\bm p'-\sigma, &\bm k'-
\end{pmatrix} f_{\bm k,+} f_{\bm k',-}.
\end{equation}

Analysis of Eqs.~\eqref{me:ee:anti} and \eqref{W:ee} demonstrates that 
\begin{subequations}
\label{W:ee:expl}
\begin{align}
&W
\begin{pmatrix}
\bm k+, &\bm p+\\
\bm k'-, &\bm p'-
\end{pmatrix} = W_0(\bm k \bm k'; \bm p \bm p')\\ 
&+
 W_1(\bm k \bm k'; \bm p \bm p')\left\{[\bm p \times \bm k]_z - [\bm p'\times \bm k']_z \right\}, \nonumber
 \\
& W
\begin{pmatrix}
\bm k+, &\bm p-\\
\bm k'-, &\bm p'+
\end{pmatrix} = W_0(\bm k \bm k'; \bm p' \bm p) \\
&+ 
 W_1(\bm k \bm k'; \bm p' \bm p)\left\{[\bm p' \times \bm k]_z - [\bm p\times \bm k']_z \right\}. \nonumber
\end{align}
\end{subequations}
where $W_0(\bm k \bm k'; \bm p' \bm p)$ and $W_1(\bm k \bm k'; \bm p' \bm p)$ are the symmetric functions of the scattering angles. Terms $\propto W_1$ describe the skew effect at the electron-electron scattering. Note that the asymmetric sum over $\bm p, \bm p'$ in Eq.~\eqref{Q:out} vanishes and the kinetic equation can be brought to the form
\begin{equation}
\label{kin:ee}
\frac{\partial f_{\bm k}^+}{\partial t} + Q_{\bm k}\{f^+,f^-\} =0, 
\end{equation}
where
\begin{multline}
\label{Q:ee}
Q_{\bm k}\{f^+, f^-\} = 2\sum_{\bm p\bm p'\bm k'} 
\frac{2\pi}{\hbar} |V_{\bm p - \bm k}|^2 \delta_{\bm k + \bm k', \bm p+ \bm p'} \\
\times \delta (E_{\bm k} + E_{\bm k'} - E_{\bm p} - E_{\bm p'})
 (f_{\bm k}^+ f_{\bm k'}^- - f_{\bm p}^+ f_{\bm p'}^-) \\
 -  2\sum_{\bm p\bm p'\bm k'}W_1(\bm k\bm k';\bm p\bm p')\left\{[\bm p \times \bm k]_z - [\bm p'\times \bm k']_z \right\} f_{\bm p}^+ f_{\bm p'}^-.
 \end{multline}
First lines in Eq.~\eqref{Q:ee} describe standard (symmetric) electron-electron scattering in agreement with Refs.~\cite{glazov02,amico02,glazovnato}. Last line in Eq.~\eqref{Q:ee} describes the skew-scattering effect at the electron-electron collisions. One can show, however, that due to the energy and momentum conservation the valley Hall effect is absent, provided that $\bm f_{\bm p}^+ = \bm f_{\bm p}^- = f_0(E_{\bm p})(1 + \bm u\cdot \bm p)$ (with $\bm u$ being a constant vector, i.e., for the current-carrying distribution). Indeed, in such a case the energy and momentum conservation laws yield  $f_{\bm p}^+ f_{\bm p'}^- = f_{\bm k}^+ f_{\bm k'}^-$ and the summation over $\bm p$ and $\bm p'$ in the asymmetric term vanishes.

\section{Properties of functions $\mathcal F_n(z)$}\label{app:Cn}

The functions $\mathcal F_n(z)$ are defined as [cf Eq.~\eqref{aux:Cn}]
\begin{equation}
\label{aux:Cn:app}
\mathcal F_n(z) = \int_{-\pi/2}^{\pi/2} \frac{\sin^2{\varphi}}{\cos^n{\varphi}} 
\exp(-\frac{z}{\cos{\varphi}}) \frac{d\varphi}{\pi}, \quad z>0.
\end{equation}
For $z>0$ the integral converges because at $\cos{\varphi}\to 0$ the exponent rapidly vanishes. 

At $z\to \infty$ the leading contribution to the integral results from $\varphi \to 0$ where the $\cos{\varphi}$ reaches its maximum. As a result it is possible to replace $\exp(-z/\cos{\varphi})$ by $\exp[-z(1+\varphi^2/2)]$ and $\sin^2{\varphi}/\cos^n{\varphi}$ by $\varphi^2$. As a result
\begin{multline}
\label{aux:Cn:large}
\mathcal F_n(z) \approx \int_{-\infty}^{\infty} \varphi^2
\exp[-{z}\left(1+\frac{\varphi^2}{2}\right)] \frac{d\varphi}{\pi}\\
= \sqrt{\frac{2}{\pi z^3}} e^{-z}, \quad z\to \infty.
\end{multline}

\begin{figure}[h]
\includegraphics[width=0.9\linewidth]{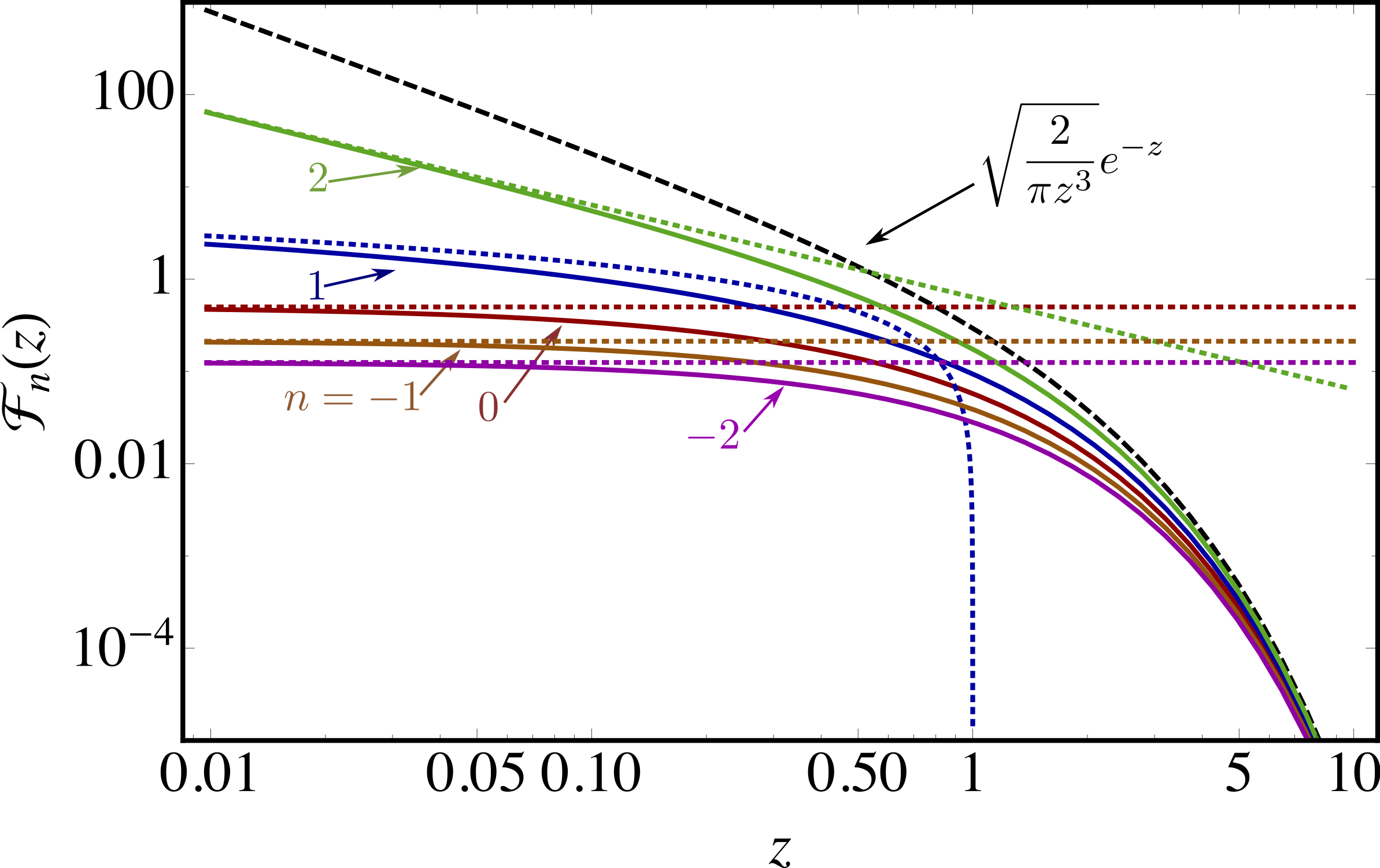}
\caption{Functions $\mathcal F_n(z)$ calculated numerically after Eq.~\eqref{aux:Cn:app} (solid lines) and their asymptotics: large-$z$, Eq.~\eqref{aux:Cn:large} (black dashed), and small-$z$, Eqs.~\eqref{aux:Cn:small:all} (dotted).}\label{fig:Cn}
\end{figure}

At $z\to 0$ for $n>0$ the main contribution to the integral comes from the points where $\cos{\varphi}\to 0$, i.e., at $\varphi \approx \pm \pi/2$. Changing the integration variable and assuming that $\alpha = |\varphi \mp \pi/2|\ll 1$ we have for $n>1$
\begin{subequations}
\label{aux:Cn:small:all}
 \begin{multline}
\label{aux:Cn:small}
\mathcal F_n(z) \approx \frac{2}{\pi}\int_{0}^{1} \frac{\exp(-\frac{z}{\alpha})}{\alpha^n} d\alpha \\
\approx  \frac{2(n-2)!}{\pi z^{n-1}}, \quad z\to 0,~~n>1.
\end{multline} 
For $n=1$ the leading contribution is logarithmic
 \begin{equation}
\label{aux:C1:small}
\mathcal F_1(z) \approx \frac{2}{\pi} \ln{1/z}, \quad z\to 0.
\end{equation} 
For $n<0$ the divergence is absent and
 \begin{equation}
\label{aux:Cn:0}
\mathcal F_n(0) = \frac{\Gamma\left[(1-n)/2\right]}{2\sqrt{\pi}\Gamma\left[(2-n)/2\right]},
\end{equation} 
\end{subequations}
where $\Gamma(x)$ is the $\Gamma$-function. The plots of relevant functions $\mathcal F_{n}(z)$ for $n=-2,\ldots,2$ are shown in Fig.~\ref{fig:Cn} together with their asymptotics.

\section{Solution of kinetic equation in impurity stripes}\label{app:kineq:imp}

Here we present the analytical solution of kinetic Eq.~\eqref{kin:AA:i} describing the valley accumulation in the impurity stripes. We consider the right stripe, $x\geqslant w/2$ for specificity.

First, we address the effect of the anomalous velocity. Let us assume, in agreement with Eq.~\eqref{v:sj:i}, that 
\begin{equation}
\label{va:imp:app}
v_a^+(x) = \int_{-\pi/2}^{\pi/2} \Phi(\varphi) \exp(-\frac{x-w/2}{l_i\cos{\varphi}}) \frac{d\varphi}{\pi},
\end{equation}
where $\Phi(\varphi)$ is an arbitrary function of the angle $\varphi$. Let us check that
\begin{equation}
\label{S:va:imp:app}
S_\varphi(x) =-\frac{2N_1}{v\cos{\varphi}} \Phi(\varphi) \exp(-\frac{x-w/2}{l_i\cos{\varphi}})\Theta(\cos{\varphi})
\end{equation}
satisfies Eq.~\eqref{kin:AA:i} (with $G^+=0$). Indeed,
\begin{multline}
\overline{S_\varphi} = -\frac{N_1}{v} \int_{-\pi/2}^{\pi/2} \frac{\Phi(\varphi)}{\cos{\varphi}} \exp(-\frac{x-w/2}{l_i\cos{\varphi}}) \frac{d\varphi}{\pi} 
\\
= \frac{N_1}{\tau_i} \frac{\partial v_a(x)}{\partial x},
\end{multline}
while 
\begin{multline}
\frac{\partial}{\partial x} v_x S_{\varphi}(x) = \frac{2N_1}{l_i\cos{\varphi}} \Phi(\varphi) \exp(-\frac{x-w/2}{l_i\cos{\varphi}})\Theta(\cos{\varphi}) \\
=-\tau_i^{-1} S_\varphi(x).
\end{multline}
Hence
\[
\frac{\partial}{\partial x} \left[v_x S_{\varphi}(x) + v_a^+(x) N_1\right] = -\frac{1}{\tau_i} \left(S_{\varphi}(x) - \overline{S_\varphi(x)}\right)
\]
and Eq.~\eqref{kin:AA:i} is fulfilled.

Second, we consider the effect of the generation. In accordance with Eq.~\eqref{G:i:sk} we take $G^+(\varphi,x) = \cos{\varphi} \mathcal G(x)$ with $\mathcal G(x)$ being an arbitrary function of coordinate. We take $S_\varphi(x)$ to be $\varphi$-independent, as a result the cosine terms in Eq.~~\eqref{kin:AA:i} cancel and we have
\[
v \frac{\partial S_\varphi}{\partial x} = \mathcal G(x),
\]
which yields Eq.~\eqref{dS:sk:i} of the main text.

The solutions obtained above vanish at $x\to \infty$ (for the right stripe). Corresponding solutions for $x\to -\infty$ can be obtained by the mirror symmetry. To obtain the solution in the whole system one has to add arbitrary constants to $S_0(x) = \overline{S_\varphi(x)}$ in order to make $S_0(x)$ continuous at $x=\pm w/2$. 

\section{Valley accumulation for arbitrary $l(x)$}\label{app:lambda}

Our goal is to solve the kinetic equation for the valley accumulation in the channel with arbitrary dependence of the impurity induced mean free path $l(x)$ [cf. Eqs.~\eqref{kin:AA} and \eqref{kin:AA:i}]:
\begin{equation}
\label{kin:AA:lambda}
\frac{\partial}{\partial x} \left[\cos{\varphi} S_{\varphi}(x) + V_a(x)\right] + \frac{ S_{\varphi} - \overline{ S_{\varphi}}}{l(x)} 
= g(\varphi,x),
\end{equation}
and determine the profile of the valley polarization $\overline S_{\varphi}(x)$. Here $V_a(x) = N_1[v_a^+ + v_{sj}^+(x)]/v$ and $g(\varphi,x) = G^+(\varphi,x)/v$ are the reduced anomalous velocity and generation rate, respectively. The electron-electron collisions are neglected. To solve Eq.~\eqref{kin:AA:lambda} for arbitrary $l(x)$, $V_a(x)$ and $g(x)$ it is convenient to pass to the new variable $y$ defined by
\begin{equation}
\label{y:new}
\frac{dy}{dx} = \frac{1}{l(x)}, \quad y(x) = \int_0^x \frac{dx'}{l(x')}.
\end{equation}
In this case the dependence of $l(x)$ vanishes. Next, we pass from Eq.~\eqref{kin:AA:lambda} to the integral equation for $S_0(y) \equiv \overline{S_\varphi(y)}$ as
\begin{equation}
\label{int:eq:S0}
S_0(y) - \int_{-\infty}^\infty S_0(y_1) Q(y-y_1) dy_1 = R(y),
\end{equation}
where the kernel 
\[
Q(y) = \int_{-\pi/2}^{\pi/2} \frac{\exp(-\frac{|y|}{\cos{\varphi}})}{\cos{\varphi}} = \frac{1}{\pi} K_0(|y|),
\]
with $K_0(y)$ being the modified Bessel function, and 
\[
R(y) = \int_{-\infty}^\infty \left[-\frac{dV_a}{dy} + g(\varphi,y) \right] Q(y-y_1) dy_1.
\]
Equation~\eqref{int:eq:S0} can be solved by the Fourier transform
\begin{equation}
\label{sol}
S_0(y) = \int_{-\infty}^\infty \frac{d\xi}{2\pi} e^{\mathrm i \xi y} S(\xi),~~S(\xi) = \frac{R(\xi)}{1-\frac{1}{\sqrt{1+\xi^2}}},
\end{equation}
where $R(\xi) = \int_{-\infty}^\infty R(y) \exp(-\mathrm i \xi y)$.


\begin{thebibliography}{100}%
\makeatletter
\providecommand \@ifxundefined [1]{%
 \@ifx{#1\undefined}
}%
\providecommand \@ifnum [1]{%
 \ifnum #1\expandafter \@firstoftwo
 \else \expandafter \@secondoftwo
 \fi
}%
\providecommand \@ifx [1]{%
 \ifx #1\expandafter \@firstoftwo
 \else \expandafter \@secondoftwo
 \fi
}%
\providecommand \natexlab [1]{#1}%
\providecommand \enquote  [1]{``#1''}%
\providecommand \bibnamefont  [1]{#1}%
\providecommand \bibfnamefont [1]{#1}%
\providecommand \citenamefont [1]{#1}%
\providecommand \href@noop [0]{\@secondoftwo}%
\providecommand \href [0]{\begingroup \@sanitize@url \@href}%
\providecommand \@href[1]{\@@startlink{#1}\@@href}%
\providecommand \@@href[1]{\endgroup#1\@@endlink}%
\providecommand \@sanitize@url [0]{\catcode `\\12\catcode `\$12\catcode
  `\&12\catcode `\#12\catcode `\^12\catcode `\_12\catcode `\%12\relax}%
\providecommand \@@startlink[1]{}%
\providecommand \@@endlink[0]{}%
\providecommand \url  [0]{\begingroup\@sanitize@url \@url }%
\providecommand \@url [1]{\endgroup\@href {#1}{\urlprefix }}%
\providecommand \urlprefix  [0]{URL }%
\providecommand \Eprint [0]{\href }%
\providecommand \doibase [0]{http://dx.doi.org/}%
\providecommand \selectlanguage [0]{\@gobble}%
\providecommand \bibinfo  [0]{\@secondoftwo}%
\providecommand \bibfield  [0]{\@secondoftwo}%
\providecommand \translation [1]{[#1]}%
\providecommand \BibitemOpen [0]{}%
\providecommand \bibitemStop [0]{}%
\providecommand \bibitemNoStop [0]{.\EOS\space}%
\providecommand \EOS [0]{\spacefactor3000\relax}%
\providecommand \BibitemShut  [1]{\csname bibitem#1\endcsname}%
\let\auto@bib@innerbib\@empty
\bibitem [{\citenamefont {Dyakonov}(2017)}]{dyakonov_book}%
  \BibitemOpen
  \bibinfo {editor} {\bibfnamefont {M.~I.}\ \bibnamefont {Dyakonov}},\ ed.,\
  \href@noop {} {\emph {\bibinfo {title} {Spin physics in semiconductors}}},\
  \bibinfo {edition} {2nd}\ ed.,\ Springer Series in Solid-State Sciences 157\
  (\bibinfo  {publisher} {Springer International Publishing},\ \bibinfo {year}
  {2017})\BibitemShut {NoStop}%
\bibitem [{\citenamefont {Dyakonov}\ and\ \citenamefont
  {Perel'}(1971{\natexlab{a}})}]{dyakonov71a}%
  \BibitemOpen
  \bibfield  {author} {\bibinfo {author} {\bibfnamefont {M.~I.}\ \bibnamefont
  {Dyakonov}}\ and\ \bibinfo {author} {\bibfnamefont {V.~I.}\ \bibnamefont
  {Perel'}},\ }\bibfield  {title} {\enquote {\bibinfo {title} {Current induced
  spin orientation of electrons in semiconductors},}\ }\href
  {https://www.sciencedirect.com/science/article/pii/0375960171901964}
  {\bibfield  {journal} {\bibinfo  {journal} {Phys. Lett. A}\ }\textbf
  {\bibinfo {volume} {35A}},\ \bibinfo {pages} {459} (\bibinfo {year}
  {1971}{\natexlab{a}})}\BibitemShut {NoStop}%
\bibitem [{\citenamefont {Dyakonov}\ and\ \citenamefont
  {Perel'}(1971{\natexlab{b}})}]{dyakonov71}%
  \BibitemOpen
  \bibfield  {author} {\bibinfo {author} {\bibfnamefont {M.I.}\ \bibnamefont
  {Dyakonov}}\ and\ \bibinfo {author} {\bibfnamefont {V.I}\ \bibnamefont
  {Perel'}},\ }\bibfield  {title} {\enquote {\bibinfo {title} {{Possibility of
  Orienting Electron Spins with Current}},}\ }\href@noop {} {\bibfield
  {journal} {\bibinfo  {journal} {JETP Lett.}\ }\textbf {\bibinfo {volume}
  {13}},\ \bibinfo {pages} {657} (\bibinfo {year}
  {1971}{\natexlab{b}})}\BibitemShut {NoStop}%
\bibitem [{\citenamefont {Hirsch}(1999)}]{hirsch99}%
  \BibitemOpen
  \bibfield  {author} {\bibinfo {author} {\bibfnamefont {J.~E.}\ \bibnamefont
  {Hirsch}},\ }\bibfield  {title} {\enquote {\bibinfo {title} {Spin {H}all
  effect},}\ }\href@noop {} {\bibfield  {journal} {\bibinfo  {journal} {Phys.
  Rev. Lett.}\ }\textbf {\bibinfo {volume} {83}},\ \bibinfo {pages} {1834}
  (\bibinfo {year} {1999})}\BibitemShut {NoStop}%
\bibitem [{\citenamefont {Murakami}\ \emph {et~al.}(2003)\citenamefont
  {Murakami}, \citenamefont {Nagaosa},\ and\ \citenamefont
  {Zhang}}]{murakami03}%
  \BibitemOpen
  \bibfield  {author} {\bibinfo {author} {\bibfnamefont {Shuichi}\ \bibnamefont
  {Murakami}}, \bibinfo {author} {\bibfnamefont {Naoto}\ \bibnamefont
  {Nagaosa}}, \ and\ \bibinfo {author} {\bibfnamefont {Shou-Cheng}\
  \bibnamefont {Zhang}},\ }\bibfield  {title} {\enquote {\bibinfo {title}
  {Dissipationless quantum spin current at room temperature},}\ }\href@noop {}
  {\bibfield  {journal} {\bibinfo  {journal} {Science}\ }\textbf {\bibinfo
  {volume} {301}},\ \bibinfo {pages} {1348} (\bibinfo {year}
  {2003})}\BibitemShut {NoStop}%
\bibitem [{\citenamefont {Sinova}\ \emph {et~al.}(2004)\citenamefont {Sinova},
  \citenamefont {Culcer}, \citenamefont {Niu}, \citenamefont {Sinitsyn},
  \citenamefont {Jungwirth},\ and\ \citenamefont {MacDonald}}]{sinova04}%
  \BibitemOpen
  \bibfield  {author} {\bibinfo {author} {\bibfnamefont {Jairo}\ \bibnamefont
  {Sinova}}, \bibinfo {author} {\bibfnamefont {Dimitrie}\ \bibnamefont
  {Culcer}}, \bibinfo {author} {\bibfnamefont {Q.}~\bibnamefont {Niu}},
  \bibinfo {author} {\bibfnamefont {N.~A.}\ \bibnamefont {Sinitsyn}}, \bibinfo
  {author} {\bibfnamefont {T.}~\bibnamefont {Jungwirth}}, \ and\ \bibinfo
  {author} {\bibfnamefont {A.~H.}\ \bibnamefont {MacDonald}},\ }\bibfield
  {title} {\enquote {\bibinfo {title} {Universal intrinsic spin {H}all
  effect},}\ }\href@noop {} {\bibfield  {journal} {\bibinfo  {journal} {Phys.
  Rev. Lett.}\ }\textbf {\bibinfo {volume} {92}},\ \bibinfo {pages} {126603}
  (\bibinfo {year} {2004})}\BibitemShut {NoStop}%
\bibitem [{\citenamefont {Wunderlich}\ \emph {et~al.}(2005)\citenamefont
  {Wunderlich}, \citenamefont {Kaestner}, \citenamefont {Sinova},\ and\
  \citenamefont {Jungwirth}}]{wunderlich05}%
  \BibitemOpen
  \bibfield  {author} {\bibinfo {author} {\bibfnamefont {J.}~\bibnamefont
  {Wunderlich}}, \bibinfo {author} {\bibfnamefont {B.}~\bibnamefont
  {Kaestner}}, \bibinfo {author} {\bibfnamefont {J.}~\bibnamefont {Sinova}}, \
  and\ \bibinfo {author} {\bibfnamefont {T.}~\bibnamefont {Jungwirth}},\
  }\bibfield  {title} {\enquote {\bibinfo {title} {Experimental observation of
  the spin-{H}all effect in a two-dimensional spin-orbit coupled semiconductor
  system},}\ }\href
  {https://journals.aps.org/prl/abstract/10.1103/PhysRevLett.94.047204}
  {\bibfield  {journal} {\bibinfo  {journal} {Phys. Rev. Lett.}\ }\textbf
  {\bibinfo {volume} {94}},\ \bibinfo {pages} {47204} (\bibinfo {year}
  {2005})}\BibitemShut {NoStop}%
\bibitem [{\citenamefont {Kavokin}\ \emph {et~al.}(2005)\citenamefont
  {Kavokin}, \citenamefont {Malpuech},\ and\ \citenamefont
  {Glazov}}]{kavokin05prl}%
  \BibitemOpen
  \bibfield  {author} {\bibinfo {author} {\bibfnamefont {Alexey}\ \bibnamefont
  {Kavokin}}, \bibinfo {author} {\bibfnamefont {Guillaume}\ \bibnamefont
  {Malpuech}}, \ and\ \bibinfo {author} {\bibfnamefont {Mikhail}\ \bibnamefont
  {Glazov}},\ }\bibfield  {title} {\enquote {\bibinfo {title} {{Optical Spin
  Hall Effect}},}\ }\href {\doibase 10.1103/PhysRevLett.95.136601} {\bibfield
  {journal} {\bibinfo  {journal} {Phys. Rev. Lett.}\ }\textbf {\bibinfo
  {volume} {95}},\ \bibinfo {pages} {136601} (\bibinfo {year}
  {2005})}\BibitemShut {NoStop}%
\bibitem [{\citenamefont {Leyder}\ \emph {et~al.}(2007)\citenamefont {Leyder},
  \citenamefont {Romanelli}, \citenamefont {Karr}, \citenamefont {Giacobino},
  \citenamefont {Liew}, \citenamefont {Glazov}, \citenamefont {Kavokin},
  \citenamefont {Malpuech},\ and\ \citenamefont {Bramati}}]{Leyder:2007ve}%
  \BibitemOpen
  \bibfield  {author} {\bibinfo {author} {\bibfnamefont {C.}~\bibnamefont
  {Leyder}}, \bibinfo {author} {\bibfnamefont {M.}~\bibnamefont {Romanelli}},
  \bibinfo {author} {\bibfnamefont {J.~Ph.}\ \bibnamefont {Karr}}, \bibinfo
  {author} {\bibfnamefont {E.}~\bibnamefont {Giacobino}}, \bibinfo {author}
  {\bibfnamefont {T.~C.~H.}\ \bibnamefont {Liew}}, \bibinfo {author}
  {\bibfnamefont {M.~M.}\ \bibnamefont {Glazov}}, \bibinfo {author}
  {\bibfnamefont {A.~V.}\ \bibnamefont {Kavokin}}, \bibinfo {author}
  {\bibfnamefont {G.}~\bibnamefont {Malpuech}}, \ and\ \bibinfo {author}
  {\bibfnamefont {A.}~\bibnamefont {Bramati}},\ }\bibfield  {title} {\enquote
  {\bibinfo {title} {Observation of the optical spin {H}all effect},}\ }\href
  {http://dx.doi.org/10.1038/nphys676} {\bibfield  {journal} {\bibinfo
  {journal} {Nat Phys}\ }\textbf {\bibinfo {volume} {3}},\ \bibinfo {pages}
  {628--631} (\bibinfo {year} {2007})}\BibitemShut {NoStop}%
\bibitem [{\citenamefont {Mak}\ \emph {et~al.}(2014)\citenamefont {Mak},
  \citenamefont {McGill}, \citenamefont {Park},\ and\ \citenamefont
  {McEuen}}]{Mak27062014}%
  \BibitemOpen
  \bibfield  {author} {\bibinfo {author} {\bibfnamefont {K.~F.}\ \bibnamefont
  {Mak}}, \bibinfo {author} {\bibfnamefont {K.~L.}\ \bibnamefont {McGill}},
  \bibinfo {author} {\bibfnamefont {J.}~\bibnamefont {Park}}, \ and\ \bibinfo
  {author} {\bibfnamefont {P.~L.}\ \bibnamefont {McEuen}},\ }\bibfield  {title}
  {\enquote {\bibinfo {title} {The valley {H}all effect in {MoS$_2$}
  transistors},}\ }\href {\doibase 10.1126/science.1250140} {\bibfield
  {journal} {\bibinfo  {journal} {Science}\ }\textbf {\bibinfo {volume}
  {344}},\ \bibinfo {pages} {1489--1492} (\bibinfo {year} {2014})}\BibitemShut
  {NoStop}%
\bibitem [{\citenamefont {Ubrig}\ \emph {et~al.}(2017)\citenamefont {Ubrig},
  \citenamefont {Jo}, \citenamefont {Philippi}, \citenamefont {Costanzo},
  \citenamefont {Berger}, \citenamefont {Kuzmenko},\ and\ \citenamefont
  {Morpurgo}}]{Ubrig:2017aa}%
  \BibitemOpen
  \bibfield  {author} {\bibinfo {author} {\bibfnamefont {Nicolas}\ \bibnamefont
  {Ubrig}}, \bibinfo {author} {\bibfnamefont {Sanghyun}\ \bibnamefont {Jo}},
  \bibinfo {author} {\bibfnamefont {Marc}\ \bibnamefont {Philippi}}, \bibinfo
  {author} {\bibfnamefont {Davide}\ \bibnamefont {Costanzo}}, \bibinfo {author}
  {\bibfnamefont {Helmuth}\ \bibnamefont {Berger}}, \bibinfo {author}
  {\bibfnamefont {Alexey~B.}\ \bibnamefont {Kuzmenko}}, \ and\ \bibinfo
  {author} {\bibfnamefont {Alberto~F.}\ \bibnamefont {Morpurgo}},\ }\bibfield
  {title} {\enquote {\bibinfo {title} {{Microscopic Origin of the Valley Hall
  Effect in Transition Metal Dichalcogenides Revealed by Wavelength-Dependent
  Mapping}},}\ }\href {\doibase 10.1021/acs.nanolett.7b02666} {\bibfield
  {journal} {\bibinfo  {journal} {Nano Letters}\ }\textbf {\bibinfo {volume}
  {17}},\ \bibinfo {pages} {5719--5725} (\bibinfo {year} {2017})}\BibitemShut
  {NoStop}%
\bibitem [{\citenamefont {Lundt}\ \emph {et~al.}(2019)\citenamefont {Lundt},
  \citenamefont {Dusanowski}, \citenamefont {Sedov}, \citenamefont {Stepanov},
  \citenamefont {Glazov}, \citenamefont {Klembt}, \citenamefont {Klaas},
  \citenamefont {Beierlein}, \citenamefont {Qin}, \citenamefont {Tongay},
  \citenamefont {Richard}, \citenamefont {Kavokin}, \citenamefont {H\"ofling},\
  and\ \citenamefont {Schneider}}]{Lundt:2019aa}%
  \BibitemOpen
  \bibfield  {author} {\bibinfo {author} {\bibfnamefont {Nils}\ \bibnamefont
  {Lundt}}, \bibinfo {author} {\bibfnamefont {\L{}ukasz}\ \bibnamefont
  {Dusanowski}}, \bibinfo {author} {\bibfnamefont {Evgeny}\ \bibnamefont
  {Sedov}}, \bibinfo {author} {\bibfnamefont {Petr}\ \bibnamefont {Stepanov}},
  \bibinfo {author} {\bibfnamefont {Mikhail~M.}\ \bibnamefont {Glazov}},
  \bibinfo {author} {\bibfnamefont {Sebastian}\ \bibnamefont {Klembt}},
  \bibinfo {author} {\bibfnamefont {Martin}\ \bibnamefont {Klaas}}, \bibinfo
  {author} {\bibfnamefont {Johannes}\ \bibnamefont {Beierlein}}, \bibinfo
  {author} {\bibfnamefont {Ying}\ \bibnamefont {Qin}}, \bibinfo {author}
  {\bibfnamefont {Sefaattin}\ \bibnamefont {Tongay}}, \bibinfo {author}
  {\bibfnamefont {Maxime}\ \bibnamefont {Richard}}, \bibinfo {author}
  {\bibfnamefont {Alexey~V.}\ \bibnamefont {Kavokin}}, \bibinfo {author}
  {\bibfnamefont {Sven}\ \bibnamefont {H\"ofling}}, \ and\ \bibinfo {author}
  {\bibfnamefont {Christian}\ \bibnamefont {Schneider}},\ }\bibfield  {title}
  {\enquote {\bibinfo {title} {Optical valley {H}all effect for highly
  valley-coherent exciton-polaritons in an atomically thin semiconductor},}\
  }\href {\doibase 10.1038/s41565-019-0492-0} {\bibfield  {journal} {\bibinfo
  {journal} {Nature Nanotechnology}\ }\textbf {\bibinfo {volume} {14}},\
  \bibinfo {pages} {770--775} (\bibinfo {year} {2019})}\BibitemShut {NoStop}%
\bibitem [{\citenamefont {Hall}(1881)}]{Hall:1881aa}%
  \BibitemOpen
  \bibfield  {author} {\bibinfo {author} {\bibfnamefont {E.~H.}\ \bibnamefont
  {Hall}},\ }\bibfield  {title} {\enquote {\bibinfo {title} {{XXXVIII.} {O}n
  the new action of magnetism on a permanent electric current},}\ }\href
  {\doibase 10.1080/14786448008626936} {\bibfield  {journal} {\bibinfo
  {journal} {The London, Edinburgh, and Dublin Philosophical Magazine and
  Journal of Science}\ }\textbf {\bibinfo {volume} {5}},\ \bibinfo {pages}
  {157} (\bibinfo {year} {1881})}\BibitemShut {NoStop}%
\bibitem [{\citenamefont {Nagaosa}\ \emph {et~al.}(2010)\citenamefont
  {Nagaosa}, \citenamefont {Sinova}, \citenamefont {Onoda}, \citenamefont
  {MacDonald},\ and\ \citenamefont {Ong}}]{RevModPhys.82.1539}%
  \BibitemOpen
  \bibfield  {author} {\bibinfo {author} {\bibfnamefont {Naoto}\ \bibnamefont
  {Nagaosa}}, \bibinfo {author} {\bibfnamefont {Jairo}\ \bibnamefont {Sinova}},
  \bibinfo {author} {\bibfnamefont {Shigeki}\ \bibnamefont {Onoda}}, \bibinfo
  {author} {\bibfnamefont {A.~H.}\ \bibnamefont {MacDonald}}, \ and\ \bibinfo
  {author} {\bibfnamefont {N.~P.}\ \bibnamefont {Ong}},\ }\bibfield  {title}
  {\enquote {\bibinfo {title} {Anomalous $\mbox{H}$all effect},}\ }\href
  {\doibase 10.1103/RevModPhys.82.1539} {\bibfield  {journal} {\bibinfo
  {journal} {Rev. Mod. Phys.}\ }\textbf {\bibinfo {volume} {82}},\ \bibinfo
  {pages} {1539--1592} (\bibinfo {year} {2010})}\BibitemShut {NoStop}%
\bibitem [{\citenamefont {Mak}\ \emph {et~al.}(2010)\citenamefont {Mak},
  \citenamefont {Lee}, \citenamefont {Hone}, \citenamefont {Shan},\ and\
  \citenamefont {Heinz}}]{Mak:2010bh}%
  \BibitemOpen
  \bibfield  {author} {\bibinfo {author} {\bibfnamefont {Kin~Fai}\ \bibnamefont
  {Mak}}, \bibinfo {author} {\bibfnamefont {Changgu}\ \bibnamefont {Lee}},
  \bibinfo {author} {\bibfnamefont {James}\ \bibnamefont {Hone}}, \bibinfo
  {author} {\bibfnamefont {Jie}\ \bibnamefont {Shan}}, \ and\ \bibinfo {author}
  {\bibfnamefont {Tony~F.}\ \bibnamefont {Heinz}},\ }\bibfield  {title}
  {\enquote {\bibinfo {title} {Atomically thin {MoS}$_{2}$: A new direct-gap
  semiconductor},}\ }\href {\doibase 10.1103/PhysRevLett.105.136805} {\bibfield
   {journal} {\bibinfo  {journal} {Phys. Rev. Lett.}\ }\textbf {\bibinfo
  {volume} {105}},\ \bibinfo {pages} {136805} (\bibinfo {year}
  {2010})}\BibitemShut {NoStop}%
\bibitem [{\citenamefont {Splendiani}\ \emph {et~al.}(2010)\citenamefont
  {Splendiani}, \citenamefont {Sun}, \citenamefont {Zhang}, \citenamefont {Li},
  \citenamefont {Kim}, \citenamefont {Chim}, \citenamefont {Galli},\ and\
  \citenamefont {Wang}}]{Splendiani:2010a}%
  \BibitemOpen
  \bibfield  {author} {\bibinfo {author} {\bibfnamefont {Andrea}\ \bibnamefont
  {Splendiani}}, \bibinfo {author} {\bibfnamefont {Liang}\ \bibnamefont {Sun}},
  \bibinfo {author} {\bibfnamefont {Yuanbo}\ \bibnamefont {Zhang}}, \bibinfo
  {author} {\bibfnamefont {Tianshu}\ \bibnamefont {Li}}, \bibinfo {author}
  {\bibfnamefont {Jonghwan}\ \bibnamefont {Kim}}, \bibinfo {author}
  {\bibfnamefont {Chi-Yung}\ \bibnamefont {Chim}}, \bibinfo {author}
  {\bibfnamefont {Giulia}\ \bibnamefont {Galli}}, \ and\ \bibinfo {author}
  {\bibfnamefont {Feng}\ \bibnamefont {Wang}},\ }\bibfield  {title} {\enquote
  {\bibinfo {title} {Emerging photoluminescence in monolayer {MoS}$_2$},}\
  }\href@noop {} {\bibfield  {journal} {\bibinfo  {journal} {Nano Letters}\
  }\textbf {\bibinfo {volume} {10}},\ \bibinfo {pages} {1271} (\bibinfo {year}
  {2010})}\BibitemShut {NoStop}%
\bibitem [{\citenamefont {Kormanyos}\ \emph {et~al.}(2015)\citenamefont
  {Kormanyos}, \citenamefont {Burkard}, \citenamefont {Gmitra}, \citenamefont
  {Fabian}, \citenamefont {Z{\'o}lyomi}, \citenamefont {Drummond},\ and\
  \citenamefont {Fal'ko}}]{2053-1583-2-2-022001}%
  \BibitemOpen
  \bibfield  {author} {\bibinfo {author} {\bibfnamefont {Andor}\ \bibnamefont
  {Kormanyos}}, \bibinfo {author} {\bibfnamefont {Guido}\ \bibnamefont
  {Burkard}}, \bibinfo {author} {\bibfnamefont {Martin}\ \bibnamefont
  {Gmitra}}, \bibinfo {author} {\bibfnamefont {Jaroslav}\ \bibnamefont
  {Fabian}}, \bibinfo {author} {\bibfnamefont {Viktor}\ \bibnamefont
  {Z{\'o}lyomi}}, \bibinfo {author} {\bibfnamefont {Neil~D}\ \bibnamefont
  {Drummond}}, \ and\ \bibinfo {author} {\bibfnamefont {Vladimir}\ \bibnamefont
  {Fal'ko}},\ }\bibfield  {title} {\enquote {\bibinfo {title} {$\bm k\cdot \bm
  p$ theory for two-dimensional transition metal dichalcogenide
  semiconductors},}\ }\href {http://stacks.iop.org/2053-1583/2/i=2/a=022001}
  {\bibfield  {journal} {\bibinfo  {journal} {2D Materials}\ }\textbf {\bibinfo
  {volume} {2}},\ \bibinfo {pages} {022001} (\bibinfo {year}
  {2015})}\BibitemShut {NoStop}%
\bibitem [{\citenamefont {Kolobov}\ and\ \citenamefont
  {Tominaga}(2016)}]{Kolobov2016book}%
  \BibitemOpen
  \bibfield  {author} {\bibinfo {author} {\bibfnamefont {Alexander~V.}\
  \bibnamefont {Kolobov}}\ and\ \bibinfo {author} {\bibfnamefont {Junji}\
  \bibnamefont {Tominaga}},\ }\href {\doibase 10.1007/978-3-319-31450-1} {\emph
  {\bibinfo {title} {Two-Dimensional Transition-Metal Dichalcogenides}}}\
  (\bibinfo  {publisher} {Springer International Publishing},\ \bibinfo {year}
  {2016})\BibitemShut {NoStop}%
\bibitem [{\citenamefont {Wang}\ \emph {et~al.}(2018)\citenamefont {Wang},
  \citenamefont {Chernikov}, \citenamefont {Glazov}, \citenamefont {Heinz},
  \citenamefont {Marie}, \citenamefont {Amand},\ and\ \citenamefont
  {Urbaszek}}]{RevModPhys.90.021001}%
  \BibitemOpen
  \bibfield  {author} {\bibinfo {author} {\bibfnamefont {Gang}\ \bibnamefont
  {Wang}}, \bibinfo {author} {\bibfnamefont {Alexey}\ \bibnamefont
  {Chernikov}}, \bibinfo {author} {\bibfnamefont {Mikhail~M.}\ \bibnamefont
  {Glazov}}, \bibinfo {author} {\bibfnamefont {Tony~F.}\ \bibnamefont {Heinz}},
  \bibinfo {author} {\bibfnamefont {Xavier}\ \bibnamefont {Marie}}, \bibinfo
  {author} {\bibfnamefont {Thierry}\ \bibnamefont {Amand}}, \ and\ \bibinfo
  {author} {\bibfnamefont {Bernhard}\ \bibnamefont {Urbaszek}},\ }\bibfield
  {title} {\enquote {\bibinfo {title} {Colloquium: Excitons in atomically thin
  transition metal dichalcogenides},}\ }\href {\doibase
  10.1103/RevModPhys.90.021001} {\bibfield  {journal} {\bibinfo  {journal}
  {Rev. Mod. Phys.}\ }\textbf {\bibinfo {volume} {90}},\ \bibinfo {pages}
  {021001} (\bibinfo {year} {2018})}\BibitemShut {NoStop}%
\bibitem [{\citenamefont {Xiao}\ \emph {et~al.}(2012)\citenamefont {Xiao},
  \citenamefont {Liu}, \citenamefont {Feng}, \citenamefont {Xu},\ and\
  \citenamefont {Yao}}]{Xiao:2012cr}%
  \BibitemOpen
  \bibfield  {author} {\bibinfo {author} {\bibfnamefont {Di}~\bibnamefont
  {Xiao}}, \bibinfo {author} {\bibfnamefont {Gui-Bin}\ \bibnamefont {Liu}},
  \bibinfo {author} {\bibfnamefont {Wanxiang}\ \bibnamefont {Feng}}, \bibinfo
  {author} {\bibfnamefont {Xiaodong}\ \bibnamefont {Xu}}, \ and\ \bibinfo
  {author} {\bibfnamefont {Wang}\ \bibnamefont {Yao}},\ }\bibfield  {title}
  {\enquote {\bibinfo {title} {Coupled spin and valley physics in monolayers of
  {MoS}$_{2}$ and other group-{VI} dichalcogenides},}\ }\href {\doibase
  10.1103/PhysRevLett.108.196802} {\bibfield  {journal} {\bibinfo  {journal}
  {Phys. Rev. Lett.}\ }\textbf {\bibinfo {volume} {108}},\ \bibinfo {pages}
  {196802} (\bibinfo {year} {2012})}\BibitemShut {NoStop}%
\bibitem [{\citenamefont {Mak}\ \emph {et~al.}(2012)\citenamefont {Mak},
  \citenamefont {He}, \citenamefont {Shan},\ and\ \citenamefont
  {Heinz}}]{Mak:2012qf}%
  \BibitemOpen
  \bibfield  {author} {\bibinfo {author} {\bibfnamefont {Kin~Fai}\ \bibnamefont
  {Mak}}, \bibinfo {author} {\bibfnamefont {Keliang}\ \bibnamefont {He}},
  \bibinfo {author} {\bibfnamefont {Jie}\ \bibnamefont {Shan}}, \ and\ \bibinfo
  {author} {\bibfnamefont {Tony~F.}\ \bibnamefont {Heinz}},\ }\bibfield
  {title} {\enquote {\bibinfo {title} {Control of valley polarization in
  monolayer {MoS}$_2$ by optical helicity},}\ }\href
  {http://dx.doi.org/10.1038/nnano.2012.96} {\bibfield  {journal} {\bibinfo
  {journal} {Nat Nano}\ }\textbf {\bibinfo {volume} {7}},\ \bibinfo {pages}
  {494--498} (\bibinfo {year} {2012})}\BibitemShut {NoStop}%
\bibitem [{\citenamefont {Xu}\ \emph {et~al.}(2014)\citenamefont {Xu},
  \citenamefont {Yao}, \citenamefont {Xiao},\ and\ \citenamefont
  {Heinz}}]{Xu:2014cr}%
  \BibitemOpen
  \bibfield  {author} {\bibinfo {author} {\bibfnamefont {Xiaodong}\
  \bibnamefont {Xu}}, \bibinfo {author} {\bibfnamefont {Wang}\ \bibnamefont
  {Yao}}, \bibinfo {author} {\bibfnamefont {Di}~\bibnamefont {Xiao}}, \ and\
  \bibinfo {author} {\bibfnamefont {Tony~F.}\ \bibnamefont {Heinz}},\
  }\bibfield  {title} {\enquote {\bibinfo {title} {Spin and pseudospins in
  layered transition metal dichalcogenides},}\ }\href
  {http://dx.doi.org/10.1038/nphys2942} {\bibfield  {journal} {\bibinfo
  {journal} {Nat Phys}\ }\textbf {\bibinfo {volume} {10}},\ \bibinfo {pages}
  {343--350} (\bibinfo {year} {2014})}\BibitemShut {NoStop}%
\bibitem [{\citenamefont {Konabe}\ and\ \citenamefont
  {Yamamoto}(2014)}]{PhysRevB.90.075430}%
  \BibitemOpen
  \bibfield  {author} {\bibinfo {author} {\bibfnamefont {Satoru}\ \bibnamefont
  {Konabe}}\ and\ \bibinfo {author} {\bibfnamefont {Takahiro}\ \bibnamefont
  {Yamamoto}},\ }\bibfield  {title} {\enquote {\bibinfo {title} {Valley
  photothermoelectric effects in transition-metal dichalcogenides},}\ }\href
  {\doibase 10.1103/PhysRevB.90.075430} {\bibfield  {journal} {\bibinfo
  {journal} {Phys. Rev. B}\ }\textbf {\bibinfo {volume} {90}},\ \bibinfo
  {pages} {075430} (\bibinfo {year} {2014})}\BibitemShut {NoStop}%
\bibitem [{\citenamefont {Jin}\ \emph {et~al.}(2018)\citenamefont {Jin},
  \citenamefont {Kim}, \citenamefont {Utama}, \citenamefont {Regan},
  \citenamefont {Kleemann}, \citenamefont {Cai}, \citenamefont {Shen},
  \citenamefont {Shinner}, \citenamefont {Sengupta}, \citenamefont {Watanabe},
  \citenamefont {Taniguchi}, \citenamefont {Tongay}, \citenamefont {Zettl},\
  and\ \citenamefont {Wang}}]{Jin893}%
  \BibitemOpen
  \bibfield  {author} {\bibinfo {author} {\bibfnamefont {Chenhao}\ \bibnamefont
  {Jin}}, \bibinfo {author} {\bibfnamefont {Jonghwan}\ \bibnamefont {Kim}},
  \bibinfo {author} {\bibfnamefont {M.~Iqbal~Bakti}\ \bibnamefont {Utama}},
  \bibinfo {author} {\bibfnamefont {Emma~C.}\ \bibnamefont {Regan}}, \bibinfo
  {author} {\bibfnamefont {Hans}\ \bibnamefont {Kleemann}}, \bibinfo {author}
  {\bibfnamefont {Hui}\ \bibnamefont {Cai}}, \bibinfo {author} {\bibfnamefont
  {Yuxia}\ \bibnamefont {Shen}}, \bibinfo {author} {\bibfnamefont
  {Matthew~James}\ \bibnamefont {Shinner}}, \bibinfo {author} {\bibfnamefont
  {Arjun}\ \bibnamefont {Sengupta}}, \bibinfo {author} {\bibfnamefont {Kenji}\
  \bibnamefont {Watanabe}}, \bibinfo {author} {\bibfnamefont {Takashi}\
  \bibnamefont {Taniguchi}}, \bibinfo {author} {\bibfnamefont {Sefaattin}\
  \bibnamefont {Tongay}}, \bibinfo {author} {\bibfnamefont {Alex}\ \bibnamefont
  {Zettl}}, \ and\ \bibinfo {author} {\bibfnamefont {Feng}\ \bibnamefont
  {Wang}},\ }\bibfield  {title} {\enquote {\bibinfo {title} {Imaging of pure
  spin-valley diffusion current in {WS}$_2$-{WSe}$_2$ heterostructures},}\
  }\href {\doibase 10.1126/science.aao3503} {\bibfield  {journal} {\bibinfo
  {journal} {Science}\ }\textbf {\bibinfo {volume} {360}},\ \bibinfo {pages}
  {893--896} (\bibinfo {year} {2018})}\BibitemShut {NoStop}%
\bibitem [{\citenamefont {Onga}\ \emph {et~al.}(2017)\citenamefont {Onga},
  \citenamefont {Zhang}, \citenamefont {Ideue},\ and\ \citenamefont
  {Iwasa}}]{Onga:2017aa}%
  \BibitemOpen
  \bibfield  {author} {\bibinfo {author} {\bibfnamefont {Masaru}\ \bibnamefont
  {Onga}}, \bibinfo {author} {\bibfnamefont {Yijin}\ \bibnamefont {Zhang}},
  \bibinfo {author} {\bibfnamefont {Toshiya}\ \bibnamefont {Ideue}}, \ and\
  \bibinfo {author} {\bibfnamefont {Yoshihiro}\ \bibnamefont {Iwasa}},\
  }\bibfield  {title} {\enquote {\bibinfo {title} {Exciton {H}all effect in
  monolayer {MoS}$_2$},}\ }\href {http://dx.doi.org/10.1038/nmat4996}
  {\bibfield  {journal} {\bibinfo  {journal} {Nature Materials}\ }\textbf
  {\bibinfo {volume} {16}},\ \bibinfo {pages} {1193} (\bibinfo {year}
  {2017})}\BibitemShut {NoStop}%
\bibitem [{\citenamefont {Unuchek}\ \emph {et~al.}(2019)\citenamefont
  {Unuchek}, \citenamefont {Ciarrocchi}, \citenamefont {Avsar}, \citenamefont
  {Sun}, \citenamefont {Watanabe}, \citenamefont {Taniguchi},\ and\
  \citenamefont {Kis}}]{Unuchek:2019aa}%
  \BibitemOpen
  \bibfield  {author} {\bibinfo {author} {\bibfnamefont {Dmitrii}\ \bibnamefont
  {Unuchek}}, \bibinfo {author} {\bibfnamefont {Alberto}\ \bibnamefont
  {Ciarrocchi}}, \bibinfo {author} {\bibfnamefont {Ahmet}\ \bibnamefont
  {Avsar}}, \bibinfo {author} {\bibfnamefont {Zhe}\ \bibnamefont {Sun}},
  \bibinfo {author} {\bibfnamefont {Kenji}\ \bibnamefont {Watanabe}}, \bibinfo
  {author} {\bibfnamefont {Takashi}\ \bibnamefont {Taniguchi}}, \ and\ \bibinfo
  {author} {\bibfnamefont {Andras}\ \bibnamefont {Kis}},\ }\bibfield  {title}
  {\enquote {\bibinfo {title} {Valley-polarized exciton currents in a van der
  {W}aals heterostructure},}\ }\href {\doibase 10.1038/s41565-019-0559-y}
  {\bibfield  {journal} {\bibinfo  {journal} {Nature Nanotechnology}\ }\textbf
  {\bibinfo {volume} {14}},\ \bibinfo {pages} {1104} (\bibinfo {year}
  {2019})}\BibitemShut {NoStop}%
\bibitem [{\citenamefont {Kulig}\ \emph {et~al.}(2018)\citenamefont {Kulig},
  \citenamefont {Zipfel}, \citenamefont {Nagler}, \citenamefont {Blanter},
  \citenamefont {Sch\"uller}, \citenamefont {Korn}, \citenamefont {Paradiso},
  \citenamefont {Glazov},\ and\ \citenamefont
  {Chernikov}}]{PhysRevLett.120.207401}%
  \BibitemOpen
  \bibfield  {author} {\bibinfo {author} {\bibfnamefont {Marvin}\ \bibnamefont
  {Kulig}}, \bibinfo {author} {\bibfnamefont {Jonas}\ \bibnamefont {Zipfel}},
  \bibinfo {author} {\bibfnamefont {Philipp}\ \bibnamefont {Nagler}}, \bibinfo
  {author} {\bibfnamefont {Sofia}\ \bibnamefont {Blanter}}, \bibinfo {author}
  {\bibfnamefont {Christian}\ \bibnamefont {Sch\"uller}}, \bibinfo {author}
  {\bibfnamefont {Tobias}\ \bibnamefont {Korn}}, \bibinfo {author}
  {\bibfnamefont {Nicola}\ \bibnamefont {Paradiso}}, \bibinfo {author}
  {\bibfnamefont {Mikhail~M.}\ \bibnamefont {Glazov}}, \ and\ \bibinfo {author}
  {\bibfnamefont {Alexey}\ \bibnamefont {Chernikov}},\ }\bibfield  {title}
  {\enquote {\bibinfo {title} {Exciton diffusion and halo effects in monolayer
  semiconductors},}\ }\href {\doibase 10.1103/PhysRevLett.120.207401}
  {\bibfield  {journal} {\bibinfo  {journal} {Phys. Rev. Lett.}\ }\textbf
  {\bibinfo {volume} {120}},\ \bibinfo {pages} {207401} (\bibinfo {year}
  {2018})}\BibitemShut {NoStop}%
\bibitem [{\citenamefont {Kalameitsev}\ \emph {et~al.}(2019)\citenamefont
  {Kalameitsev}, \citenamefont {Kovalev},\ and\ \citenamefont
  {Savenko}}]{PhysRevLett.122.256801}%
  \BibitemOpen
  \bibfield  {author} {\bibinfo {author} {\bibfnamefont {A.~V.}\ \bibnamefont
  {Kalameitsev}}, \bibinfo {author} {\bibfnamefont {V.~M.}\ \bibnamefont
  {Kovalev}}, \ and\ \bibinfo {author} {\bibfnamefont {I.~G.}\ \bibnamefont
  {Savenko}},\ }\bibfield  {title} {\enquote {\bibinfo {title} {Valley
  acoustoelectric effect},}\ }\href {\doibase 10.1103/PhysRevLett.122.256801}
  {\bibfield  {journal} {\bibinfo  {journal} {Phys. Rev. Lett.}\ }\textbf
  {\bibinfo {volume} {122}},\ \bibinfo {pages} {256801} (\bibinfo {year}
  {2019})}\BibitemShut {NoStop}%
\bibitem [{\citenamefont {Glazov}\ and\ \citenamefont
  {Golub}(2020{\natexlab{a}})}]{2020arXiv200405091G}%
  \BibitemOpen
  \bibfield  {author} {\bibinfo {author} {\bibfnamefont {M.~M.}\ \bibnamefont
  {Glazov}}\ and\ \bibinfo {author} {\bibfnamefont {L.~E.}\ \bibnamefont
  {Golub}},\ }\bibfield  {title} {\enquote {\bibinfo {title} {Valley {H}all
  effect caused by the phonon and photon drag},}\ }\href {\doibase
  10.1103/PhysRevB.102.155302} {\bibfield  {journal} {\bibinfo  {journal}
  {Phys. Rev. B}\ }\textbf {\bibinfo {volume} {102}},\ \bibinfo {pages}
  {155302} (\bibinfo {year} {2020}{\natexlab{a}})}\BibitemShut {NoStop}%
\bibitem [{\citenamefont {Glazov}\ and\ \citenamefont
  {Golub}(2020{\natexlab{b}})}]{Glazov2020b}%
  \BibitemOpen
  \bibfield  {author} {\bibinfo {author} {\bibfnamefont {M.~M.}\ \bibnamefont
  {Glazov}}\ and\ \bibinfo {author} {\bibfnamefont {L.~E.}\ \bibnamefont
  {Golub}},\ }\bibfield  {title} {\enquote {\bibinfo {title} {{Skew Scattering
  and Side Jump Drive Exciton Valley Hall Effect in Two-Dimensional
  Crystals}},}\ }\href {\doibase 10.1103/PhysRevLett.125.157403} {\bibfield
  {journal} {\bibinfo  {journal} {Physical Review Letters}\ }\textbf {\bibinfo
  {volume} {125}},\ \bibinfo {pages} {157403} (\bibinfo {year}
  {2020}{\natexlab{b}})}\BibitemShut {NoStop}%
\bibitem [{\citenamefont {Xiao}\ \emph {et~al.}(2010)\citenamefont {Xiao},
  \citenamefont {Chang},\ and\ \citenamefont {Niu}}]{RevModPhys.82.1959}%
  \BibitemOpen
  \bibfield  {author} {\bibinfo {author} {\bibfnamefont {Di}~\bibnamefont
  {Xiao}}, \bibinfo {author} {\bibfnamefont {Ming-Che}\ \bibnamefont {Chang}},
  \ and\ \bibinfo {author} {\bibfnamefont {Qian}\ \bibnamefont {Niu}},\
  }\bibfield  {title} {\enquote {\bibinfo {title} {Berry phase effects on
  electronic properties},}\ }\href {\doibase 10.1103/RevModPhys.82.1959}
  {\bibfield  {journal} {\bibinfo  {journal} {Rev. Mod. Phys.}\ }\textbf
  {\bibinfo {volume} {82}},\ \bibinfo {pages} {1959--2007} (\bibinfo {year}
  {2010})}\BibitemShut {NoStop}%
\bibitem [{\citenamefont {Kuga}\ \emph {et~al.}(2008)\citenamefont {Kuga},
  \citenamefont {Murakami},\ and\ \citenamefont
  {Nagaosa}}]{PhysRevB.78.205201}%
  \BibitemOpen
  \bibfield  {author} {\bibinfo {author} {\bibfnamefont {Shun-ichi}\
  \bibnamefont {Kuga}}, \bibinfo {author} {\bibfnamefont {Shuichi}\
  \bibnamefont {Murakami}}, \ and\ \bibinfo {author} {\bibfnamefont {Naoto}\
  \bibnamefont {Nagaosa}},\ }\bibfield  {title} {\enquote {\bibinfo {title}
  {Spin {H}all effect of excitons},}\ }\href {\doibase
  10.1103/PhysRevB.78.205201} {\bibfield  {journal} {\bibinfo  {journal} {Phys.
  Rev. B}\ }\textbf {\bibinfo {volume} {78}},\ \bibinfo {pages} {205201}
  (\bibinfo {year} {2008})}\BibitemShut {NoStop}%
\bibitem [{\citenamefont {Yao}\ and\ \citenamefont
  {Niu}(2008)}]{PhysRevLett.101.106401}%
  \BibitemOpen
  \bibfield  {author} {\bibinfo {author} {\bibfnamefont {Wang}\ \bibnamefont
  {Yao}}\ and\ \bibinfo {author} {\bibfnamefont {Qian}\ \bibnamefont {Niu}},\
  }\bibfield  {title} {\enquote {\bibinfo {title} {Berry phase effect on the
  exciton transport and on the exciton {B}ose-{E}instein condensate},}\ }\href
  {\doibase 10.1103/PhysRevLett.101.106401} {\bibfield  {journal} {\bibinfo
  {journal} {Phys. Rev. Lett.}\ }\textbf {\bibinfo {volume} {101}},\ \bibinfo
  {pages} {106401} (\bibinfo {year} {2008})}\BibitemShut {NoStop}%
\bibitem [{\citenamefont {Li}\ \emph {et~al.}(2015)\citenamefont {Li},
  \citenamefont {Li}, \citenamefont {Shi}, \citenamefont {Zhang}, \citenamefont
  {Yang},\ and\ \citenamefont {Chang}}]{PhysRevLett.115.166804}%
  \BibitemOpen
  \bibfield  {author} {\bibinfo {author} {\bibfnamefont {Yun-Mei}\ \bibnamefont
  {Li}}, \bibinfo {author} {\bibfnamefont {Jian}\ \bibnamefont {Li}}, \bibinfo
  {author} {\bibfnamefont {Li-Kun}\ \bibnamefont {Shi}}, \bibinfo {author}
  {\bibfnamefont {Dong}\ \bibnamefont {Zhang}}, \bibinfo {author}
  {\bibfnamefont {Wen}\ \bibnamefont {Yang}}, \ and\ \bibinfo {author}
  {\bibfnamefont {Kai}\ \bibnamefont {Chang}},\ }\bibfield  {title} {\enquote
  {\bibinfo {title} {{Light-Induced Exciton Spin Hall Effect in van der Waals
  Heterostructures}},}\ }\href {\doibase 10.1103/PhysRevLett.115.166804}
  {\bibfield  {journal} {\bibinfo  {journal} {Phys. Rev. Lett.}\ }\textbf
  {\bibinfo {volume} {115}},\ \bibinfo {pages} {166804} (\bibinfo {year}
  {2015})}\BibitemShut {NoStop}%
\bibitem [{\citenamefont {Kovalev}\ and\ \citenamefont
  {Savenko}(2019)}]{PhysRevB.100.121405}%
  \BibitemOpen
  \bibfield  {author} {\bibinfo {author} {\bibfnamefont {V.~M.}\ \bibnamefont
  {Kovalev}}\ and\ \bibinfo {author} {\bibfnamefont {I.~G.}\ \bibnamefont
  {Savenko}},\ }\bibfield  {title} {\enquote {\bibinfo {title} {Quantum
  anomalous valley {H}all effect for bosons},}\ }\href {\doibase
  10.1103/PhysRevB.100.121405} {\bibfield  {journal} {\bibinfo  {journal}
  {Phys. Rev. B}\ }\textbf {\bibinfo {volume} {100}},\ \bibinfo {pages}
  {121405} (\bibinfo {year} {2019})}\BibitemShut {NoStop}%
\bibitem [{\citenamefont {Gianfrate}\ \emph {et~al.}(2020)\citenamefont
  {Gianfrate}, \citenamefont {Bleu}, \citenamefont {Dominici}, \citenamefont
  {Ardizzone}, \citenamefont {De~Giorgi}, \citenamefont {Ballarini},
  \citenamefont {Lerario}, \citenamefont {West}, \citenamefont {Pfeiffer},
  \citenamefont {Solnyshkov}, \citenamefont {Sanvitto},\ and\ \citenamefont
  {Malpuech}}]{Gianfrate:2020aa}%
  \BibitemOpen
  \bibfield  {author} {\bibinfo {author} {\bibfnamefont {A.}~\bibnamefont
  {Gianfrate}}, \bibinfo {author} {\bibfnamefont {O.}~\bibnamefont {Bleu}},
  \bibinfo {author} {\bibfnamefont {L.}~\bibnamefont {Dominici}}, \bibinfo
  {author} {\bibfnamefont {V.}~\bibnamefont {Ardizzone}}, \bibinfo {author}
  {\bibfnamefont {M.}~\bibnamefont {De~Giorgi}}, \bibinfo {author}
  {\bibfnamefont {D.}~\bibnamefont {Ballarini}}, \bibinfo {author}
  {\bibfnamefont {G.}~\bibnamefont {Lerario}}, \bibinfo {author} {\bibfnamefont
  {K.~W.}\ \bibnamefont {West}}, \bibinfo {author} {\bibfnamefont {L.~N.}\
  \bibnamefont {Pfeiffer}}, \bibinfo {author} {\bibfnamefont {D.~D.}\
  \bibnamefont {Solnyshkov}}, \bibinfo {author} {\bibfnamefont
  {D.}~\bibnamefont {Sanvitto}}, \ and\ \bibinfo {author} {\bibfnamefont
  {G.}~\bibnamefont {Malpuech}},\ }\bibfield  {title} {\enquote {\bibinfo
  {title} {Measurement of the quantum geometric tensor and of the anomalous
  {H}all drift},}\ }\href {\doibase 10.1038/s41586-020-1989-2} {\bibfield
  {journal} {\bibinfo  {journal} {Nature}\ }\textbf {\bibinfo {volume} {578}},\
  \bibinfo {pages} {381--385} (\bibinfo {year} {2020})}\BibitemShut {NoStop}%
\bibitem [{\citenamefont {Sinitsyn}\ \emph {et~al.}(2007)\citenamefont
  {Sinitsyn}, \citenamefont {MacDonald}, \citenamefont {Jungwirth},
  \citenamefont {Dugaev},\ and\ \citenamefont {Sinova}}]{PhysRevB.75.045315}%
  \BibitemOpen
  \bibfield  {author} {\bibinfo {author} {\bibfnamefont {N.~A.}\ \bibnamefont
  {Sinitsyn}}, \bibinfo {author} {\bibfnamefont {A.~H.}\ \bibnamefont
  {MacDonald}}, \bibinfo {author} {\bibfnamefont {T.}~\bibnamefont
  {Jungwirth}}, \bibinfo {author} {\bibfnamefont {V.~K.}\ \bibnamefont
  {Dugaev}}, \ and\ \bibinfo {author} {\bibfnamefont {Jairo}\ \bibnamefont
  {Sinova}},\ }\bibfield  {title} {\enquote {\bibinfo {title} {Anomalous {H}all
  effect in a two-dimensional {D}irac band: {T}he link between the
  {K}ubo-{S}treda formula and the semiclassical {B}oltzmann equation
  approach},}\ }\href {\doibase 10.1103/PhysRevB.75.045315} {\bibfield
  {journal} {\bibinfo  {journal} {Phys. Rev. B}\ }\textbf {\bibinfo {volume}
  {75}},\ \bibinfo {pages} {045315} (\bibinfo {year} {2007})}\BibitemShut
  {NoStop}%
\bibitem [{\citenamefont {Sinitsyn}(2007)}]{Sinitsyn_2007}%
  \BibitemOpen
  \bibfield  {author} {\bibinfo {author} {\bibfnamefont {N.~A.}\ \bibnamefont
  {Sinitsyn}},\ }\bibfield  {title} {\enquote {\bibinfo {title} {Semiclassical
  theories of the anomalous {H}all effect},}\ }\href {\doibase
  10.1088/0953-8984/20/02/023201} {\bibfield  {journal} {\bibinfo  {journal}
  {Journal of Physics: Condensed Matter}\ }\textbf {\bibinfo {volume} {20}},\
  \bibinfo {pages} {023201} (\bibinfo {year} {2007})}\BibitemShut {NoStop}%
\bibitem [{\citenamefont {Ado}\ \emph {et~al.}(2015)\citenamefont {Ado},
  \citenamefont {Dmitriev}, \citenamefont {Ostrovsky},\ and\ \citenamefont
  {Titov}}]{Ado_2015}%
  \BibitemOpen
  \bibfield  {author} {\bibinfo {author} {\bibfnamefont {I.~A.}\ \bibnamefont
  {Ado}}, \bibinfo {author} {\bibfnamefont {I.~A.}\ \bibnamefont {Dmitriev}},
  \bibinfo {author} {\bibfnamefont {P.~M.}\ \bibnamefont {Ostrovsky}}, \ and\
  \bibinfo {author} {\bibfnamefont {M.}~\bibnamefont {Titov}},\ }\bibfield
  {title} {\enquote {\bibinfo {title} {Anomalous {H}all effect with massive
  {D}irac fermions},}\ }\href {\doibase 10.1209/0295-5075/111/37004} {\bibfield
   {journal} {\bibinfo  {journal} {{EPL}}\ }\textbf {\bibinfo {volume} {111}},\
  \bibinfo {pages} {37004} (\bibinfo {year} {2015})}\BibitemShut {NoStop}%
\bibitem [{\citenamefont {de~Jong}\ and\ \citenamefont
  {Molenkamp}(1995)}]{PhysRevB.51.13389}%
  \BibitemOpen
  \bibfield  {author} {\bibinfo {author} {\bibfnamefont {M.~J.~M.}\
  \bibnamefont {de~Jong}}\ and\ \bibinfo {author} {\bibfnamefont {L.~W.}\
  \bibnamefont {Molenkamp}},\ }\bibfield  {title} {\enquote {\bibinfo {title}
  {Hydrodynamic electron flow in high-mobility wires},}\ }\href {\doibase
  10.1103/PhysRevB.51.13389} {\bibfield  {journal} {\bibinfo  {journal} {Phys.
  Rev. B}\ }\textbf {\bibinfo {volume} {51}},\ \bibinfo {pages} {13389--13402}
  (\bibinfo {year} {1995})}\BibitemShut {NoStop}%
\bibitem [{\citenamefont {Titov}\ \emph {et~al.}(2013)\citenamefont {Titov},
  \citenamefont {Gorbachev}, \citenamefont {Narozhny}, \citenamefont
  {Tudorovskiy}, \citenamefont {Sch\"utt}, \citenamefont {Ostrovsky},
  \citenamefont {Gornyi}, \citenamefont {Mirlin}, \citenamefont {Katsnelson},
  \citenamefont {Novoselov}, \citenamefont {Geim},\ and\ \citenamefont
  {Ponomarenko}}]{PhysRevLett.111.166601}%
  \BibitemOpen
  \bibfield  {author} {\bibinfo {author} {\bibfnamefont {M.}~\bibnamefont
  {Titov}}, \bibinfo {author} {\bibfnamefont {R.~V.}\ \bibnamefont
  {Gorbachev}}, \bibinfo {author} {\bibfnamefont {B.~N.}\ \bibnamefont
  {Narozhny}}, \bibinfo {author} {\bibfnamefont {T.}~\bibnamefont
  {Tudorovskiy}}, \bibinfo {author} {\bibfnamefont {M.}~\bibnamefont
  {Sch\"utt}}, \bibinfo {author} {\bibfnamefont {P.~M.}\ \bibnamefont
  {Ostrovsky}}, \bibinfo {author} {\bibfnamefont {I.~V.}\ \bibnamefont
  {Gornyi}}, \bibinfo {author} {\bibfnamefont {A.~D.}\ \bibnamefont {Mirlin}},
  \bibinfo {author} {\bibfnamefont {M.~I.}\ \bibnamefont {Katsnelson}},
  \bibinfo {author} {\bibfnamefont {K.~S.}\ \bibnamefont {Novoselov}}, \bibinfo
  {author} {\bibfnamefont {A.~K.}\ \bibnamefont {Geim}}, \ and\ \bibinfo
  {author} {\bibfnamefont {L.~A.}\ \bibnamefont {Ponomarenko}},\ }\bibfield
  {title} {\enquote {\bibinfo {title} {Giant magnetodrag in graphene at charge
  neutrality},}\ }\href {\doibase 10.1103/PhysRevLett.111.166601} {\bibfield
  {journal} {\bibinfo  {journal} {Phys. Rev. Lett.}\ }\textbf {\bibinfo
  {volume} {111}},\ \bibinfo {pages} {166601} (\bibinfo {year}
  {2013})}\BibitemShut {NoStop}%
\bibitem [{\citenamefont {Bandurin}\ \emph {et~al.}(2016)\citenamefont
  {Bandurin}, \citenamefont {Torre}, \citenamefont {Kumar}, \citenamefont
  {Ben~Shalom}, \citenamefont {Tomadin}, \citenamefont {Principi},
  \citenamefont {Auton}, \citenamefont {Khestanova}, \citenamefont {Novoselov},
  \citenamefont {Grigorieva}, \citenamefont {Ponomarenko}, \citenamefont
  {Geim},\ and\ \citenamefont {Polini}}]{Bandurin1055}%
  \BibitemOpen
  \bibfield  {author} {\bibinfo {author} {\bibfnamefont {D.~A.}\ \bibnamefont
  {Bandurin}}, \bibinfo {author} {\bibfnamefont {I.}~\bibnamefont {Torre}},
  \bibinfo {author} {\bibfnamefont {R.~Krishna}\ \bibnamefont {Kumar}},
  \bibinfo {author} {\bibfnamefont {M.}~\bibnamefont {Ben~Shalom}}, \bibinfo
  {author} {\bibfnamefont {A.}~\bibnamefont {Tomadin}}, \bibinfo {author}
  {\bibfnamefont {A.}~\bibnamefont {Principi}}, \bibinfo {author}
  {\bibfnamefont {G.~H.}\ \bibnamefont {Auton}}, \bibinfo {author}
  {\bibfnamefont {E.}~\bibnamefont {Khestanova}}, \bibinfo {author}
  {\bibfnamefont {K.~S.}\ \bibnamefont {Novoselov}}, \bibinfo {author}
  {\bibfnamefont {I.~V.}\ \bibnamefont {Grigorieva}}, \bibinfo {author}
  {\bibfnamefont {L.~A.}\ \bibnamefont {Ponomarenko}}, \bibinfo {author}
  {\bibfnamefont {A.~K.}\ \bibnamefont {Geim}}, \ and\ \bibinfo {author}
  {\bibfnamefont {M.}~\bibnamefont {Polini}},\ }\bibfield  {title} {\enquote
  {\bibinfo {title} {Negative local resistance caused by viscous electron
  backflow in graphene},}\ }\href {\doibase 10.1126/science.aad0201} {\bibfield
   {journal} {\bibinfo  {journal} {Science}\ }\textbf {\bibinfo {volume}
  {351}},\ \bibinfo {pages} {1055--1058} (\bibinfo {year} {2016})}\BibitemShut
  {NoStop}%
\bibitem [{\citenamefont {Moll}\ \emph {et~al.}(2016)\citenamefont {Moll},
  \citenamefont {Kushwaha}, \citenamefont {Nandi}, \citenamefont {Schmidt},\
  and\ \citenamefont {Mackenzie}}]{Moll1061}%
  \BibitemOpen
  \bibfield  {author} {\bibinfo {author} {\bibfnamefont {Philip J.~W.}\
  \bibnamefont {Moll}}, \bibinfo {author} {\bibfnamefont {Pallavi}\
  \bibnamefont {Kushwaha}}, \bibinfo {author} {\bibfnamefont {Nabhanila}\
  \bibnamefont {Nandi}}, \bibinfo {author} {\bibfnamefont {Burkhard}\
  \bibnamefont {Schmidt}}, \ and\ \bibinfo {author} {\bibfnamefont {Andrew~P.}\
  \bibnamefont {Mackenzie}},\ }\bibfield  {title} {\enquote {\bibinfo {title}
  {{Evidence for hydrodynamic electron flow in PdCoO$_2$}},}\ }\href {\doibase
  10.1126/science.aac8385} {\bibfield  {journal} {\bibinfo  {journal}
  {Science}\ }\textbf {\bibinfo {volume} {351}},\ \bibinfo {pages} {1061--1064}
  (\bibinfo {year} {2016})}\BibitemShut {NoStop}%
\bibitem [{\citenamefont {Gusev}\ \emph {et~al.}(2018)\citenamefont {Gusev},
  \citenamefont {Levin}, \citenamefont {Levinson},\ and\ \citenamefont
  {Bakarov}}]{Gusev:2018tg}%
  \BibitemOpen
  \bibfield  {author} {\bibinfo {author} {\bibfnamefont {G.~M.}\ \bibnamefont
  {Gusev}}, \bibinfo {author} {\bibfnamefont {A.~D.}\ \bibnamefont {Levin}},
  \bibinfo {author} {\bibfnamefont {E.~V.}\ \bibnamefont {Levinson}}, \ and\
  \bibinfo {author} {\bibfnamefont {A.~K.}\ \bibnamefont {Bakarov}},\
  }\bibfield  {title} {\enquote {\bibinfo {title} {Viscous electron flow in
  mesoscopic two-dimensional electron gas},}\ }\href {\doibase
  10.1063/1.5020763} {\bibfield  {journal} {\bibinfo  {journal} {AIP Advances}\
  }\textbf {\bibinfo {volume} {8}},\ \bibinfo {pages} {025318} (\bibinfo {year}
  {2018})}\BibitemShut {NoStop}%
\bibitem [{\citenamefont {Krishna~Kumar}\ \emph {et~al.}(2017)\citenamefont
  {Krishna~Kumar}, \citenamefont {Bandurin}, \citenamefont {Pellegrino},
  \citenamefont {Cao}, \citenamefont {Principi}, \citenamefont {Guo},
  \citenamefont {Auton}, \citenamefont {Ben~Shalom}, \citenamefont
  {Ponomarenko}, \citenamefont {Falkovich}, \citenamefont {Watanabe},
  \citenamefont {Taniguchi}, \citenamefont {Grigorieva}, \citenamefont
  {Levitov}, \citenamefont {Polini},\ and\ \citenamefont
  {Geim}}]{Krishna-Kumar:2017wn}%
  \BibitemOpen
  \bibfield  {author} {\bibinfo {author} {\bibfnamefont {R.}~\bibnamefont
  {Krishna~Kumar}}, \bibinfo {author} {\bibfnamefont {D.~A.}\ \bibnamefont
  {Bandurin}}, \bibinfo {author} {\bibfnamefont {F.~M.~D.}\ \bibnamefont
  {Pellegrino}}, \bibinfo {author} {\bibfnamefont {Y.}~\bibnamefont {Cao}},
  \bibinfo {author} {\bibfnamefont {A.}~\bibnamefont {Principi}}, \bibinfo
  {author} {\bibfnamefont {H.}~\bibnamefont {Guo}}, \bibinfo {author}
  {\bibfnamefont {G.~H.}\ \bibnamefont {Auton}}, \bibinfo {author}
  {\bibfnamefont {M.}~\bibnamefont {Ben~Shalom}}, \bibinfo {author}
  {\bibfnamefont {L.~A.}\ \bibnamefont {Ponomarenko}}, \bibinfo {author}
  {\bibfnamefont {G.}~\bibnamefont {Falkovich}}, \bibinfo {author}
  {\bibfnamefont {K.}~\bibnamefont {Watanabe}}, \bibinfo {author}
  {\bibfnamefont {T.}~\bibnamefont {Taniguchi}}, \bibinfo {author}
  {\bibfnamefont {I.~V.}\ \bibnamefont {Grigorieva}}, \bibinfo {author}
  {\bibfnamefont {L.~S.}\ \bibnamefont {Levitov}}, \bibinfo {author}
  {\bibfnamefont {M.}~\bibnamefont {Polini}}, \ and\ \bibinfo {author}
  {\bibfnamefont {A.~K.}\ \bibnamefont {Geim}},\ }\bibfield  {title} {\enquote
  {\bibinfo {title} {Superballistic flow of viscous electron fluid through
  graphene constrictions},}\ }\href {\doibase 10.1038/nphys4240} {\bibfield
  {journal} {\bibinfo  {journal} {Nature Physics}\ }\textbf {\bibinfo {volume}
  {13}},\ \bibinfo {pages} {1182--1185} (\bibinfo {year} {2017})}\BibitemShut
  {NoStop}%
\bibitem [{\citenamefont {Ku}\ \emph {et~al.}(2020)\citenamefont {Ku},
  \citenamefont {Zhou}, \citenamefont {Li}, \citenamefont {Shin}, \citenamefont
  {Shi}, \citenamefont {Burch}, \citenamefont {Anderson}, \citenamefont
  {Pierce}, \citenamefont {Xie}, \citenamefont {Hamo}, \citenamefont {Vool},
  \citenamefont {Zhang}, \citenamefont {Casola}, \citenamefont {Taniguchi},
  \citenamefont {Watanabe}, \citenamefont {Fogler}, \citenamefont {Kim},
  \citenamefont {Yacoby},\ and\ \citenamefont {Walsworth}}]{Ku:2020ue}%
  \BibitemOpen
  \bibfield  {author} {\bibinfo {author} {\bibfnamefont {Mark J.~H.}\
  \bibnamefont {Ku}}, \bibinfo {author} {\bibfnamefont {Tony~X.}\ \bibnamefont
  {Zhou}}, \bibinfo {author} {\bibfnamefont {Qing}\ \bibnamefont {Li}},
  \bibinfo {author} {\bibfnamefont {Young~J.}\ \bibnamefont {Shin}}, \bibinfo
  {author} {\bibfnamefont {Jing~K.}\ \bibnamefont {Shi}}, \bibinfo {author}
  {\bibfnamefont {Claire}\ \bibnamefont {Burch}}, \bibinfo {author}
  {\bibfnamefont {Laurel~E.}\ \bibnamefont {Anderson}}, \bibinfo {author}
  {\bibfnamefont {Andrew~T.}\ \bibnamefont {Pierce}}, \bibinfo {author}
  {\bibfnamefont {Yonglong}\ \bibnamefont {Xie}}, \bibinfo {author}
  {\bibfnamefont {Assaf}\ \bibnamefont {Hamo}}, \bibinfo {author}
  {\bibfnamefont {Uri}\ \bibnamefont {Vool}}, \bibinfo {author} {\bibfnamefont
  {Huiliang}\ \bibnamefont {Zhang}}, \bibinfo {author} {\bibfnamefont
  {Francesco}\ \bibnamefont {Casola}}, \bibinfo {author} {\bibfnamefont
  {Takashi}\ \bibnamefont {Taniguchi}}, \bibinfo {author} {\bibfnamefont
  {Kenji}\ \bibnamefont {Watanabe}}, \bibinfo {author} {\bibfnamefont
  {Michael~M.}\ \bibnamefont {Fogler}}, \bibinfo {author} {\bibfnamefont
  {Philip}\ \bibnamefont {Kim}}, \bibinfo {author} {\bibfnamefont {Amir}\
  \bibnamefont {Yacoby}}, \ and\ \bibinfo {author} {\bibfnamefont {Ronald~L.}\
  \bibnamefont {Walsworth}},\ }\bibfield  {title} {\enquote {\bibinfo {title}
  {Imaging viscous flow of the {D}irac fluid in graphene},}\ }\href {\doibase
  10.1038/s41586-020-2507-2} {\bibfield  {journal} {\bibinfo  {journal}
  {Nature}\ }\textbf {\bibinfo {volume} {583}},\ \bibinfo {pages} {537--541}
  (\bibinfo {year} {2020})}\BibitemShut {NoStop}%
\bibitem [{\citenamefont {Sulpizio}\ \emph {et~al.}(2019)\citenamefont
  {Sulpizio}, \citenamefont {Ella}, \citenamefont {Rozen}, \citenamefont
  {Birkbeck}, \citenamefont {Perello}, \citenamefont {Dutta}, \citenamefont
  {Ben-Shalom}, \citenamefont {Taniguchi}, \citenamefont {Watanabe},
  \citenamefont {Holder}, \citenamefont {Queiroz}, \citenamefont {Principi},
  \citenamefont {Stern}, \citenamefont {Scaffidi}, \citenamefont {Geim},\ and\
  \citenamefont {Ilani}}]{Sulpizio:2019uc}%
  \BibitemOpen
  \bibfield  {author} {\bibinfo {author} {\bibfnamefont {Joseph~A.}\
  \bibnamefont {Sulpizio}}, \bibinfo {author} {\bibfnamefont {Lior}\
  \bibnamefont {Ella}}, \bibinfo {author} {\bibfnamefont {Asaf}\ \bibnamefont
  {Rozen}}, \bibinfo {author} {\bibfnamefont {John}\ \bibnamefont {Birkbeck}},
  \bibinfo {author} {\bibfnamefont {David~J.}\ \bibnamefont {Perello}},
  \bibinfo {author} {\bibfnamefont {Debarghya}\ \bibnamefont {Dutta}}, \bibinfo
  {author} {\bibfnamefont {Moshe}\ \bibnamefont {Ben-Shalom}}, \bibinfo
  {author} {\bibfnamefont {Takashi}\ \bibnamefont {Taniguchi}}, \bibinfo
  {author} {\bibfnamefont {Kenji}\ \bibnamefont {Watanabe}}, \bibinfo {author}
  {\bibfnamefont {Tobias}\ \bibnamefont {Holder}}, \bibinfo {author}
  {\bibfnamefont {Raquel}\ \bibnamefont {Queiroz}}, \bibinfo {author}
  {\bibfnamefont {Alessandro}\ \bibnamefont {Principi}}, \bibinfo {author}
  {\bibfnamefont {Ady}\ \bibnamefont {Stern}}, \bibinfo {author} {\bibfnamefont
  {Thomas}\ \bibnamefont {Scaffidi}}, \bibinfo {author} {\bibfnamefont
  {Andre~K.}\ \bibnamefont {Geim}}, \ and\ \bibinfo {author} {\bibfnamefont
  {Shahal}\ \bibnamefont {Ilani}},\ }\bibfield  {title} {\enquote {\bibinfo
  {title} {Visualizing {P}oiseuille flow of hydrodynamic electrons},}\ }\href
  {\doibase 10.1038/s41586-019-1788-9} {\bibfield  {journal} {\bibinfo
  {journal} {Nature}\ }\textbf {\bibinfo {volume} {576}},\ \bibinfo {pages}
  {75--79} (\bibinfo {year} {2019})}\BibitemShut {NoStop}%
\bibitem [{\citenamefont {Berdyugin}\ \emph {et~al.}(2019)\citenamefont
  {Berdyugin}, \citenamefont {Xu}, \citenamefont {Pellegrino}, \citenamefont
  {Krishna~Kumar}, \citenamefont {Principi}, \citenamefont {Torre},
  \citenamefont {Ben~Shalom}, \citenamefont {Taniguchi}, \citenamefont
  {Watanabe}, \citenamefont {Grigorieva}, \citenamefont {Polini}, \citenamefont
  {Geim},\ and\ \citenamefont {Bandurin}}]{Berdyugin162}%
  \BibitemOpen
  \bibfield  {author} {\bibinfo {author} {\bibfnamefont {A.~I.}\ \bibnamefont
  {Berdyugin}}, \bibinfo {author} {\bibfnamefont {S.~G.}\ \bibnamefont {Xu}},
  \bibinfo {author} {\bibfnamefont {F.~M.~D.}\ \bibnamefont {Pellegrino}},
  \bibinfo {author} {\bibfnamefont {R.}~\bibnamefont {Krishna~Kumar}}, \bibinfo
  {author} {\bibfnamefont {A.}~\bibnamefont {Principi}}, \bibinfo {author}
  {\bibfnamefont {I.}~\bibnamefont {Torre}}, \bibinfo {author} {\bibfnamefont
  {M.}~\bibnamefont {Ben~Shalom}}, \bibinfo {author} {\bibfnamefont
  {T.}~\bibnamefont {Taniguchi}}, \bibinfo {author} {\bibfnamefont
  {K.}~\bibnamefont {Watanabe}}, \bibinfo {author} {\bibfnamefont {I.~V.}\
  \bibnamefont {Grigorieva}}, \bibinfo {author} {\bibfnamefont
  {M.}~\bibnamefont {Polini}}, \bibinfo {author} {\bibfnamefont {A.~K.}\
  \bibnamefont {Geim}}, \ and\ \bibinfo {author} {\bibfnamefont {D.~A.}\
  \bibnamefont {Bandurin}},\ }\bibfield  {title} {\enquote {\bibinfo {title}
  {Measuring {H}all viscosity of graphene{\textquoteright}s electron fluid},}\
  }\href {\doibase 10.1126/science.aau0685} {\bibfield  {journal} {\bibinfo
  {journal} {Science}\ }\textbf {\bibinfo {volume} {364}},\ \bibinfo {pages}
  {162--165} (\bibinfo {year} {2019})}\BibitemShut {NoStop}%
\bibitem [{\citenamefont {Gusev}\ \emph {et~al.}(2020)\citenamefont {Gusev},
  \citenamefont {Jaroshevich}, \citenamefont {Levin}, \citenamefont {Kvon},\
  and\ \citenamefont {Bakarov}}]{Gusev:2020vd}%
  \BibitemOpen
  \bibfield  {author} {\bibinfo {author} {\bibfnamefont {G.~M.}\ \bibnamefont
  {Gusev}}, \bibinfo {author} {\bibfnamefont {A.~S.}\ \bibnamefont
  {Jaroshevich}}, \bibinfo {author} {\bibfnamefont {A.~D.}\ \bibnamefont
  {Levin}}, \bibinfo {author} {\bibfnamefont {Z.~D.}\ \bibnamefont {Kvon}}, \
  and\ \bibinfo {author} {\bibfnamefont {A.~K.}\ \bibnamefont {Bakarov}},\
  }\bibfield  {title} {\enquote {\bibinfo {title} {Stokes flow around an
  obstacle in viscous two-dimensional electron liquid},}\ }\href {\doibase
  10.1038/s41598-020-64807-6} {\bibfield  {journal} {\bibinfo  {journal}
  {Scientific Reports}\ }\textbf {\bibinfo {volume} {10}},\ \bibinfo {pages}
  {7860} (\bibinfo {year} {2020})}\BibitemShut {NoStop}%
\bibitem [{\citenamefont {Gurzhi}(1963)}]{gurzhi63}%
  \BibitemOpen
  \bibfield  {author} {\bibinfo {author} {\bibfnamefont {R.~N.}\ \bibnamefont
  {Gurzhi}},\ }\bibfield  {title} {\enquote {\bibinfo {title} {Minimum of
  resistance in impurity-free conductors},}\ }\href@noop {} {\bibfield
  {journal} {\bibinfo  {journal} {JETP}\ }\textbf {\bibinfo {volume} {17}},\
  \bibinfo {pages} {521} (\bibinfo {year} {1963})}\BibitemShut {NoStop}%
\bibitem [{\citenamefont {Gurzhi}(1968)}]{Gurzhi_1968}%
  \BibitemOpen
  \bibfield  {author} {\bibinfo {author} {\bibfnamefont {R.~N.}\ \bibnamefont
  {Gurzhi}},\ }\bibfield  {title} {\enquote {\bibinfo {title} {Hydrodynamic
  effects in solids at low temperatures},}\ }\href {\doibase
  10.1070/pu1968v011n02abeh003815} {\bibfield  {journal} {\bibinfo  {journal}
  {Soviet Physics Uspekhi}\ }\textbf {\bibinfo {volume} {11}},\ \bibinfo
  {pages} {255--270} (\bibinfo {year} {1968})}\BibitemShut {NoStop}%
\bibitem [{\citenamefont {M\"uller}\ \emph {et~al.}(2009)\citenamefont
  {M\"uller}, \citenamefont {Schmalian},\ and\ \citenamefont
  {Fritz}}]{PhysRevLett.103.025301}%
  \BibitemOpen
  \bibfield  {author} {\bibinfo {author} {\bibfnamefont {Markus}\ \bibnamefont
  {M\"uller}}, \bibinfo {author} {\bibfnamefont {J\"org}\ \bibnamefont
  {Schmalian}}, \ and\ \bibinfo {author} {\bibfnamefont {Lars}\ \bibnamefont
  {Fritz}},\ }\bibfield  {title} {\enquote {\bibinfo {title} {Graphene: A
  nearly perfect fluid},}\ }\href {\doibase 10.1103/PhysRevLett.103.025301}
  {\bibfield  {journal} {\bibinfo  {journal} {Phys. Rev. Lett.}\ }\textbf
  {\bibinfo {volume} {103}},\ \bibinfo {pages} {025301} (\bibinfo {year}
  {2009})}\BibitemShut {NoStop}%
\bibitem [{\citenamefont {Andreev}\ \emph {et~al.}(2011)\citenamefont
  {Andreev}, \citenamefont {Kivelson},\ and\ \citenamefont
  {Spivak}}]{PhysRevLett.106.256804}%
  \BibitemOpen
  \bibfield  {author} {\bibinfo {author} {\bibfnamefont {A.~V.}\ \bibnamefont
  {Andreev}}, \bibinfo {author} {\bibfnamefont {Steven~A.}\ \bibnamefont
  {Kivelson}}, \ and\ \bibinfo {author} {\bibfnamefont {B.}~\bibnamefont
  {Spivak}},\ }\bibfield  {title} {\enquote {\bibinfo {title} {Hydrodynamic
  description of transport in strongly correlated electron systems},}\ }\href
  {\doibase 10.1103/PhysRevLett.106.256804} {\bibfield  {journal} {\bibinfo
  {journal} {Phys. Rev. Lett.}\ }\textbf {\bibinfo {volume} {106}},\ \bibinfo
  {pages} {256804} (\bibinfo {year} {2011})}\BibitemShut {NoStop}%
\bibitem [{\citenamefont {Torre}\ \emph {et~al.}(2015)\citenamefont {Torre},
  \citenamefont {Tomadin}, \citenamefont {Geim},\ and\ \citenamefont
  {Polini}}]{PhysRevB.92.165433}%
  \BibitemOpen
  \bibfield  {author} {\bibinfo {author} {\bibfnamefont {Iacopo}\ \bibnamefont
  {Torre}}, \bibinfo {author} {\bibfnamefont {Andrea}\ \bibnamefont {Tomadin}},
  \bibinfo {author} {\bibfnamefont {Andre~K.}\ \bibnamefont {Geim}}, \ and\
  \bibinfo {author} {\bibfnamefont {Marco}\ \bibnamefont {Polini}},\ }\bibfield
   {title} {\enquote {\bibinfo {title} {Nonlocal transport and the hydrodynamic
  shear viscosity in graphene},}\ }\href {\doibase 10.1103/PhysRevB.92.165433}
  {\bibfield  {journal} {\bibinfo  {journal} {Phys. Rev. B}\ }\textbf {\bibinfo
  {volume} {92}},\ \bibinfo {pages} {165433} (\bibinfo {year}
  {2015})}\BibitemShut {NoStop}%
\bibitem [{\citenamefont {Pellegrino}\ \emph {et~al.}(2017)\citenamefont
  {Pellegrino}, \citenamefont {Torre},\ and\ \citenamefont
  {Polini}}]{PhysRevB.96.195401}%
  \BibitemOpen
  \bibfield  {author} {\bibinfo {author} {\bibfnamefont {Francesco M.~D.}\
  \bibnamefont {Pellegrino}}, \bibinfo {author} {\bibfnamefont {Iacopo}\
  \bibnamefont {Torre}}, \ and\ \bibinfo {author} {\bibfnamefont {Marco}\
  \bibnamefont {Polini}},\ }\bibfield  {title} {\enquote {\bibinfo {title}
  {Nonlocal transport and the hall viscosity of two-dimensional hydrodynamic
  electron liquids},}\ }\href {\doibase 10.1103/PhysRevB.96.195401} {\bibfield
  {journal} {\bibinfo  {journal} {Phys. Rev. B}\ }\textbf {\bibinfo {volume}
  {96}},\ \bibinfo {pages} {195401} (\bibinfo {year} {2017})}\BibitemShut
  {NoStop}%
\bibitem [{\citenamefont {Narozhny}\ \emph {et~al.}(2015)\citenamefont
  {Narozhny}, \citenamefont {Gornyi}, \citenamefont {Titov}, \citenamefont
  {Sch\"utt},\ and\ \citenamefont {Mirlin}}]{PhysRevB.91.035414}%
  \BibitemOpen
  \bibfield  {author} {\bibinfo {author} {\bibfnamefont {B.~N.}\ \bibnamefont
  {Narozhny}}, \bibinfo {author} {\bibfnamefont {I.~V.}\ \bibnamefont
  {Gornyi}}, \bibinfo {author} {\bibfnamefont {M.}~\bibnamefont {Titov}},
  \bibinfo {author} {\bibfnamefont {M.}~\bibnamefont {Sch\"utt}}, \ and\
  \bibinfo {author} {\bibfnamefont {A.~D.}\ \bibnamefont {Mirlin}},\ }\bibfield
   {title} {\enquote {\bibinfo {title} {Hydrodynamics in graphene:
  Linear-response transport},}\ }\href {\doibase 10.1103/PhysRevB.91.035414}
  {\bibfield  {journal} {\bibinfo  {journal} {Phys. Rev. B}\ }\textbf {\bibinfo
  {volume} {91}},\ \bibinfo {pages} {035414} (\bibinfo {year}
  {2015})}\BibitemShut {NoStop}%
\bibitem [{\citenamefont {Alekseev}(2016)}]{PhysRevLett.117.166601}%
  \BibitemOpen
  \bibfield  {author} {\bibinfo {author} {\bibfnamefont {P.~S.}\ \bibnamefont
  {Alekseev}},\ }\bibfield  {title} {\enquote {\bibinfo {title} {Negative
  magnetoresistance in viscous flow of two-dimensional electrons},}\ }\href
  {\doibase 10.1103/PhysRevLett.117.166601} {\bibfield  {journal} {\bibinfo
  {journal} {Phys. Rev. Lett.}\ }\textbf {\bibinfo {volume} {117}},\ \bibinfo
  {pages} {166601} (\bibinfo {year} {2016})}\BibitemShut {NoStop}%
\bibitem [{\citenamefont {Levitov}\ and\ \citenamefont
  {Falkovich}(2016)}]{Levitov:2016aa}%
  \BibitemOpen
  \bibfield  {author} {\bibinfo {author} {\bibfnamefont {Leonid}\ \bibnamefont
  {Levitov}}\ and\ \bibinfo {author} {\bibfnamefont {Gregory}\ \bibnamefont
  {Falkovich}},\ }\bibfield  {title} {\enquote {\bibinfo {title} {Electron
  viscosity, current vortices and negative nonlocal resistance in graphene},}\
  }\href {\doibase 10.1038/nphys3667} {\bibfield  {journal} {\bibinfo
  {journal} {Nature Physics}\ }\textbf {\bibinfo {volume} {12}},\ \bibinfo
  {pages} {672--676} (\bibinfo {year} {2016})}\BibitemShut {NoStop}%
\bibitem [{\citenamefont {Kashuba}\ \emph {et~al.}(2018)\citenamefont
  {Kashuba}, \citenamefont {Trauzettel},\ and\ \citenamefont
  {Molenkamp}}]{PhysRevB.97.205129}%
  \BibitemOpen
  \bibfield  {author} {\bibinfo {author} {\bibfnamefont {Oleksiy}\ \bibnamefont
  {Kashuba}}, \bibinfo {author} {\bibfnamefont {Bj\"orn}\ \bibnamefont
  {Trauzettel}}, \ and\ \bibinfo {author} {\bibfnamefont {Laurens~W.}\
  \bibnamefont {Molenkamp}},\ }\bibfield  {title} {\enquote {\bibinfo {title}
  {{Relativistic Gurzhi effect in channels of Dirac materials}},}\ }\href
  {\doibase 10.1103/PhysRevB.97.205129} {\bibfield  {journal} {\bibinfo
  {journal} {Phys. Rev. B}\ }\textbf {\bibinfo {volume} {97}},\ \bibinfo
  {pages} {205129} (\bibinfo {year} {2018})}\BibitemShut {NoStop}%
\bibitem [{\citenamefont {Lucas}(2017)}]{PhysRevB.95.115425}%
  \BibitemOpen
  \bibfield  {author} {\bibinfo {author} {\bibfnamefont {Andrew}\ \bibnamefont
  {Lucas}},\ }\bibfield  {title} {\enquote {\bibinfo {title} {Stokes paradox in
  electronic {F}ermi liquids},}\ }\href {\doibase 10.1103/PhysRevB.95.115425}
  {\bibfield  {journal} {\bibinfo  {journal} {Phys. Rev. B}\ }\textbf {\bibinfo
  {volume} {95}},\ \bibinfo {pages} {115425} (\bibinfo {year}
  {2017})}\BibitemShut {NoStop}%
\bibitem [{\citenamefont {Narozhny}\ \emph {et~al.}(2017)\citenamefont
  {Narozhny}, \citenamefont {Gornyi}, \citenamefont {Mirlin},\ and\
  \citenamefont {Schmalian}}]{Narozhny:2017vc}%
  \BibitemOpen
  \bibfield  {author} {\bibinfo {author} {\bibfnamefont {Boris~N.}\
  \bibnamefont {Narozhny}}, \bibinfo {author} {\bibfnamefont {Igor~V.}\
  \bibnamefont {Gornyi}}, \bibinfo {author} {\bibfnamefont {Alexander~D.}\
  \bibnamefont {Mirlin}}, \ and\ \bibinfo {author} {\bibfnamefont {J{\"o}rg}\
  \bibnamefont {Schmalian}},\ }\bibfield  {title} {\enquote {\bibinfo {title}
  {Hydrodynamic approach to electronic transport in graphene},}\ }\href
  {\doibase https://doi.org/10.1002/andp.201700043} {\bibfield  {journal}
  {\bibinfo  {journal} {Annalen der Physik}\ }\textbf {\bibinfo {volume}
  {529}},\ \bibinfo {pages} {1700043} (\bibinfo {year} {2017})}\BibitemShut
  {NoStop}%
\bibitem [{\citenamefont {Lucas}\ and\ \citenamefont {Fong}(2018)}]{Lucas2018}%
  \BibitemOpen
  \bibfield  {author} {\bibinfo {author} {\bibfnamefont {Andrew}\ \bibnamefont
  {Lucas}}\ and\ \bibinfo {author} {\bibfnamefont {Kin~Chung}\ \bibnamefont
  {Fong}},\ }\bibfield  {title} {\enquote {\bibinfo {title} {Hydrodynamics of
  electrons in graphene},}\ }\href {\doibase 10.1088/1361-648x/aaa274}
  {\bibfield  {journal} {\bibinfo  {journal} {Journal of Physics: Condensed
  Matter}\ }\textbf {\bibinfo {volume} {30}},\ \bibinfo {pages} {053001}
  (\bibinfo {year} {2018})}\BibitemShut {NoStop}%
\bibitem [{\citenamefont {Alekseev}\ \emph {et~al.}(2018)\citenamefont
  {Alekseev}, \citenamefont {Dmitriev}, \citenamefont {Gornyi}, \citenamefont
  {Kachorovskii}, \citenamefont {Narozhny},\ and\ \citenamefont
  {Titov}}]{PhysRevB.97.085109}%
  \BibitemOpen
  \bibfield  {author} {\bibinfo {author} {\bibfnamefont {P.~S.}\ \bibnamefont
  {Alekseev}}, \bibinfo {author} {\bibfnamefont {A.~P.}\ \bibnamefont
  {Dmitriev}}, \bibinfo {author} {\bibfnamefont {I.~V.}\ \bibnamefont
  {Gornyi}}, \bibinfo {author} {\bibfnamefont {V.~Yu.}\ \bibnamefont
  {Kachorovskii}}, \bibinfo {author} {\bibfnamefont {B.~N.}\ \bibnamefont
  {Narozhny}}, \ and\ \bibinfo {author} {\bibfnamefont {M.}~\bibnamefont
  {Titov}},\ }\bibfield  {title} {\enquote {\bibinfo {title} {Nonmonotonic
  magnetoresistance of a two-dimensional viscous electron-hole fluid in a
  confined geometry},}\ }\href {\doibase 10.1103/PhysRevB.97.085109} {\bibfield
   {journal} {\bibinfo  {journal} {Phys. Rev. B}\ }\textbf {\bibinfo {volume}
  {97}},\ \bibinfo {pages} {085109} (\bibinfo {year} {2018})}\BibitemShut
  {NoStop}%
\bibitem [{\citenamefont {Alekseev}\ and\ \citenamefont
  {Semina}(2018)}]{PhysRevB.98.165412}%
  \BibitemOpen
  \bibfield  {author} {\bibinfo {author} {\bibfnamefont {P.~S.}\ \bibnamefont
  {Alekseev}}\ and\ \bibinfo {author} {\bibfnamefont {M.~A.}\ \bibnamefont
  {Semina}},\ }\bibfield  {title} {\enquote {\bibinfo {title} {Ballistic flow
  of two-dimensional interacting electrons},}\ }\href {\doibase
  10.1103/PhysRevB.98.165412} {\bibfield  {journal} {\bibinfo  {journal} {Phys.
  Rev. B}\ }\textbf {\bibinfo {volume} {98}},\ \bibinfo {pages} {165412}
  (\bibinfo {year} {2018})}\BibitemShut {NoStop}%
\bibitem [{\citenamefont {Alekseev}\ and\ \citenamefont
  {Semina}(2019)}]{PhysRevB.100.125419}%
  \BibitemOpen
  \bibfield  {author} {\bibinfo {author} {\bibfnamefont {P.~S.}\ \bibnamefont
  {Alekseev}}\ and\ \bibinfo {author} {\bibfnamefont {M.~A.}\ \bibnamefont
  {Semina}},\ }\bibfield  {title} {\enquote {\bibinfo {title} {Hall effect in a
  ballistic flow of two-dimensional interacting particles},}\ }\href {\doibase
  10.1103/PhysRevB.100.125419} {\bibfield  {journal} {\bibinfo  {journal}
  {Phys. Rev. B}\ }\textbf {\bibinfo {volume} {100}},\ \bibinfo {pages}
  {125419} (\bibinfo {year} {2019})}\BibitemShut {NoStop}%
\bibitem [{\citenamefont {Apostolov}\ \emph {et~al.}(2019)\citenamefont
  {Apostolov}, \citenamefont {Pesin},\ and\ \citenamefont
  {Levchenko}}]{PhysRevB.100.115401}%
  \BibitemOpen
  \bibfield  {author} {\bibinfo {author} {\bibfnamefont {S.~S.}\ \bibnamefont
  {Apostolov}}, \bibinfo {author} {\bibfnamefont {D.~A.}\ \bibnamefont
  {Pesin}}, \ and\ \bibinfo {author} {\bibfnamefont {A.}~\bibnamefont
  {Levchenko}},\ }\bibfield  {title} {\enquote {\bibinfo {title} {Magnetodrag
  in the hydrodynamic regime: Effects of magnetoplasmon resonance and {H}all
  viscosity},}\ }\href {\doibase 10.1103/PhysRevB.100.115401} {\bibfield
  {journal} {\bibinfo  {journal} {Phys. Rev. B}\ }\textbf {\bibinfo {volume}
  {100}},\ \bibinfo {pages} {115401} (\bibinfo {year} {2019})}\BibitemShut
  {NoStop}%
\bibitem [{\citenamefont {Hasdeo}\ \emph {et~al.}(2021)\citenamefont {Hasdeo},
  \citenamefont {Ekstr\"om}, \citenamefont {Idrisov},\ and\ \citenamefont
  {Schmidt}}]{PhysRevB.103.125106}%
  \BibitemOpen
  \bibfield  {author} {\bibinfo {author} {\bibfnamefont {Eddwi~H.}\
  \bibnamefont {Hasdeo}}, \bibinfo {author} {\bibfnamefont {Johan}\
  \bibnamefont {Ekstr\"om}}, \bibinfo {author} {\bibfnamefont {Edvin~G.}\
  \bibnamefont {Idrisov}}, \ and\ \bibinfo {author} {\bibfnamefont {Thomas~L.}\
  \bibnamefont {Schmidt}},\ }\bibfield  {title} {\enquote {\bibinfo {title}
  {Electron hydrodynamics of two-dimensional anomalous {H}all materials},}\
  }\href {\doibase 10.1103/PhysRevB.103.125106} {\bibfield  {journal} {\bibinfo
   {journal} {Phys. Rev. B}\ }\textbf {\bibinfo {volume} {103}},\ \bibinfo
  {pages} {125106} (\bibinfo {year} {2021})}\BibitemShut {NoStop}%
\bibitem [{\citenamefont {Funaki}\ \emph {et~al.}(2021)\citenamefont {Funaki},
  \citenamefont {Toshio},\ and\ \citenamefont
  {Tatara}}]{funaki2021vorticityinduced}%
  \BibitemOpen
  \bibfield  {author} {\bibinfo {author} {\bibfnamefont {Hiroshi}\ \bibnamefont
  {Funaki}}, \bibinfo {author} {\bibfnamefont {Riki}\ \bibnamefont {Toshio}}, \
  and\ \bibinfo {author} {\bibfnamefont {Gen}\ \bibnamefont {Tatara}},\
  }\href@noop {} {\enquote {\bibinfo {title} {Vorticity-induced anomalous Hall
  effect in electron fluid},}\ } \bibinfo {year} preprint arXiv:2103.00861 ({2021})
\bibitem{tatara1} {Gen Tatara, Hydrodynamic theory of vorticity-induced spin transport,
Phys. Rev. B {\bf 104}, 184414 (2021).}
  \BibitemShut {NoStop}%
\bibitem [{\citenamefont {Fuchs}(1938)}]{Fuchs:1938tn}%
  \BibitemOpen
  \bibfield  {author} {\bibinfo {author} {\bibfnamefont {K.}~\bibnamefont
  {Fuchs}},\ }\bibfield  {title} {\enquote {\bibinfo {title} {The conductivity
  of thin metallic films according to the electron theory of metals},}\
  }\bibfield  {booktitle} {\emph {\bibinfo {booktitle} {Mathematical
  Proceedings of the Cambridge Philosophical Society}},\ }\href {\doibase DOI:
  10.1017/S0305004100019952} {\ \textbf {\bibinfo {volume} {34}},\ \bibinfo
  {pages} {100--108} (\bibinfo {year} {1938})}\BibitemShut {NoStop}%
\bibitem [{\citenamefont {Reuter}\ and\ \citenamefont
  {Sondheimer}(1948)}]{sondheimer48}%
  \BibitemOpen
  \bibfield  {author} {\bibinfo {author} {\bibfnamefont {G.E.H.}\ \bibnamefont
  {Reuter}}\ and\ \bibinfo {author} {\bibfnamefont {E.H.}\ \bibnamefont
  {Sondheimer}},\ }\bibfield  {title} {\enquote {\bibinfo {title} {The theory
  of the anomalous skin effect in metals},}\ }\href@noop {} {\bibfield
  {journal} {\bibinfo  {journal} {Proc. Roy. Soc. A}\ }\textbf {\bibinfo
  {volume} {195}},\ \bibinfo {pages} {336} (\bibinfo {year}
  {1948})}\BibitemShut {NoStop}%
\bibitem [{\citenamefont {{Fal'kovskii}}(1970)}]{Falkovskii70}%
  \BibitemOpen
  \bibfield  {author} {\bibinfo {author} {\bibfnamefont {L.~A.}\ \bibnamefont
  {{Fal'kovskii}}},\ }\bibfield  {title} {\enquote {\bibinfo {title} {{Diffuse
  Boundary Condition for Conduction Electrons}},}\ }\href@noop {} {\bibfield
  {journal} {\bibinfo  {journal} {JETP Lett.}\ }\textbf {\bibinfo {volume}
  {11}},\ \bibinfo {pages} {138} (\bibinfo {year} {1970})}\BibitemShut
  {NoStop}%
%
\bibitem{ll10_eng} {L.D.~Landau and E.M.~Lifshitz,
\newblock {\em Physical Kinetics},
\newblock (Butterworth-Heinemann, Oxford, 1981).}%
%
\bibitem [{\citenamefont {Greene}(1966)}]{PhysRev.141.687}%
  \BibitemOpen
  \bibfield  {author} {\bibinfo {author} {\bibfnamefont {R.~F.}\ \bibnamefont
  {Greene}},\ }\bibfield  {title} {\enquote {\bibinfo {title} {Boundary
  conditions for electron distributions at crystal surfaces},}\ }\href
  {\doibase 10.1103/PhysRev.141.687} {\bibfield  {journal} {\bibinfo  {journal}
  {Phys. Rev.}\ }\textbf {\bibinfo {volume} {141}},\ \bibinfo {pages}
  {687--689} (\bibinfo {year} {1966})}\BibitemShut {NoStop}%
\bibitem [{\citenamefont {Andreev}(1972)}]{Andreev:1972wt}%
  \BibitemOpen
  \bibfield  {author} {\bibinfo {author} {\bibfnamefont {A.~F.}\ \bibnamefont
  {Andreev}},\ }\bibfield  {title} {\enquote {\bibinfo {title} {Interaction of
  conduction electrons with a metal surface},}\ }\bibfield  {journal} {
  {\bibinfo {journal} {{Sov. Phys.-Uspekhi}}},\ }\href {\doibase
  10.1070/pu1972v014n05abeh004680} {\ \textbf {\bibinfo {volume} {14}},\
  \bibinfo {pages} {609--615} (\bibinfo {year} {1972})}\BibitemShut {NoStop}%
\bibitem [{\citenamefont {Kiselev}\ and\ \citenamefont
  {Schmalian}(2019)}]{PhysRevB.99.035430}%
  \BibitemOpen
  \bibfield  {author} {\bibinfo {author} {\bibfnamefont {Egor~I.}\ \bibnamefont
  {Kiselev}}\ and\ \bibinfo {author} {\bibfnamefont {J\"org}\ \bibnamefont
  {Schmalian}},\ }\bibfield  {title} {\enquote {\bibinfo {title} {Boundary
  conditions of viscous electron flow},}\ }\href {\doibase
  10.1103/PhysRevB.99.035430} {\bibfield  {journal} {\bibinfo  {journal} {Phys.
  Rev. B}\ }\textbf {\bibinfo {volume} {99}},\ \bibinfo {pages} {035430}
  (\bibinfo {year} {2019})}\BibitemShut {NoStop}%
 %
\bibitem{ivchenko05a}
{E.~L. Ivchenko,
\newblock {\em Optical spectroscopy of semiconductor nanostructures},
\newblock (Alpha Science, Harrow UK, 2005).}
 %
 \bibitem{ll3_eng}
{L.~D. Landau and E.~M. Lifshitz,
\newblock {\em Quantum Mechanics: Non-Relativistic Theory},
\newblock (Butterworth-Heinemann, Oxford, 1977).}
 %
 \bibitem{daviesBOOK}
{J.~Davies,
\newblock {\em The physics of low-dimensional semiconductors}.
\newblock (Cambridge University Press, 1998).}
%
\bibitem [{\citenamefont {Karplus}\ and\ \citenamefont
  {Luttinger}(1954)}]{PhysRev.95.1154}%
  \BibitemOpen
  \bibfield  {author} {\bibinfo {author} {\bibfnamefont {Robert}\ \bibnamefont
  {Karplus}}\ and\ \bibinfo {author} {\bibfnamefont {J.~M.}\ \bibnamefont
  {Luttinger}},\ }\bibfield  {title} {\enquote {\bibinfo {title} {Hall effect
  in ferromagnetics},}\ }\href {\doibase 10.1103/PhysRev.95.1154} {\bibfield
  {journal} {\bibinfo  {journal} {Phys. Rev.}\ }\textbf {\bibinfo {volume}
  {95}},\ \bibinfo {pages} {1154--1160} (\bibinfo {year} {1954})}\BibitemShut
  {NoStop}%
\bibitem [{\citenamefont {Belinicher}\ \emph {et~al.}(1982)\citenamefont
  {Belinicher}, \citenamefont {Ivchenko},\ and\ \citenamefont
  {Sturman}}]{belinicher82}%
  \BibitemOpen
  \bibfield  {author} {\bibinfo {author} {\bibfnamefont {V.~I.}\ \bibnamefont
  {Belinicher}}, \bibinfo {author} {\bibfnamefont {E.~L.}\ \bibnamefont
  {Ivchenko}}, \ and\ \bibinfo {author} {\bibfnamefont {B.~I.}\ \bibnamefont
  {Sturman}},\ }\bibfield  {title} {\enquote {\bibinfo {title} {Kinetic theory
  of the displacement photovoltaic effect in piezoelectrics},}\ }\href
  {http://www.jetp.ac.ru/cgi-bin/e/index/e/56/2/p359?a=list} {\bibfield
  {journal} {\bibinfo  {journal} {JETP}\ }\textbf {\bibinfo {volume} {56}},\
  \bibinfo {pages} {359} (\bibinfo {year} {1982})}\BibitemShut {NoStop}%
\bibitem [{\citenamefont {Sturman}(2019)}]{Sturman2019}%
  \BibitemOpen
  \bibfield  {author} {\bibinfo {author} {\bibfnamefont {B.~I.}\ \bibnamefont
  {Sturman}},\ }\bibfield  {title} {\enquote {\bibinfo {title} {Ballistic and
  shift currents in the bulk photovoltaic effect theory},}\ }\href
  {https://doi.org/10.3367/ufne.2019.06.038578} {\bibfield  {journal} {\bibinfo
   {journal} {Physics-Uspekhi}\ }\textbf {\bibinfo {volume} {63}} (\bibinfo
  {year} {2019})}\BibitemShut {NoStop}%
\bibitem [{\citenamefont {Sundaram}\ and\ \citenamefont
  {Niu}(1999)}]{PhysRevB.59.14915}%
  \BibitemOpen
  \bibfield  {author} {\bibinfo {author} {\bibfnamefont {Ganesh}\ \bibnamefont
  {Sundaram}}\ and\ \bibinfo {author} {\bibfnamefont {Qian}\ \bibnamefont
  {Niu}},\ }\bibfield  {title} {\enquote {\bibinfo {title} {Wave-packet
  dynamics in slowly perturbed crystals: Gradient corrections and berry-phase
  effects},}\ }\href {\doibase 10.1103/PhysRevB.59.14915} {\bibfield  {journal}
  {\bibinfo  {journal} {Phys. Rev. B}\ }\textbf {\bibinfo {volume} {59}},\
  \bibinfo {pages} {14915--14925} (\bibinfo {year} {1999})}\BibitemShut
  {NoStop}%
\bibitem [{\citenamefont {Sekine}\ and\ \citenamefont
  {Nagaosa}(2020)}]{PhysRevB.101.155204}%
  \BibitemOpen
  \bibfield  {author} {\bibinfo {author} {\bibfnamefont {Akihiko}\ \bibnamefont
  {Sekine}}\ and\ \bibinfo {author} {\bibfnamefont {Naoto}\ \bibnamefont
  {Nagaosa}},\ }\bibfield  {title} {\enquote {\bibinfo {title} {Quantum kinetic
  theory of thermoelectric and thermal transport in a magnetic field},}\ }\href
  {\doibase 10.1103/PhysRevB.101.155204} {\bibfield  {journal} {\bibinfo
  {journal} {Phys. Rev. B}\ }\textbf {\bibinfo {volume} {101}},\ \bibinfo
  {pages} {155204} (\bibinfo {year} {2020})}\BibitemShut {NoStop}%
\bibitem [{\citenamefont {{K\"onig}}\ and\ \citenamefont
  {{Levchenko}}(2021)}]{2021arXiv210205675K}%
  \BibitemOpen
  \bibfield  {author} {\bibinfo {author} {\bibfnamefont {Elio~J.}\ \bibnamefont
  {{K\"onig}}}\ and\ \bibinfo {author} {\bibfnamefont {Alex}\ \bibnamefont
  {{Levchenko}}},\ }\href@noop {} {\enquote {\bibinfo {title} {{Quantum
  kinetics of anomalous and nonlinear Hall effects in topological
  semimetals}},}\ }\bibinfo {howpublished} {preprint arXiv:2102.05675}
  (\bibinfo {year} {2021})\BibitemShut {NoStop}%
\bibitem [{\citenamefont {Boguslawski}(1980)}]{boguslawski}%
  \BibitemOpen
  \bibfield  {author} {\bibinfo {author} {\bibfnamefont {P.}~\bibnamefont
  {Boguslawski}},\ }\bibfield  {title} {\enquote {\bibinfo {title}
  {Electron-electron spin-flip scattering and spin relaxation in {III-V} and
  {II-VI} semiconductors},}\ }\href@noop {} {\bibfield  {journal} {\bibinfo
  {journal} {Solid State Commun.}\ }\textbf {\bibinfo {volume} {33}},\ \bibinfo
  {pages} {389} (\bibinfo {year} {1980})}\BibitemShut {NoStop}%
\bibitem [{\citenamefont {\c{S}. C.~B\u{a}descu}\ \emph
  {et~al.}(2005)\citenamefont {\c{S}. C.~B\u{a}descu}, \citenamefont
  {Lyanda-Geller},\ and\ \citenamefont {Reinecke}}]{badescu:161304}%
  \BibitemOpen
  \bibfield  {author} {\bibinfo {author} {\bibnamefont {\c{S}.
  C.~B\u{a}descu}}, \bibinfo {author} {\bibfnamefont {Y.~B.}\ \bibnamefont
  {Lyanda-Geller}}, \ and\ \bibinfo {author} {\bibfnamefont {T.~L.}\
  \bibnamefont {Reinecke}},\ }\bibfield  {title} {\enquote {\bibinfo {title}
  {Asymmetric exchange between electron spins in coupled semiconductor quantum
  dots},}\ }\href {\doibase 10.1103/PhysRevB.72.161304} {\bibfield  {journal}
  {\bibinfo  {journal} {Phys. Rev. B}\ }\textbf {\bibinfo {volume} {72}},\
  \bibinfo {eid} {161304} (\bibinfo {year} {2005})}\BibitemShut {NoStop}%
\bibitem [{\citenamefont {Glazov}\ and\ \citenamefont
  {Kulakovskii}(2009)}]{glazov2009}%
  \BibitemOpen
  \bibfield  {author} {\bibinfo {author} {\bibfnamefont {M.~M.}\ \bibnamefont
  {Glazov}}\ and\ \bibinfo {author} {\bibfnamefont {V.~D.}\ \bibnamefont
  {Kulakovskii}},\ }\bibfield  {title} {\enquote {\bibinfo {title} {Spin-orbit
  effect on electron-electron interaction and the fine structure of electron
  complexes in quantum dots},}\ }\href {\doibase 10.1103/PhysRevB.79.195305}
  {\bibfield  {journal} {\bibinfo  {journal} {Phys. Rev. B}\ }\textbf {\bibinfo
  {volume} {79}},\ \bibinfo {eid} {195305} (\bibinfo {year}
  {2009})}\BibitemShut {NoStop}%
\bibitem [{\citenamefont {Badalyan}\ and\ \citenamefont
  {Vignale}(2009)}]{badalyan-2009}%
  \BibitemOpen
  \bibfield  {author} {\bibinfo {author} {\bibfnamefont {S.~M.}\ \bibnamefont
  {Badalyan}}\ and\ \bibinfo {author} {\bibfnamefont {G.}~\bibnamefont
  {Vignale}},\ }\bibfield  {title} {\enquote {\bibinfo {title} {{Spin Hall Drag
  in Electronic Bilayers}},}\ }\href {\doibase 10.1103/PhysRevLett.103.196601}
  {\bibfield  {journal} {\bibinfo  {journal} {Phys. Rev. Lett.}\ }\textbf
  {\bibinfo {volume} {103}},\ \bibinfo {eid} {196601} (\bibinfo {year}
  {2009})}\BibitemShut {NoStop}%
\bibitem [{\citenamefont {Glazov}(2010)}]{glazov2010}%
  \BibitemOpen
  \bibfield  {author} {\bibinfo {author} {\bibfnamefont {M~M}\ \bibnamefont
  {Glazov}},\ }\bibfield  {title} {\enquote {\bibinfo {title} {The fine
  structure of two-electron states in single and double quantum dots},}\ }\href
  {http://stacks.iop.org/0953-8984/22/025301} {\bibfield  {journal} {\bibinfo
  {journal} {Journal of Physics: Condensed Matter}\ }\textbf {\bibinfo {volume}
  {22}},\ \bibinfo {pages} {025301 (9pp)} (\bibinfo {year} {2010})}\BibitemShut
  {NoStop}%
\bibitem [{\citenamefont {Glazov}\ \emph {et~al.}(2011)\citenamefont {Glazov},
  \citenamefont {Semina}, \citenamefont {Badalyan},\ and\ \citenamefont
  {Vignale}}]{PhysRevB.84.033305}%
  \BibitemOpen
  \bibfield  {author} {\bibinfo {author} {\bibfnamefont {M.~M.}\ \bibnamefont
  {Glazov}}, \bibinfo {author} {\bibfnamefont {M.~A.}\ \bibnamefont {Semina}},
  \bibinfo {author} {\bibfnamefont {S.~M.}\ \bibnamefont {Badalyan}}, \ and\
  \bibinfo {author} {\bibfnamefont {G.}~\bibnamefont {Vignale}},\ }\bibfield
  {title} {\enquote {\bibinfo {title} {{Spin-current generation from
  Coulomb-Rashba interaction in semiconductor bilayers}},}\ }\href {\doibase
  10.1103/PhysRevB.84.033305} {\bibfield  {journal} {\bibinfo  {journal} {Phys.
  Rev. B}\ }\textbf {\bibinfo {volume} {84}},\ \bibinfo {pages} {033305}
  (\bibinfo {year} {2011})}\BibitemShut {NoStop}%
\bibitem [{\citenamefont {Pesin}(2018)}]{PhysRevLett.121.226601}%
  \BibitemOpen
  \bibfield  {author} {\bibinfo {author} {\bibfnamefont {D.~A.}\ \bibnamefont
  {Pesin}},\ }\bibfield  {title} {\enquote {\bibinfo {title} {{Two-Particle
  Collisional Coordinate Shifts and Hydrodynamic Anomalous Hall Effect in
  Systems without Lorentz Invariance}},}\ }\href {\doibase
  10.1103/PhysRevLett.121.226601} {\bibfield  {journal} {\bibinfo  {journal}
  {Phys. Rev. Lett.}\ }\textbf {\bibinfo {volume} {121}},\ \bibinfo {pages}
  {226601} (\bibinfo {year} {2018})}\BibitemShut {NoStop}%
\bibitem [{\citenamefont {Chen}\ \emph {et~al.}(2014)\citenamefont {Chen},
  \citenamefont {Son}, \citenamefont {Stephanov}, \citenamefont {Yee},\ and\
  \citenamefont {Yin}}]{PhysRevLett.113.182302}%
  \BibitemOpen
  \bibfield  {author} {\bibinfo {author} {\bibfnamefont {Jing-Yuan}\
  \bibnamefont {Chen}}, \bibinfo {author} {\bibfnamefont {Dam~T.}\ \bibnamefont
  {Son}}, \bibinfo {author} {\bibfnamefont {Mikhail~A.}\ \bibnamefont
  {Stephanov}}, \bibinfo {author} {\bibfnamefont {Ho-Ung}\ \bibnamefont {Yee}},
  \ and\ \bibinfo {author} {\bibfnamefont {Yi}~\bibnamefont {Yin}},\ }\bibfield
   {title} {\enquote {\bibinfo {title} {Lorentz invariance in chiral kinetic
  theory},}\ }\href {\doibase 10.1103/PhysRevLett.113.182302} {\bibfield
  {journal} {\bibinfo  {journal} {Phys. Rev. Lett.}\ }\textbf {\bibinfo
  {volume} {113}},\ \bibinfo {pages} {182302} (\bibinfo {year}
  {2014})}\BibitemShut {NoStop}%
\bibitem [{\citenamefont {Glazov}\ and\ \citenamefont
  {Ivchenko}(2002)}]{glazov02}%
  \BibitemOpen
  \bibfield  {author} {\bibinfo {author} {\bibfnamefont {M.~M.}\ \bibnamefont
  {Glazov}}\ and\ \bibinfo {author} {\bibfnamefont {E.~L.}\ \bibnamefont
  {Ivchenko}},\ }\bibfield  {title} {\enquote {\bibinfo {title} {Precession
  spin relaxation mechanism caused by frequent electron--electron
  collisions},}\ }\href@noop {} {\bibfield  {journal} {\bibinfo  {journal}
  {JETP Letters}\ }\textbf {\bibinfo {volume} {75}},\ \bibinfo {pages} {403}
  (\bibinfo {year} {2002})}\BibitemShut {NoStop}%
\bibitem [{\citenamefont {D\char39{}Amico}\ and\ \citenamefont
  {Vignale}(2002)}]{amico02}%
  \BibitemOpen
  \bibfield  {author} {\bibinfo {author} {\bibfnamefont {Irene}\ \bibnamefont
  {D\char39{}Amico}}\ and\ \bibinfo {author} {\bibfnamefont {Giovanni}\
  \bibnamefont {Vignale}},\ }\bibfield  {title} {\enquote {\bibinfo {title}
  {Coulomb interaction effects in spin-polarized transport},}\ }\href {\doibase
  10.1103/PhysRevB.65.085109} {\bibfield  {journal} {\bibinfo  {journal} {Phys.
  Rev. B}\ }\textbf {\bibinfo {volume} {65}},\ \bibinfo {pages} {085109}
  (\bibinfo {year} {2002})}\BibitemShut {NoStop}%
\bibitem [{\citenamefont {D'Amico}\ and\ \citenamefont
  {Vignale}(2003)}]{amico:045307}%
  \BibitemOpen
  \bibfield  {author} {\bibinfo {author} {\bibfnamefont {Irene}\ \bibnamefont
  {D'Amico}}\ and\ \bibinfo {author} {\bibfnamefont {Giovanni}\ \bibnamefont
  {Vignale}},\ }\bibfield  {title} {\enquote {\bibinfo {title} {Spin {C}oulomb
  drag in the two-dimensional electron liquid},}\ }\href
  {http://link.aps.org/abstract/PRB/v68/e045307} {\bibfield  {journal}
  {\bibinfo  {journal} {Phys. Rev. B}\ }\textbf {\bibinfo {volume} {68}},\
  \bibinfo {eid} {045307} (\bibinfo {year} {2003})}\BibitemShut {NoStop}%
\bibitem [{\citenamefont {Glazov}\ and\ \citenamefont
  {Ivchenko}(2004)}]{glazov04a}%
  \BibitemOpen
  \bibfield  {author} {\bibinfo {author} {\bibfnamefont {M.~M.}\ \bibnamefont
  {Glazov}}\ and\ \bibinfo {author} {\bibfnamefont {E.~L.}\ \bibnamefont
  {Ivchenko}},\ }\bibfield  {title} {\enquote {\bibinfo {title} {Effect of
  electron-electron interaction on spin relaxation of charge carriers in
  semiconductors},}\ }\href@noop {} {\bibfield  {journal} {\bibinfo  {journal}
  {JETP}\ }\textbf {\bibinfo {volume} {99}},\ \bibinfo {pages} {1279} (\bibinfo
  {year} {2004})}\BibitemShut {NoStop}%
\bibitem [{\citenamefont {Weber}\ \emph {et~al.}(2005)\citenamefont {Weber},
  \citenamefont {Gedik}, \citenamefont {Moore}, \citenamefont {Orenstein},
  \citenamefont {Stephens},\ and\ \citenamefont {Awschalom}}]{weber05}%
  \BibitemOpen
  \bibfield  {author} {\bibinfo {author} {\bibfnamefont {C.~P.}\ \bibnamefont
  {Weber}}, \bibinfo {author} {\bibfnamefont {N.}~\bibnamefont {Gedik}},
  \bibinfo {author} {\bibfnamefont {J.~E.}\ \bibnamefont {Moore}}, \bibinfo
  {author} {\bibfnamefont {J.}~\bibnamefont {Orenstein}}, \bibinfo {author}
  {\bibfnamefont {J.}~\bibnamefont {Stephens}}, \ and\ \bibinfo {author}
  {\bibfnamefont {D.~D.}\ \bibnamefont {Awschalom}},\ }\bibfield  {title}
  {\enquote {\bibinfo {title} {Observation of spin {C}oulomb drag in a two
  dimensional electron gas},}\ }\href@noop {} {\bibfield  {journal} {\bibinfo
  {journal} {Nature}\ }\textbf {\bibinfo {volume} {437}},\ \bibinfo {pages}
  {1330} (\bibinfo {year} {2005})}\BibitemShut {NoStop}%
\bibitem [{\citenamefont {Glazov}\ and\ \citenamefont
  {Ivchenko}(2003)}]{glazovnato}%
  \BibitemOpen
  \bibfield  {author} {\bibinfo {author} {\bibfnamefont {M.~M.}\ \bibnamefont
  {Glazov}}\ and\ \bibinfo {author} {\bibfnamefont {E.~L.}\ \bibnamefont
  {Ivchenko}},\ }\enquote {\bibinfo {title} {{D'yakonov-Perel' Spin Relaxation
  under Electron-Electron Collisions In QWs}},}\ in\ \href@noop {} {\emph
  {\bibinfo {booktitle} {Optical Properties of 2D Systems with Interacting
  Electrons}}},\ \bibinfo {editor} {edited by\ \bibinfo {editor} {\bibfnamefont
  {W.~J.}\ \bibnamefont {Ossau}}\ and\ \bibinfo {editor} {\bibfnamefont
  {R.}~\bibnamefont {Suris}}}\ (\bibinfo  {publisher} {Springer},\ \bibinfo
  {year} {2003})\ p.\ \bibinfo {pages} {181}\BibitemShut {NoStop}%
\bibitem [{\citenamefont {Gantmakher}\ and\ \citenamefont
  {Levinson}(1987)}]{gantmakher87}%
  \BibitemOpen
  \bibfield  {author} {\bibinfo {author} {\bibfnamefont {V.~F.}\ \bibnamefont
  {Gantmakher}}\ and\ \bibinfo {author} {\bibfnamefont {Y.~B.}\ \bibnamefont
  {Levinson}},\ }\href@noop {} {\emph {\bibinfo {title} {Carrier Scattering in
  Metals and Semiconductors}}}\ (\bibinfo  {publisher} {North-Holland
  Publishing Company},\ \bibinfo {year} {1987})\BibitemShut {NoStop}%
\bibitem [{\citenamefont {Chaplik}(1971)}]{chaplik71eng}%
  \BibitemOpen
  \bibfield  {author} {\bibinfo {author} {\bibfnamefont {A.~V.}\ \bibnamefont
  {Chaplik}},\ }\bibfield  {title} {\enquote {\bibinfo {title} {Energy spectrum
  and electron scattering processes in inversion layers},}\ }\href@noop {}
  {\bibfield  {journal} {\bibinfo  {journal} {JETP}\ }\textbf {\bibinfo
  {volume} {33}},\ \bibinfo {pages} {997} (\bibinfo {year} {1971})}\BibitemShut
  {NoStop}%
\bibitem [{\citenamefont {Giuliani}\ and\ \citenamefont
  {Quinn}(1982)}]{Giuliani82}%
  \BibitemOpen
  \bibfield  {author} {\bibinfo {author} {\bibfnamefont {Gabriele~F.}\
  \bibnamefont {Giuliani}}\ and\ \bibinfo {author} {\bibfnamefont {John~J.}\
  \bibnamefont {Quinn}},\ }\bibfield  {title} {\enquote {\bibinfo {title}
  {Lifetime of a quasiparticle in a two-dimensional electron gas},}\ }\href
  {\doibase 10.1103/PhysRevB.26.4421} {\bibfield  {journal} {\bibinfo
  {journal} {Phys. Rev. B}\ }\textbf {\bibinfo {volume} {26}},\ \bibinfo
  {pages} {4421--4428} (\bibinfo {year} {1982})}\BibitemShut {NoStop}%
\bibitem [{\citenamefont {Zheng}\ and\ \citenamefont
  {Das~Sarma}(1996)}]{PhysRevB.53.9964}%
  \BibitemOpen
  \bibfield  {author} {\bibinfo {author} {\bibfnamefont {Lian}\ \bibnamefont
  {Zheng}}\ and\ \bibinfo {author} {\bibfnamefont {S.}~\bibnamefont
  {Das~Sarma}},\ }\bibfield  {title} {\enquote {\bibinfo {title} {Coulomb
  scattering lifetime of a two-dimensional electron gas},}\ }\href {\doibase
  10.1103/PhysRevB.53.9964} {\bibfield  {journal} {\bibinfo  {journal} {Phys.
  Rev. B}\ }\textbf {\bibinfo {volume} {53}},\ \bibinfo {pages} {9964--9967}
  (\bibinfo {year} {1996})}\BibitemShut {NoStop}%
\bibitem [{\citenamefont {Alekseev}\ and\ \citenamefont
  {Dmitriev}(2020)}]{PhysRevB.102.241409}%
  \BibitemOpen
  \bibfield  {author} {\bibinfo {author} {\bibfnamefont {P.~S.}\ \bibnamefont
  {Alekseev}}\ and\ \bibinfo {author} {\bibfnamefont {A.~P.}\ \bibnamefont
  {Dmitriev}},\ }\bibfield  {title} {\enquote {\bibinfo {title} {Viscosity of
  two-dimensional electrons},}\ }\href {\doibase 10.1103/PhysRevB.102.241409}
  {\bibfield  {journal} {\bibinfo  {journal} {Phys. Rev. B}\ }\textbf {\bibinfo
  {volume} {102}},\ \bibinfo {pages} {241409} (\bibinfo {year}
  {2020})}\BibitemShut {NoStop}%
\bibitem [{\citenamefont {Sturman}(1984)}]{Sturman_1984}%
  \BibitemOpen
  \bibfield  {author} {\bibinfo {author} {\bibfnamefont {B.~I.}\ \bibnamefont
  {Sturman}},\ }\bibfield  {title} {\enquote {\bibinfo {title} {Collision
  integral for elastic scattering of electrons and phonons},}\ }\href {\doibase
  10.1070/pu1984v027n11abeh004122} {\bibfield  {journal} {\bibinfo  {journal}
  {Soviet Physics Uspekhi}\ }\textbf {\bibinfo {volume} {27}},\ \bibinfo
  {pages} {881--884} (\bibinfo {year} {1984})}\BibitemShut {NoStop}%
%
\bibitem{de2013non}
{S.~De~Groot and P.~Mazur,
\newblock {\em Non-Equilibrium Thermodynamics},
\newblock (Dover Books on Physics. Dover Publications, 2013).}
%
\bibitem{Doornenbal_2019}
{R.~J. Doornenbal, M.~Polini, and R.~A. Duine,
\newblock Spin{\textendash}vorticity coupling in viscous electron fluids,
\newblock {Journal of Physics: Materials} {\bf 2}, 015006 (2019).}
%
\bibitem{PhysRevB.96.020401}
{M.~Matsuo, Y.~Ohnuma, and S.~Maekawa,
\newblock Theory of spin hydrodynamic generation,
\newblock {Phys. Rev. B} {\bf 96}, 020401 (2017).}
%
\bibitem{Takahashi:2020um}
{R.~Takahashi, H.~Chudo, M.~Matsuo, K.~Harii, Y.~Ohnuma, S.~Maekawa, and
  E.~Saitoh,
\newblock Giant spin hydrodynamic generation in laminar flow.
\newblock {Nature Communications} {\bf 11}, 3009 (2020).}

\end{thebibliography}
\end{document}